\documentclass[fleqn,usenatbib]{mnras}

\usepackage{newtxtext,newtxmath}

\usepackage[T1]{fontenc}
\usepackage{subcaption}
\DeclareRobustCommand{\VAN}[3]{#2}
\let\VANthebibliography\thebibliography
\def\thebibliography{\DeclareRobustCommand{\VAN}[3]{##3}\VANthebibliography}

\usepackage{graphicx}	
\usepackage{amsmath}	

\title[Cosmological constraining power of WL peaks]{Cosmological constraining power of the redshifts, heights, and angular clustering of weak gravitational lensing peaks}

\author[J. C. Broxterman et al.]{
Jeger C. Broxterman,$^{1,2}$\thanks{E-mail: broxterman@lorentz.leidenuniv.nl}
Matthieu Schaller,$^{1,2}$
Ian G. McCarthy,$^{3}$
Willem Elbers,$^{4}$
John Helly,$^{4}$
\newauthor
Henk Hoekstra,$^{2}$
Konrad Kuijken,$^{2}$
Jaime Salcido,$^{3}$
Joop Schaye,$^{2}$
Naomi Schutte$^{2}$
and Elena Sellentin$^{2,5}$
\\
$^{1}$Lorentz Institute for Theoretical Physics, Leiden University, PO Box 9506, NL-2300 RA Leiden, the Netherlands \\
$^{2}$Leiden Observatory, Leiden University, PO Box 9513, NL-2300 RA Leiden, The Netherlands \\
$^{3}$Astrophysics Research Institute, Liverpool John Moores University, Liverpool L3 5RF, UK \\
$^{4}$Institute for Computational Cosmology, Department of Physics, University of Durham, South Road, Durham DH1 3LE, UK \\
$^{5}$Mathematical Institute, Leiden University, PO Box 9512, NL-2300 RA Leiden, The Netherlands \\
}

\date{Accepted XXX. Received YYY; in original form ZZZ}

\pubyear{\the\year{}}

\begin{document}
\label{firstpage}
\pagerange{\pageref{firstpage}--\pageref{lastpage}}
\maketitle

\begin{abstract}
Weak gravitational lensing (WL) peaks probe non-Gaussian information of the large-scale distribution of matter that is not captured by two-point lensing statistics. We study the cosmological potential of the height distribution, redshift distribution, and angular clustering of high-valued WL peaks using a Bayesian inference approach that mimics a $\textit{Euclid}$ analysis. We use a forthcoming dark-matter-only hypercube, which varies cosmology in a ten-dimensional space, including evolving dark energy, neutrino mass, decaying dark matter, and the running of the scalar spectral index. We find the individual WL peak statistics to be complementary, as the redshift distribution best constrains the matter density, $\Omega_\mathrm{m}$, and the dark energy equation-of-state parameters, $w_0$, and $w_a$; the height distribution and angular clustering are most sensitive to the amplitude of the primordial power spectrum, $\ln(10^{10}A_\mathrm{s})$; while combining the three statistics allows us to also probe the baryon density, $\Omega_\mathrm{b}$, and the Hubble parameter, $h$. A comparison to the shear two-point correlation function demonstrates that WL peaks alone outperform the commonly used statistics, while even tighter constraints are obtained when combining all. Considering the 2-dimensional $\Omega_\mathrm{m}-\ln(10^{10}A_\mathrm{s})$ and $w_0-w_a$ parameter planes, we find that the redshift distribution outperforms the angular clustering and peak-height distributions. We study the impact of the smoothing scale and find that, typically, the smallest scales yield the best results and the figure of merit improves by a factor $\approx2$ when combining multiple scales.
\end{abstract}



\section{Introduction}

The concordance model of cosmology, $\Lambda$CDM, is able to describe a plethora of observations with only a very limited number of free parameters and main model ingredients, namely dark energy (DE, $\Lambda$ if its energy density is constant with cosmic time), cold dark matter (CDM), baryons, and inflation \citep[for a recent review, see e.g.][]{Lahav2024}. However, as current surveys become more precise, hints of deviations from this model are emerging. Specifically, recent DESI results, which constrain cosmology through baryonic acoustic oscillations (BAO) and the clustering of galaxies, quasars, and the Lyman-alpha forest, provide hints for evolving DE \citep[][]{DESI2025_DR1_fullshape,DESI2025_DR2cosmo}. 

Similarly, in its recent Year 6 inference, the Dark Energy Survey (DES), following the Chevallier-Polarski-Linder (CPL) $w_0-w_a$ parametrisation \citep[][]{Chevallier2001,Linder2003}, finds that the best-fitting equation of state of dark energy slightly deviates from the $\Lambda$CDM expectation. Although their best-fit measurement from the combination of weak gravitational lensing (WL) and galaxy clustering (3\texttimes2pt) is still consistent with no evolving dark energy, when they add their BAO and supernova (SNe) measurements, they find a 2.2$\sigma$ deviation from a cosmological constant \citep[][]{DES2026_wowa}.

One of the main cosmological probes for studying the time evolution of DE is cosmic shear, the coherent distortion of the images of background galaxies caused by gravitational lensing by the large-scale structure of the Universe \citep[see e.g.][]{Hoekstra2008}. As the stage III lensing surveys, the Kilo-degree Survey \citep[KiDS;][]{deJong2013}, DES \citep[][]{DES2005}, and Hyper Suprime-Cam \citep[HSC;][]{HSC2012}, are in the final stages of their (extended) cosmological analyses \citep[e.g.][]{Reischke2026}, the next generation is already underway as \textit{Euclid} \citep[][]{Mellier2025_EuclidSkyOverview} is carrying out its first cosmological inference, \textit{Rubin}'s LSST has started \citep[][]{LSST2009}, and \textit{Roman} has launched \citep[][]{Spergel2015}. Combined, these instruments are expected to exhaust the sky best suited to cosmic shear, while combining deep observations with high-quality imaging. 

Stage III surveys have consistently measured lower $S_8 \equiv \sigma_8(\Omega_\mathrm{m}/0.3)^{1/2}$ values than inferred by \textit{Planck}. Although the extent of these differences is smaller than previously reported \citep[for a recent review, see][]{Pantos2026}, next-generation surveys are needed to conclusively address these discrepancies and identify their origins. \textit{Euclid} alone is expected to provide per cent-level constraints on the equation-of-state parameters of DE \citep[][]{Canas2025}. Comparing the results of these different probes and surveys should help us understand the minor discrepancies currently observed. Extracting all cosmological information from the observed data will be key in resolving these differences.

The common statistics used by cosmic shear surveys to constrain cosmology are 2-point projections of the matter power spectrum \citep[see e.g.][]{Wright2025,DESyr62026}, which quantify the variance of the matter density contrast field as a function of scale. However, more precise constraints may be obtained by including additional observables that probe non-Gaussian information not captured by traditional two-point measurements \citep[e.g.][]{Ajani_2023_HOWLS,Vinciguerra2025}. Examples of these statistics are the WL scattering transform \citep[e.g.][]{Cheng2020}, Minkowski functionals \citep[e.g. ][]{Marques2019}, Betti numbers \citep[e.g.][]{Feldbrugge2019}, wavelet phase harmonics \citep[][]{Allys2020}, persistent homology \citep[e.g.][]{Heydenreich2021}, or three-point statistics \citep[e.g.][]{Linke2026}.

This paper focuses on another higher-order weak lensing statistic, WL peaks, the local maxima in WL convergence fields \citep[e.g.][]{Jain2000,Wang2009}, which typically correspond to the most massive structures in the Universe that have experienced the strongest non-linear collapse \citep[][]{Hamana2004,Yang2011}. It has been shown that these peaks contain cosmological information in their height distribution \citep[][]{Dietrich2010,Kratochvil2010,Broxteman2024_WLpeaks}, steepness \citep[][]{Li2023}, angular clustering \citep[][]{Marian2013,Davies2019}, and redshift distribution \citep[][]{Broxterman2025_WLpeaks}. At the same time, the measurement is only first-order in the observed shear and could therefore respond differently to systematics than two-point statistics. Combined, the different WL peak properties could probe information similar to that from traditional galaxy cluster cosmological inferences, as they are sensitive to the mass and location of massive structures. One major advantage is that the peaks are WL- and thus mass-selected and therefore do not suffer from selection effects that require assumptions about the dynamical state of massive structures, or the form and scatter of mass-to-observable scaling relations, which typically limit X-ray- or thermal-Sunyaev Zel'dovich-selected cluster abundance inferences \citep[see e.g.][]{Kugel2024,Marini2024,Kugel2025}. 

In a cosmological forecast, \citet{Zurhcer2021} showed that for a stage-III-like WL survey, including the peak height distribution can already increase the figure of merit (FoM) in the $\Omega_\mathrm{m}-\sigma_8$ plane by a factor of 4 compared to the angular power spectrum alone. Similarly, \citet{Davies2022} find that the angular clustering of WL peaks is significantly more sensitive to the cosmological parameters $h$ and $w_0$ than the peak number density distribution, and when the two properties are combined, the constraining power on $\Omega_\mathrm{m}, S_\mathrm{8}, h,$ and $w_\mathrm{0}$ more than doubles. In this paper, we aim to extend this work by conducting a cosmological forecast for high-signal-to-noise ratio (SNR) WL peaks for a Stage IV WL cosmological inference, combining different WL peak properties to study their complementarity. We do this by using a new set of simulations in a ten-dimensional (10D) cosmological parameter space that includes evolving dark energy, decaying dark matter (DDM), and a running scalar spectral index in (700~cMpc)$^3$ volumes (McCarthy~et~al.,~in~prep). We include the number densities, redshift distribution, and angular clustering of high-valued WL peaks. 

In Section~\ref{sec:theory}, we introduce the relevant WL theory to infer WL signals from a general matter distribution. In Section~\ref{sec:sims}, we introduce the hypercube simulations. In Section~\ref{sec:methods}, we demonstrate how WL signals are inferred from the simulation, which peak signals are selected, and present the details of the Bayesian likelihood forecast. The results and discussion are given in Section~\ref{sec:rend}, and the final conclusions in Section~\ref{sec:conclusions}.

\section{Theory} \label{sec:theory}
In this section, we summarise the main equations to estimate the WL signal from a matter density distribution. For extended reviews of WL and cosmic shear, see \citet{Bartelmann2001} and \citet{Kilbinger2015}. We assume a spatially flat universe such that the comoving angular diameter distance, $f_\mathrm{K}(\chi)$, reduces to the comoving distance, $\chi$. In WL, photon deflections are described via the lensing equation, which relates the unlensed angular position, $\boldsymbol{\beta}$, to the observed angular position $\boldsymbol{\theta}$ as
\begin{align}
    \boldsymbol{\beta} = \boldsymbol{\theta} - \nabla_{\boldsymbol{\theta}}\,\psi(\boldsymbol{\theta}, \chi),
\end{align}
where $\psi$ is the lensing potential. We assume the Born approximation, i.e., we evaluate the lensing quantities on unperturbed photon paths, such that $\psi$ is given by 
\begin{align}\label{eqn:lensing_pot}
    \psi(\boldsymbol{\theta},\chi) = \frac{2}{c^2} \int_0^\chi \mathrm{d}\chi' \, \frac{\chi-\chi'}{\chi\chi'}\,\phi(\boldsymbol{\theta}\,\chi',\chi'),
\end{align}
where $c$ is the speed of light in vacuum and $\phi$ is the Newtonian gravitational potential. Conventionally, the lens equation is linearised to first order in the lensing potential, and the effect is quantified in terms of the magnification matrix, $A$, as 
\begin{align}
    \frac{\partial\beta_i}{\partial\theta_j} \equiv A_{ij} = \delta_{ij} - \partial_i\partial_j\psi,
\end{align}
where the magnification matrix is decomposed as 
\begin{align}\label{eqn:magmat_matrix}
\begin{split}
    A = \begin{pmatrix}
1-\kappa - \gamma_1 & -\gamma_2  \\
-\gamma_2  & 1 - \kappa + \gamma_1 
\end{pmatrix},
\end{split}
\end{align}
where $\gamma = \gamma_1 + i\gamma_2$ is the WL shear and $\kappa$ is the WL convergence, which is used here to quantify the cosmic shear strength. Equations~\ref{eqn:lensing_pot}--\ref{eqn:magmat_matrix} can be combined to express the WL convergence as
\begin{equation}\label{eqn:kappa_with_theta}
    \kappa(\boldsymbol{\theta},\chi) = \frac{1}{c^2} \int_0^\chi\mathrm{d}\chi' \frac{\chi - \chi'}{\chi'\chi} \nabla_{\boldsymbol{\theta}}^2\phi(\chi'\boldsymbol{\theta},\chi').
\end{equation}
We use the 3D Poisson equation to estimate the deflection strength 
\begin{equation}
    \nabla^2\phi = 4\pi \overline{\rho_\mathrm{m}} a^2 G \delta \approx \chi^{-2} \nabla_{\boldsymbol{\theta}}^2\phi,
\end{equation}
where $G$ is the gravitational constant, $\overline{\rho_\mathrm{m}}$ is the mean matter density, $a$ the scale factor, and $\delta$ the matter density contrast, and, in the final step, we assume that the contributions along the line of sight cancel, i.e. the Limber approximation \citep[][]{Limber1953}. To account for a population of source galaxies, we integrate Eq.~\ref{eqn:kappa_with_theta} as
\begin{equation}\label{eqn:kappa_int_cont}
    \kappa(\boldsymbol{\theta}) = \int_0^{z_\mathrm{hor}} \mathrm{d}z \, n_\mathrm{s}(z)\, \kappa[\boldsymbol{\theta},\chi(z)],
\end{equation}
where the normalised source redshift distribution, $n_\mathrm{s}$, extends to the edge of the survey, $z_\mathrm{hor} = 3$ in our case. We assume a \textit{Euclid}-like source redshift distribution given by \citep[][]{Blanchard2020_Euclid}
\begin{align}
    n(z) \propto \bigg(\frac{z}{z_0}\bigg)^2 \exp\bigg[-\bigg(\frac{z}{z_0}\bigg)^{3/2}\bigg],
\end{align}
with $z_0 = 0.9/\sqrt2$, such that the median redshift is 0.9.

Our analysis ignores several higher-order WL effects, contaminants, and the possibly non-negligible impact of approximations, such as the observable being shear rather than reduced shear \citep[e.g.][]{Seitz1995,Bradac2005}, source clustering \citep[e.g.][]{Gatti2024}, intrinsic alignments \citep[IA; e.g.][]{Lee2026,Vedder2026}, and the Born approximation \citep[e.g.][]{Lu2021}. We also do not marginalise over nuisance parameters such as magnification bias \citep[e.g.][]{Liu2014} or shifts in the mean or shape of the redshift distribution \citep[e.g.][]{Hildebrandt2020}. However, we assume these effects do not affect the relative differences we study in our idealised inference. Nevertheless, we stress that any actual inference should consider these effects in greater detail.  
While a quantitative assessment of these systematics requires a more elaborate study, we can draw some qualitative expectations from their known dependencies. Baryonic feedback suppresses the amplitude and abundance of the highest-convergence WL peaks, since these are sourced by the most massive collapsed structures. Because the redshift and height distributions in this work are built from the same underlying peak population, they are likely to respond to baryonic suppression in a correlated way, as a systematic reduction in the number and amplitude of the most massive haloes would suppress both statistics coherently. However, as shown in \citet{Broxterman2025_WLpeaks}, baryonic feedback primarily changes the amplitude of the redshift distribution and not its shape, which is affected by cosmology. Similarly, angular clustering at fixed WL peak number density is primarily sensitive to the spatial distribution of peaks rather than their amplitude and is thus naively expected to be less sensitive to baryonic feedback. Because we work directly with the convergence field from mass maps, IA, which arises from correlated intrinsic source-galaxy shapes, does not enter our analysis. For the height distribution, \citet{Zhang2022} show that IA impacts the height distribution in a non-negligible way. Given that our redshift assignment distribution is based on properties from the observed KiDS bright sample (as explained in more detail in Section~4.2), which includes IA, the redshift distribution is expected to be less impacted by this systematic.

\section{Simulations} \label{sec:sims}
In our analysis, we use a new set of simulations, which will be presented in detail in McCarthy~et~al.~(in preparation). We use two sets of 50 spatially flat simulations in 10D and 9D cosmological parameter spaces. The initial conditions have identical phases, and the amplitudes of the large-scale modes are fixed to the mean variance. The parameters that are varied are detailed in Table~\ref{tab:hypercube}, and include the present-day matter density parameter in the limit of no DDM, $\Omega_\mathrm{m,noDDM}$, i.e. the value that  $\Omega_\mathrm{m}$ would have been if the simulation had no DDM, the cosmic baryon fraction given by the ratio of the redshift zero baryon density to matter density parameters, $f_\mathrm{b}$, the Hubble parameter, $h$, the scalar spectral index, $n_\mathrm{s}$, the amplitude  $\ln (10^{10}A_\mathrm{s})$, and the running of the scalar spectral index, $\alpha_\mathrm{s} \equiv \mathrm{d}n_\mathrm{s} /\mathrm{d} \ln k$, of the primordial power spectrum, the dark energy equation-of-state parameters following the CPL parametrization, $w_0$ and $w_a$, the sum of 3 massive neutrino species, $\sum m_\nu c^2\, [\mathrm{eV}]$, and the decay rate of decaying cold dark matter, $\Gamma_\mathrm{dcdm} [\mathrm{km/s/Mpc}]$ following the model introduced in \citet{Elbers2025}. In the DDM extension, a fraction of the cold dark matter is treated as an unstable species with mean lifetime $\tau$, decaying into a relativistic dark radiation component. $\Gamma_{\mathrm{dcdm}} = \tau^{-1}$ is the decay rate. The decay is implemented by adjusting the particle masses in proportion to the fraction of dark matter that has decayed. 

The sampling within the 10D hypercube follows a Latin hypercube design, which is approximately uniform in the marginal distribution of each parameter individually, with the exception of $w_0$ and $w_a$. $w_0$ and $w_a$ are not sampled uniformly independently, as large parts of the $w_0-w_a$ parameter plane are ruled out by various observations. Instead, they are sampled along a degeneracy direction that is approximately parallel to the geometric degeneracy from the cosmic microwave background (CMB), DESI BAO and SNe \citep[][]{DESI2025_DR2cosmo}. The displacement along the degeneracy direction is varied by $\pm$ 0.5, which brackets all current constraints from the individual probes. The 9D hypercube can be viewed as a projection of the 10D hypercube onto the subspace where the DDM rate is fixed to zero, while the remaining nine parameter values are identical in both hypercubes. Combining the two hypercubes results in a joint 100-node training set in which the $\Gamma_\mathrm{dcdm}$ = 0 hyperplane is sampled more densely than any other value of $\Gamma_\mathrm{dcdm}$. In our forecasts, we always jointly use these two hypercube sets to train the emulators. 

Each hypercube node corresponds to a single simulation realisation, with no repeated realisations at fixed cosmology to sample cosmic variance or shape noise. While we find this is sufficient to obtain unbiased constraints in our simulation forecast using mode fixing and inferring the cosmology from a simulation run within the same framework, it may not be sufficient for an inference on real data.

\begin{table}
	\centering
	\caption{Varied parameters and their ranges in the hypercube. From top to bottom, the present-day total matter density parameter in the limit of no DDM, $\Omega_\mathrm{m,noDDM}$, the baryon fraction, $f_\mathrm{b}$; the dimensionless Hubble parameter, $h$; the scalar spectral index, $n_\mathrm{s}$, and the amplitude, $\ln(10^{10}A_\mathrm{s})$, of the primordial power spectrum; the CPL equation-of-state parameters of dark energy $w_0$ and $w_a$; the sum of neutrino masses, $\sum m_\nu c^2$ [eV]; the running of the spectral index, $\alpha_\mathrm{s}$ of the primordial power spectrum; and the DDM decay rate, $\Gamma_{\mathrm{dcdm}}$ [km/s/Mpc].}
	\begin{tabular}{llll} 
		\hline
		Parameter & Hypercube range \\
		\hline
		$\Omega_\mathrm{m,noDDM}$ & [0.2,\,0.4] \\
		$f_\mathrm{b}$ & [0.14,\,0.175] \\
		$h$ & [0.6,\,0.8] \\
        $n_\mathrm{s}$ & [0.9,\,1.0] \\
        $\ln(10^{10}A_\mathrm{s})$ & [2.965,\,3.115] \\
        $w_0$ & [-1.2,\,-0.3]\\
        $w_a$ & [-3.7($w_0$+1.0),\,-3.7($w_0$+1.0)+0.5] \\
        $\sum m_\nu c^2~ \mathrm{[eV]}$ & [0,\,0.45] \\
        $\alpha_\mathrm{s}$ & [-0.03,\,0.03] \\
        $\Gamma_{\mathrm{dcdm}}~\mathrm{[km/s/Mpc]}$ & [0,\,20] \\         
		\hline
	\end{tabular}\label{tab:hypercube}
\end{table}

The hypercube simulations were run using the same set-up as the original FLAMINGO simulations \citep[][]{Schaye2023_flamingo,Kugel2023_flamingo}, only in smaller volumes. The simulations were run using \textsc{swift} \citep[][]{Schaller2024_swift} and each contains $1260^3$ DM particles and $700^3$ massive neutrino particles in a box with a side length of $700$~cMpc, which results in a typical initial DM particle mass of a few $\times\, 10^9~\mathrm{M_\odot}$, depending on the exact values of the cosmological parameters. The primordial spectrum, depending on $\ln(10^{10}A_\mathrm{s}$), $n_\mathrm{s}$, and $\alpha_\mathrm{s}$ together with the transfer functions computed with \textsc{class} \citep[][]{Lesgourgues2011}, sets the linear matter power spectrum used to generate the initial conditions using \textsc{monofonic} \citep[][]{Hahn2020_monofonic,Michaux2021_monofonic}. Each simulation has a virtual observer with a spherical full-sky light-cone from which we use the matter shells.

We ignore the impact of baryonic physics, i.e., the redistribution of matter due to black hole feedback, supernova explosions, or other physical processes pertaining to baryons that impact the distribution of matter on (mildly) non-linear scales, even though this is one of the dominant systematics in WL inferences \citep[see e.g.,][]{Semboloni2011,Semboloni2013}. In \citet{Broxteman2024_WLpeaks,Broxterman2025_WLpeaks}, we found that the WL peak height distribution and redshift distribution depend differently on cosmology and baryonic physics, and combining the two statistics helps to simultaneously constrain cosmology and calibrate baryonic physics. In this forecast, we also include the angular clustering of these peaks, which will depend in its own way on baryonic physics, and therefore may improve the baryonic feedback calibration. At the same time, we stress that this should be tested using hydrodynamical simulations and that we do not expect this to affect the relative differences between our results for the constraining power of the WL peak properties.

\section{Methods} \label{sec:methods}
In this section, we present the pipeline for deriving full-sky WL convergence maps from maps of the matter density, and the subsequent measurements of the WL peak statistics and WL shear two-point correlation functions from these maps. Afterwards, we present the inference pipeline consisting of the emulator, likelihood, and covariance matrix, which will be used to infer cosmological constraints from an unseen simulation run. 

The construction of the WL convergence map closely follows the method described in \citet{Broxterman2025_WLpeaks}. We use the concentric \textsc{Healpix} \citep[][]{Gorski2005} light-cones at $N_\mathrm{side} =8192$ that are centred around a virtual observer at the same position in each simulation. The shells have a thickness of $\Delta z = 0.05$. Full-sky WL convergence maps are constructed from spherical matter shells by discretising Eq.~\ref{eqn:kappa_int_cont} as
\begin{align} \label{eq:kappa_from_sim}
   \kappa(\boldsymbol{\theta},\chi) = \frac{4\pi G}{c^2} \sum_{i} [1+z(\chi_i)]^{-2} \Delta \chi_i\, \delta_i(\boldsymbol{\theta})\, \Omega_{\mathrm{m}}(\chi_i)\, \rho_\mathrm{crit}(\chi_i) W(\chi_i),
\end{align}
$\rho_\mathrm{crit}$ is the critical energy density of the universe, $\Omega_{\mathrm{m}}(\chi_i) = \rho_\mathrm{m}(\chi_i)/\rho_\mathrm{crit}(\chi_i)$ is the value of the matter density parameter at $\chi_i$, whose time evolution is different from $w_0w_a$CDM as the simulations include DDM, and $\Delta \chi_i$ is the comoving thickness of light-cone shell $i$. We evaluate the matter density contrast, $\delta$, directly from the total matter shells. Each time the light-cone diameter exceeds an integer multiple of the box size, we apply a random rotation to all shells within the next integer multiple using the same rotation angles, thereby avoiding spurious discontinuities at the box boundary while still suppressing repeated large-scale structure along the line of sight. The WL kernel is given by
\begin{align}\label{eq:WL_kernel}
    W(\chi_i) = \chi_i \sum_{j=i}^{N_\mathrm{shells}} n_\mathrm{s}(\chi_j)\bigg(1-\frac{\chi_i}{\chi_j}\bigg)\Delta\chi_j,
\end{align}
with $n_\mathrm{s}(\chi) \mathrm{d} \chi = n_\mathrm{s}(z)\mathrm{d}z$ and which runs until $N_\mathrm{shells} = 60$. We mimic shape noise by assigning a randomly drawn value from a Gaussian with mean $\mu$ and standard deviation $\sigma$
\begin{align}
    \mathcal{N}\bigg\{\mu=0, \sigma = \frac{\sigma_\epsilon}{\sqrt{2n_\mathrm{gal} A_\mathrm{pix}}}\bigg\},
\end{align}
where $A_\mathrm{pix}$ is the area of one pixel ($0.18$ arcmin$^2$ for $N_\mathrm{side} = 8192$), $n_\mathrm{gal}$ is the source galaxy number density, and $\sigma_\epsilon$ is the root mean square total intrinsic ellipticity of source galaxies. To mimic a \textit{Euclid} survey, we take $n_\mathrm{gal}=30$~arcmin$^{-2}$, and $\sigma_\epsilon = 0.26\sqrt{2}$ \citep[][]{Mellier2025_EuclidSkyOverview}.

\subsection{Smoothing}
We explore the impact and complementarity of combining different smoothing scales. In our fiducial setup, similarly to \citet{Broxteman2024_WLpeaks,Broxterman2025_WLpeaks}, we smooth the integrated WL convergence map using a Gaussian kernel with a full width at half maximum of 1 arcmin, corresponding to 2.3 pixels. This was identified by \citet{Liu2015} as close to the optimal smoothing scale for their cosmological inference from CFHTLenS data, based solely on the WL peak number density. However, they already highlight that combining multiple smoothing scales may improve constraining power, as different smoothing scales correspond to different spatial scales. As we, instead, mimic a \textit{Euclid}-like analysis, with a different resolving power, and we combine multiple peak statistics, we reevaluate the impact of varying the smoothing scale on the constraining power. When selections are made based on the SNR, we recompute the noise standard deviation from the smoothed, randomly generated noise realisation for each smoothing scale. In Section~\ref{sec:smoothing_results}, we compare the fiducial results at 1~arcmin smoothing to those obtained with a smoothing scale that is a factor of 2 larger or smaller, or when combining the three smoothing scales.

\subsection{Peak statistics} \label{sec:peak_stats}
We use three different WL peak statistics in our forecast, the redshift distribution, $\mathrm{d}n(z)/\mathrm{d}z$; height distribution, $\mathrm{d}n(\kappa)/\mathrm{d}\kappa$; and angular clustering, $\omega(\theta)$. More details on the binning are given in Appendix~\ref{app:emulator}. The peaks are selected from the integrated WL convergence maps after noise and smoothing have been applied. Peaks are defined as pixels with values greater than those of their eight nearest neighbours, as commonly used in WL peak studies \citep[e.g.,][]{Weiss2019,Grandon2024}. For each peak, we determine its WL convergence value, subsequently referred to as the `peak height'. For SNR > 3 peaks, we identify the primary redshift that generates each peak, following the procedure of \citet{Broxterman2025_WLpeaks}. In the simulations, using the mass maps, the redshift of the overdensity that contributes most to the WL convergence at each WL peak, i.e. the dominant contribution to Eq.~\ref{eq:kappa_from_sim}, is assigned to that peak. 

The redshifts are distorted, mimicking the observational uncertainty in photometric redshift estimation for massive galaxy clusters. As the high-value WL peaks generally correspond to massive clusters, the redshifts of these objects are expected to be measured with much greater accuracy than those of the average \textit{Euclid} galaxy. Whereas the \textit{Euclid} requirement on the precision of the photo-$z$ estimates is $\sigma_z < 0.05(1+z)$ with catastrophic failures rate of less than
10 per cent \citep[][]{Mellier2025_EuclidSkyOverview}, these objects are expected to be measured more accurately using regular photo-$z$ estimation procedures or red-sequence cluster finders \citep[see e.g.][]{Rykoff2014,Chiu2024,Chen2025}. Based on the properties of the KiDS-1000 bright sample \citep[][]{Bilicki2021}, which were estimated using a similar number of photometric bands to those that will be available for \textit{Euclid}, the simulation redshifts are distorted by, for each peak, drawing a random redshift from a normal distribution centred on the true redshift, $z_\mathrm{true}$, and with a standard deviation of $\sigma_z = 0.02(1 + z_\mathrm{true})$, without any catastrophic outliers. The WL peak redshift distribution thus aims to exploit these precise redshift measurements of these massive objects for cosmological purposes.

As a third statistic, for the SNR > 3 peaks, we estimate the WL peak angular two-point correlation function. We use the Landy-Szalay estimator \citep[][]{LandySzalay1993}
\begin{align}
    \omega(\theta) = \frac{\mathrm{DD}(\theta)-2\mathrm{DR}(\theta)+\mathrm{RR}(\theta)}{\mathrm{RR}(\theta)},
\end{align}
to estimate the clustering signal. Here, $\theta$ is the angular separation between two WL peaks, and DD, RR, and DR represent the number of pairs of peaks separated by $\theta$ in the data sample (DD), a uniformly randomly generated sample (RR), and their cross-catalogue (DR). In practice, we use the package \textsc{Corrfunc} \citep[][]{Manodeep2020_corrfunc} to estimate the angular correlation function. The random set is chosen to be 8 times larger than the data catalogue in order to suppress Poisson noise originating from the random catalogue.

Although we probe the angular clustering and redshift distribution of the high-SNR peaks, these could, in theory, also be combined into a single statistic measuring the 3D clustering. This is, however, beyond the scope of this work.

\subsection{Two-point correlation functions}
To compare our forecast with those similar to a typical WL inference, we additionally estimate the WL angular two-point correlation functions (2PCFs), $\xi_{+/-}$. We do so by computing the WL convergence angular correlation function, $\mathcal{C}^{\kappa\kappa}$, from the full-sky WL convergence maps, which in the absence of $B$-modes relates to $\xi_{+/-}$ through a Hankel transform as \citep[][]{Schneider2002}
\begin{align}
    \xi_{+/-}(\theta) = \int_0^\infty \frac{\mathrm{d}\ell\, \ell}{2\pi} \mathcal{C}^{\kappa\kappa}(\ell) J_{0/4}(\ell\theta),
\end{align}
where $J_{0/4}$ are Bessel functions of the first kind, and $\ell$ is the multipole moment. We measure the 2PCFs in 9 logarithmic bins between 0.1 and $3~\deg$, which roughly corresponds to the ranges used by current cosmic shear inferences \citep[see e.g.][]{DESyr62026}, and is the same angular range as used for the WL peak angular clustering.

\subsection{Inference forecast}
In this section, we provide details of the forecasting pipeline we use to generate our forecasts. As explained in Section~\ref{sec:sims}, the WL peak statistics are extracted through forward modelling the hypercube simulations. We then train an emulator to predict the peak statistics as a function of the 10 cosmological parameters that are varied in the hypercube (Section~\ref{sec:emulator}). Finally, we estimate a covariance matrix from additional simulations (Section~\ref{sec:cov_mat}) and perform a Bayesian likelihood inference (Section~\ref{sec:likelihood}).

\subsubsection{Emulation}\label{sec:emulator}
To predict the WL peak statistics across the 10D parameter space spanned by the hypercube, we use Gaussian process (GP) emulators trained on statistics computed from the $2\times50$ simulations. All emulators are implemented with the \texttt{GaussianProcessRegressor} class from \textsc{scikit-learn} \citep{scikit-learn}. We train separate emulators for the different statistics and smoothing scales, and concatenate them to produce the combined forecasts. The details of the emulator construction as well as a leave-one-out test are presented in Appendix~\ref{app:emulator}. In general, we find sub-per-cent-level relative emulator accuracy across the ranges on which the statistics are used, which is sufficient for our idealised forecast.

\subsubsection{Covariance matrix}\label{sec:cov_mat}
To estimate the covariance, we run an additional suite of flat $\Lambda$CDM simulations. This set uses the same framework as the hypercube, but assumes the fiducial D3A cosmology of \citet{Schaye2023_flamingo}, i.e. a flat $\Lambda$CDM ($w_0 = -1, w_a = 0$) without DDM ($\Gamma_{\mathrm{dcdm}} = 0\,$~km~s$^{-1}$~Mpc$^{-1}$), no running of the scalar spectral index ($\alpha_\mathrm{s} = 0$), only a single massive neutrino species with $m_\nu c^2 = 0.06$~eV, and the same cosmology as the reference cosmology we are trying to infer. We assume that the covariance over the entire cosmological parameter space can be approximated by the covariance generated from these $\Lambda$CDM simulations. This assumption is expected to be a subdominant effect for cosmic shear 2pt inferences \citep[see e.g.][]{Kodwani2019}, but has not yet been studied for the non-Gaussian statistics we consider. For each simulation, we use a different random seed to generate the initial conditions with \textsc{monofonic} \citep[][]{Hahn2020_monofonic,Michaux2021_monofonic}, and we do not fix any modes, unlike the simulations that are part of the hypercube. As such, our covariance matrix should reflect the full cosmic variance that is inherent to the inference.

For each forecast and for realisation $i$, we compute the relevant WL peak statistics, $\mathbf{x}_i$. The sample covariance, $\hat{\mathbf{C}}$, is then computed as
\begin{equation}
    \hat{\mathbf{C}} = \frac{1}{N_\mathrm{cov} - 1} \sum_{i=1}^{N_\mathrm{cov}} 
    \left(\mathbf{x}_i - \bar{\mathbf{x}}\right)
    \left(\mathbf{x}_i - \bar{\mathbf{x}}\right)^\top,
\end{equation}
where $\bar{\mathbf{x}} = N_\mathrm{cov}^{-1} \sum_i \mathbf{x}_i$ is the sample mean. Since our analysis mimics a \textit{Euclid}-like survey covering $f_\mathrm{sky} = 15\,000 / 41\,253$ of the sky, the covariance is rescaled as
\begin{equation} \label{eqn:covmat}
    \mathbf{C} = \frac{\hat{\mathbf{C}}}{f_\mathrm{sky}}.
\end{equation}
When multiple statistics or smoothing scales are combined into a joint data vector, the covariance matrix is estimated from the same realisations, thereby capturing all cross-correlations among them.
In Appendix~\ref{app:corr_matrix}, we show the correlation matrix for our fiducial combined WL peak forecast, i.e. the height distribution, redshift distribution and angular clustering at 1~arcmin smoothing, illustrating that the angular correlation function exhibits strong correlations between different angular bins, whereas the cross-correlations between the different statistics and the correlations between individual bins of the height distribution and the redshift distribution are weaker.

\subsubsection{Likelihood}\label{sec:likelihood}
For our forecast, we assume an explicit likelihood. Whereas simulation-based inferences that learn the relationship between cosmological parameters and observables directly could, in theory, provide more precise constraints \citep[][]{Simone2026,Williamson2026}, we use an explicit likelihood because it allows us to gain more physical insight into the constraining properties of different statistics. We assume the likelihood follows a multivariate $t$-distribution to account for the fact that the covariance matrix $\mathbf{C}$ is estimated from $N_\mathrm{cov} = 80$ simulations rather than being known exactly \citep[][]{Sellentin2016}. The likelihood is given by
\begin{equation}
    \mathcal{L}(\boldsymbol{x}| \boldsymbol{\theta} ) \propto
    \left\{
        1 + \frac{[\boldsymbol{x}-M(\boldsymbol{\theta})]^{\top} \mathbf{C}^{-1} [\boldsymbol{x}-M(\boldsymbol{\theta})]}{N_\mathrm{cov} - 1}
    \right\}^{-N_\mathrm{cov}/2},
    \label{eq:sh_likelihood}
\end{equation}
where the data vector \textbf{x} consists of (a combination of) the different WL peak statistics, $\textbf{C}^{-1}$ is the inverse sample covariance matrix, and $M(\boldsymbol{\theta})$ is the emulator prediction. In practice, we evaluate the log-likelihood, which is given by 
\begin{equation}
\begin{split}
    \ln \mathcal{L} =
    & \ln\Gamma\!\left(\frac{N_\mathrm{cov}}{2}\right)
    - \ln\Gamma\!\left(\frac{N_\mathrm{cov}-N_{\mathrm{dv}}}{2}\right)
    - \frac{N_{\mathrm{dv}}}{2}\ln\!\bigl[\pi(N_\mathrm{cov}-1)\bigr] \\
    & - \frac{1}{2}\ln|\mathbf{C}|
    -\frac{N_\mathrm{cov}}{2}
    \ln\!\left(
        1 + \frac{[\boldsymbol{x}-M(\boldsymbol{\theta})]^{\top} \mathbf{C}^{-1} [\boldsymbol{x}-M(\boldsymbol{\theta})]}{N_\mathrm{cov} - 1}
    \right),
    \label{eq:sh_loglike_norm}
\end{split}
\end{equation}
where $\Gamma$ is the gamma function and $N_{\mathrm{dv}}$ is the dimension of the data vector $\boldsymbol{x}$. In the analysis, we infer the cosmology of the reference data vector, $\boldsymbol{x}$, for which we run an additional simulation using the hypercube infrastructure but at the fiducial cosmology of the FLAMINGO simulations \citep[][]{Schaye2023_flamingo}, which is the best-fitting Dark Energy Survey year 3 ‘3\texttimes2pt + All Ext.’ $\Lambda$CDM cosmology \citep[D3A;][]{Abbott2022} with a single massive neutrino species, from which we extract the same statistics through forward modelling. We infer the best-fitting parameters by sampling the log-likelihood with the Markov chain Monte Carlo sampler \textsc{emcee} \citep[][]{Foreman2012}, and our results are plotted with \textsc{getdist} \citep[][]{Lewis2025_getdist}.

For most parameters, we apply flat priors. For the two beyond $w_0w_a$CDM parameters, $\alpha_\mathrm{s}$ and $\Gamma_\mathrm{dcdm}$, we apply, respectively, a Gaussian and half-Gaussian prior centred on the truth with widths 0.02 and 1, to penalise strong deviations from these less common extensions.

We quantify the FoM for a 2D cosmological parameter combination of $\theta_1$ and $\theta_2$ as
\begin{align}
    \mathrm{FoM} = \frac{1}{\sqrt{|C(\theta_1,\theta_2)|}},
\end{align}
where $C$ is the 2×2 parameter covariance matrix of $\theta_1$ and $\theta_2$. Similarly, we quantify the figure of bias (FoB) as
\begin{align}
    \mathrm{FoB} = \sqrt{(\theta_\mathrm{truth}-\theta_{\mathrm{inferred}})^{\top} C(\theta_\mathrm{inferred})^{-1}(\theta_\mathrm{truth}-\theta_{\mathrm{inferred}})},
\end{align}
where $\theta_\mathrm{truth}$ is the value of the cosmological parameters in the D3A cosmology and $\theta_{\mathrm{inferred}}$ is the best-fit emulator prediction.

\section{Results and Discussion}\label{sec:rend}
  
\begin{figure*}
  \centering
  \includegraphics[width=\textwidth]{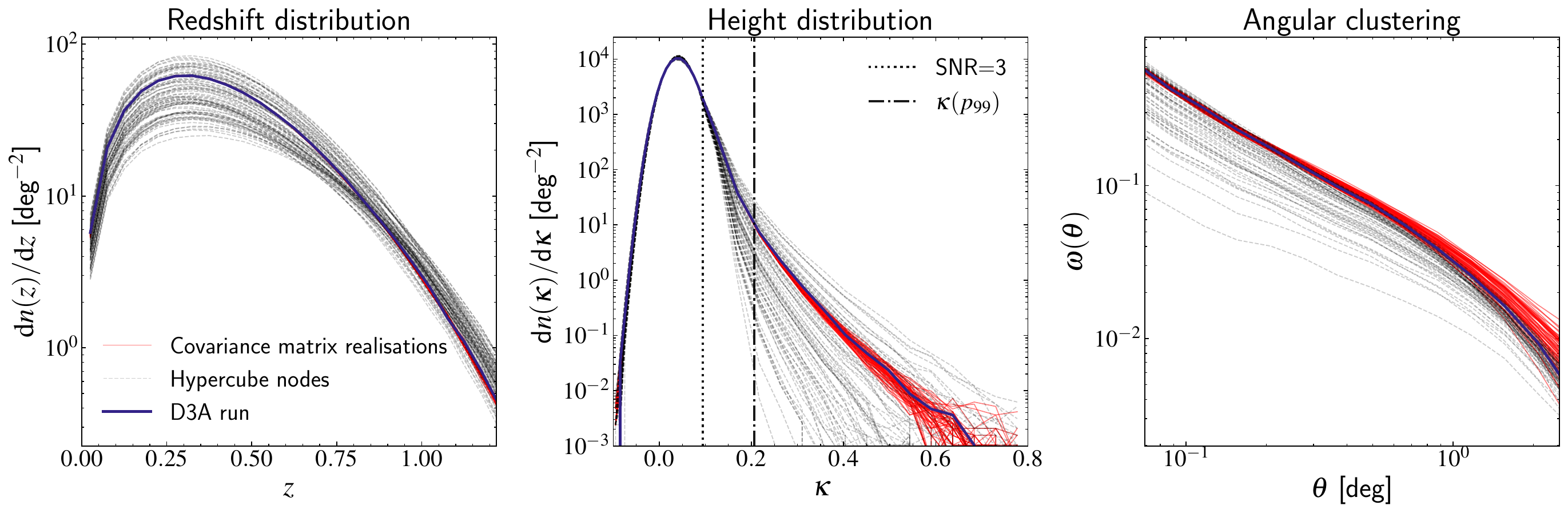}
  \caption{WL peak statistics used in the fiducial forecast with 1~arcmin smoothing. Left: redshift distribution of SNR $> 3$ peaks. Middle: height distribution. Right: angular correlation function of SNR $> 3$ peaks. The measurements from the hypercube simulations are shown as dashed light-grey curves, the covariance matrix realisations as red lines and the measurements from the additional simulation at the FLAMINGO D3A cosmology, which is inferred in this paper, as a purple curve. The dotted and dashed-dotted lines in the central panel indicate the SNR~=~3 threshold used for all statistics and the $\kappa$ value corresponding to the 99th-percentile peak, which is used as an upper bound for the peak-height distribution.}
  \label{fig:fiducial_statistics}
\end{figure*}

In Fig.~\ref{fig:fiducial_statistics}, we illustrate the measurements of the three different WL peak statistics across the different parts of the forecast pipeline. The left panel shows the redshift distribution of the SNR $> 3$ peaks, the central panel shows the height distribution, and the right panel shows the angular two-point correlation function of the SNR $> 3$ peaks. These measurements correspond to our fiducial smoothing scale of 1 arcmin. In each panel, the solid red curves indicate the measurements from the covariance-matrix realisations, the solid purple curve indicates the target distribution measured from the D3A simulation, and the dashed grey curves show the measurements in the 100 hypercube nodes. The hypercube nodes span a wide range of predictions, demonstrating sensitivity across the full 10D parameter space. The central panel also includes the SNR = 3 cut as a vertical dotted line. 

For each statistic, the D3A measurement falls within the range of the covariance realisations. The spread of the covariance-matrix realisations relative to the D3A target varies across the three statistics, reflecting the distinct cosmic-variance properties of each observable. For the redshift distribution, the covariance realisations show very little scatter around the D3A measurement across the entire redshift range, reflecting the small sample variance on the shape and amplitude of the redshift distribution of high-SNR WL peaks expected for the forecasted \textit{Euclid} measurements. For the height distribution, the spread of the covariance realisations increases for $\kappa$ values above the SNR cut. This is expected, as the number density of the peaks drops steeply with increasing $\kappa$, and the Poisson noise from the rare high-mass haloes that generate these peaks thus increases. Additionally, for the highest mass objects, the relative location of the (virtual) observer also matters, as it determines the exact peak height via the lensing kernel. As discussed in more detail in Appendix~\ref{app:emulator}, we do not use the highest 1 per cent $\kappa$ peaks during emulator training and inference for the height distribution, as we found the emulator performed better without them, likely due to increased stochasticity. For the angular clustering, the covariance realisations show a scatter of a factor of a few in the signal across the entire angular range, with the relative spread increasing toward larger angular separations.

\subsection{The impact of cosmology}
Before examining the actual cosmological forecast, we first discuss the possible ways in which the changes in cosmology may impact our WL peak forecast. The 10 cosmological parameters varied in the hypercube can affect the WL peak statistics in three distinct physical ways.

The first possibility is through the direct impact on the physical matter density. From  Eq.~\ref{eq:kappa_from_sim}, $\kappa$ is directly proportional to $\Omega_\mathrm{m}(\chi_i)\rho_\mathrm{crit}(\chi_i)$, i.e. the physical total matter density over time. Within the cosmologies in the hypercube, $\rho_\mathrm{crit}(z) = 3H(z)^2/(8\pi G)$ and the physical matter density thus scales as $\propto\Omega_\mathrm{m} h^2$, illustrating that $\Omega_\mathrm{m}$ and $h$ directly impact the WL signals through this channel. Additionally, $\Gamma_\mathrm{dcdm}$ impacts the total matter density, and therefore directly rescales the amplitude of each peak, albeit with a small magnitude as the DDM rates are small.

The second way in which the WL peaks can be impacted by cosmology is through a change in the growth of structures, which sets the matter density contrast $\delta_i(\boldsymbol{\theta})$ in Eq.~\ref{eq:kappa_from_sim}. Structure growth controls the masses of the structures that generate the WL peaks, thereby also controlling the number density and height of the peaks. Parameters that impact the structure growth directly are the amplitude of the primordial power spectrum, $\ln(10^{10}A_\mathrm{s})$, which sets the overall strength of initial density fluctuations, the sum of neutrino masses, $\sum m_\nu c^2$, whose free-streaming suppresses structure growth, but also DE ($w_0$, $w_a$), as it may suppress the growth of matter perturbations through its impact on the growth factor.

Finally, the cosmological parameters may impact the WL peak properties through changes in the geometry, i.e. by changing the comoving distance--redshift relation, $\chi(z)$. Therefore, any parameter that alters the expansion history through the Friedmann equation changes the WL kernel (Eq.~\ref{eq:WL_kernel}) and the conversion between physical separations and angles on the sky, which is relevant for the angular clustering signal. 

Although it is not possible to cleanly separate the impact of these different channels on the parameter posteriors, in the next sections, we discuss possibilities for how individual parameters may explain the differences in constraining power we observe.

\subsection{Cosmological constraints forecast}\label{sec:param_constraints}

\begin{figure*}
  \centering
  \includegraphics[width=\textwidth]{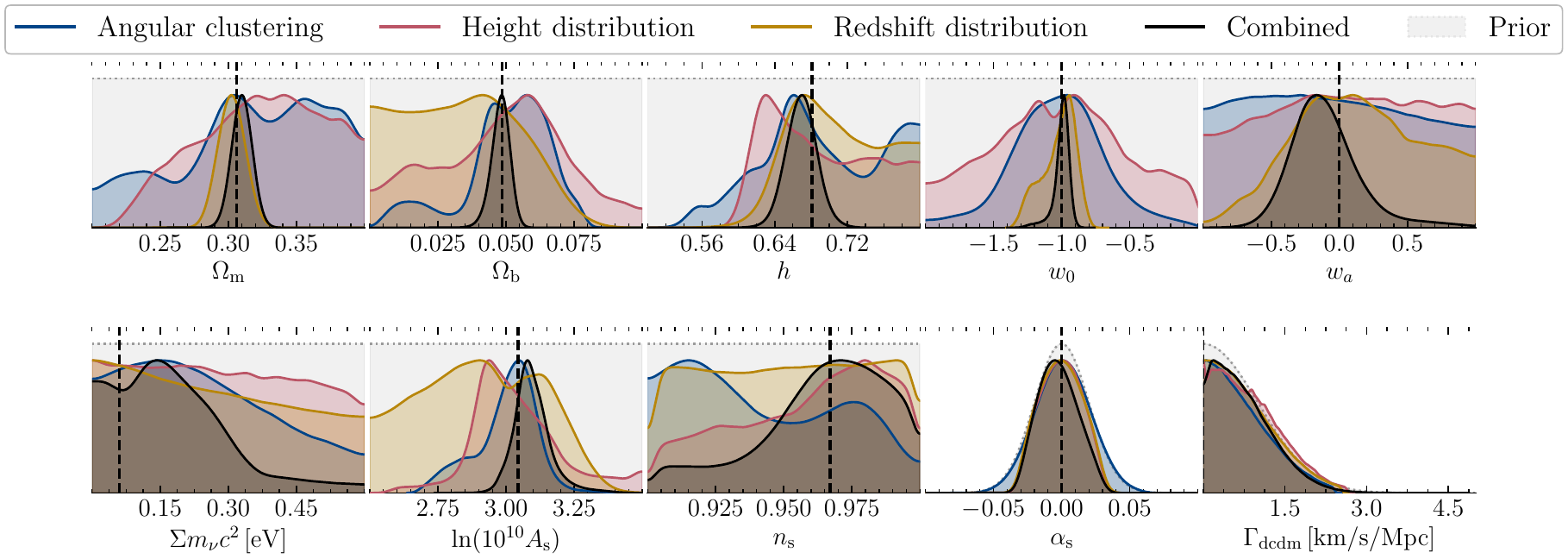}
  \caption{One-dimensional posteriors of the fiducial forecast using WL peak angular clustering (blue), height distribution (red), redshift distribution (orange), or the three statistics combined (black) for SNR > 3 WL peaks with 1 arcmin smoothing. The prior is indicated by the dotted grey curve and the shaded grey area, which uniformly spans the parameter range for all panels except the final two. The true values are the D3A cosmology indicated by the dashed black lines. For the combined inference, informative unbiased constraints are recovered for the matter and baryon densities, the Hubble parameter, the DE equation of state, and the amplitude of the primordial power spectrum, whereas the neutrino mass, the running and amplitude of the scalar spectral index and the DDM rate are either poorly constrained or largely recover the prior.}
  \label{fig:1d_1ss_2stat}
\end{figure*}

Next, we forecast the cosmological constraints from our Bayesian analysis for our fiducial set-up for each of the peak statistics individually, and for the forecast that combines the three statistics. In Fig.~\ref{fig:1d_1ss_2stat}, we show the marginalised one-dimensional posterior distributions for all 10 cosmological parameters for each of the three individual statistics and their combination, at the fiducial smoothing scale of 1 arcmin. The forecast using the height distribution, redshift distribution, angular clustering, or the combined approach yields the red, orange, blue, and black posteriors, respectively. The shapes of the priors are indicated by the grey dotted line and grey shaded area. In Appendix~\ref{app:corner_peaks}, we show the full corner plot corresponding to forecasts using the 3 WL peak statistics together, the redshift distribution individually, and the angular clustering and height distributions together. This comparison shows the value of just adding the redshift distribution over the other two statistics.

We first assess whether the inference correctly recovers the fiducial D3A cosmology. The combined inference of the three peak statistics yields strong constraints on $\Omega_\mathrm{m}$, $\Omega_\mathrm{b}$, $\ln(10^{10}A_\mathrm{s}$), $w_0$, $w_a$, and $h$ whereas the parameters $\sum m_\nu c^2$, $n_\mathrm{s}$, $\alpha_\mathrm{s}$ and $\Gamma_\mathrm{dcdm}$ are either weakly constrained or recover the prior. The combined inference (black posteriors) recovers the input parameter values well for those parameters that are well-constrained, as the marginalised posteriors are consistent with the true values (dashed black curves) within the 68 per cent credible intervals, as shown in Fig.~\ref{fig:1d_1ss_2stat}. The average FoB of the maximum a posteriori (MAP) of the parameters that are well constrained, $\Omega_\mathrm{m}, \Omega_\mathrm{b}, h, w_0, w_a$, and $\ln(10^{10}A_\mathrm{s})$ is 0.59$\sigma$. The correspondence demonstrates that the entire pipeline of emulator, likelihood, and covariance matrix together form a self-consistent and proper framework for our cosmological forecasts, and that the added noise does not severely bias the parameter constraints.

Although we expect our idealised inference to recover all parameters well, minor offsets from the truth can be due to (a combination of) noise in the likelihood arising from the finite number of covariance realisations, emulation prediction error, the covariance matrix not varying with cosmology, the added noise components,  or because of cosmic variance in the D3A run whose cosmology is inferred. We find some minor offsets from the truth, particularly when a parameter is not well constrained, as is visible in the multi-modal blue angular clustering posteriors for $\Omega_\mathrm{m}$, $h$, and $n_\mathrm{s}$. 

Next, we discuss the constraining power on each cosmological parameter and the complementarity between the different WL peak statistics in probing different parts of the cosmological parameter space. 

\subsubsection{The matter density and amplitude of the primordial power spectrum}\label{sec:omegam_as}
The matter density parameter, $\Omega_\mathrm{m}$, and the amplitude of the primordial power spectrum, $\ln(10^{10}A_\mathrm{s})$, are among the best-constrained parameters in our forecast, which is expected for $\Omega_\mathrm{m}$ since it impacts the inference through each channel as discussed before, whereas $\ln(10^{10}A_\mathrm{s})$ only acts directly through the second channel. For the combined inference we find $\Omega_\mathrm{m} = 0.310 \pm 0.006$ and $\ln(10^{10}A_\mathrm{s}) = 3.09 \pm 0.06$, while the input values were $\Omega_\mathrm{m} = 0.306$ and $\ln(10^{10}A_\mathrm{s}) = 3.04$. The corresponding best-fitting $S_8$ is $0.819\pm0.011$, compared to the true value of 0.815. Since we do not marginalise over any systematics, we limit our interpretation to checking that the inferred values are consistent with the truth, rather than drawing conclusions about the absolute constraining power.

The strong constraining power for $\Omega_\mathrm{m}$ and $\ln (10^{10}A_\mathrm{s})$ is consistent with the results of cosmic shear surveys that typically report their cosmology constraints in terms of the $S_8$ parameter \citep[e.g.][]{Abbott2022,Wright2025}, since this is the combination in the $\Omega_\mathrm{m}$--$\sigma_8$ plane that current cosmic shear surveys are, approximately, most sensitive to. In our forecast, we find that the constraint on $\Omega_\mathrm{m}$ is driven by the redshift distribution, whereas the angular clustering and height distribution both contribute significantly to the constraint on $\ln (10^{10}A_\mathrm{s})$.

\begin{figure*}
  \centering
  \includegraphics[width=\textwidth]{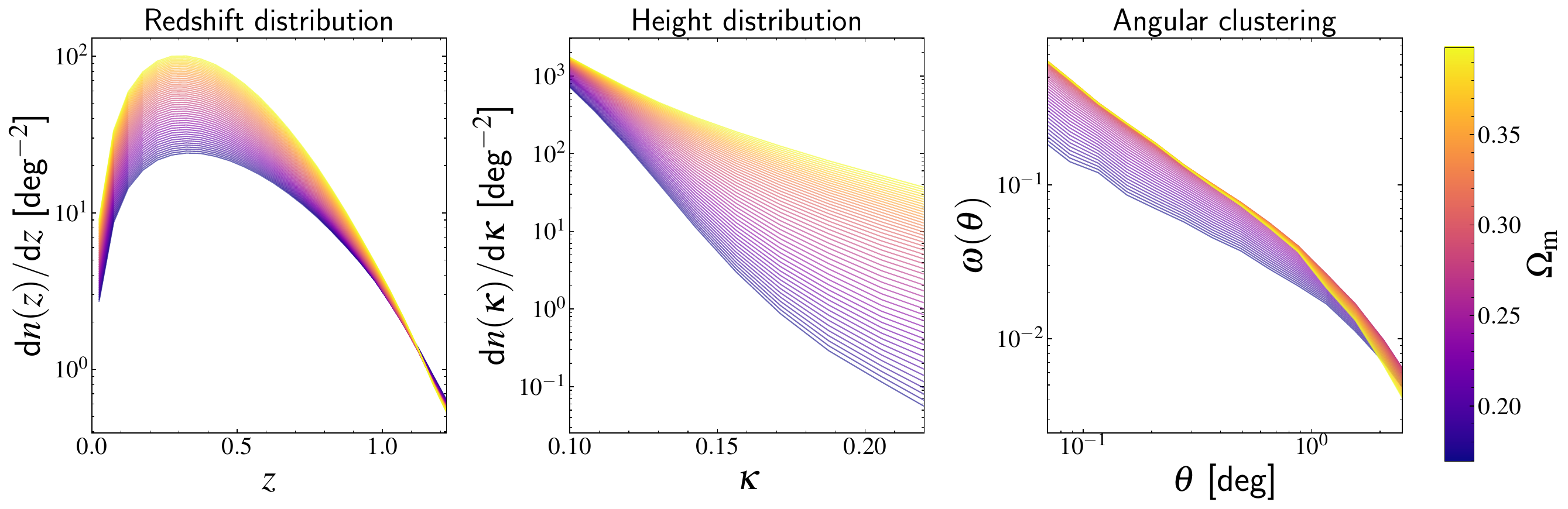}
  \caption{Emulator predictions for the redshift distribution (left), height distribution (middle), and angular clustering (right) of SNR > 3 WL peaks at 1~arcmin smoothing for varying $\Omega_\mathrm{m}$, as indicated by the colour bar, while keeping all other parameters fixed to their mean values within the hypercube. All three statistics are impacted. Larger values of $\Omega_\mathrm{m}$ increase the overall number of peaks at fixed WL convergence and give rise to peaks preferentially corresponding to redshifts where the WL kernel peaks. Below 1~deg the angular clustering signal shows a monotonic trend of larger $\Omega_\mathrm{m}$ resulting in a stronger clustering signal, but at larger angular scales this trend reverses.}
  \label{fig:sweep_omegam}
\end{figure*}

The sensitivity of the three WL peak statistics to $\Omega_\mathrm{m}$ is illustrated in Fig.~\ref{fig:sweep_omegam}, which shows the emulator predictions for each statistic as $ \Omega_\mathrm{m}$ is varied individually while all other parameters are held fixed at their mean values. The colour indicates the values of the cosmological parameters, which range from the minimum to the maximum of the hypercube, as also indicated by the colour bar. In Appendix~\ref{app:param_sweep}, we show similar figures for $h$, $w_0$, $w_a$, and $\Omega_\mathrm{b}$. Fig.~\ref{fig:sweep_omegam} illustrates different dependencies on $\Omega_\mathrm{m}$. The emulator predictions across different $\Omega_\mathrm{m}$ values result in changes of nearly two orders of magnitude in the amplitudes of the redshift and height distributions, and a factor of a few in the angular clustering amplitude, with the curves spanning the full dynamic range of the axes in all three panels. Out of the ten cosmological parameters, within the hypercube range, $\Omega_\mathrm{m}$ shows the largest deviations across each of the three statistics. The amplitude of the redshift distribution is set by the number of peaks above the SNR > 3 threshold that is applied to the height distributions. Additionally, the redshift distribution shows a secondary dependence, as $\Omega_\mathrm{m}$ also changes its shape. Although the distribution always peaks at the same redshift, which is close to the redshift at which the lensing kernel peaks given the chosen source redshift distribution, smaller values of $\Omega_\mathrm{m}$ preferentially lead to a smaller fraction of lower redshift peaks, which could be due to the fact that as $\Omega_\mathrm{m}$ decreases, $\Omega_\Lambda$ increases to maintain a spatially flat universe. Therefore, late-time density perturbations are increasingly suppressed, as in the limit of DE domination, the linear growth factor asymptotically approaches a constant, halting further structure growth.

For low values of $\Omega_\mathrm{m}$, the angular clustering signal increases monotonically with $\Omega_\mathrm{m}$. However, as $\Omega_\mathrm{m}$ is increased beyond 0.3, the clustering signal shows smaller relative changes than the other two statistics and the signal at $\theta > 0.3$~deg even starts to decrease, while the smaller scales still show the same monotonic behaviour. This non-monotonic behaviour could explain why the clustering signal is not able to provide precise constraints on $\Omega_\mathrm{m}$ with the 1D posterior even being multimodal. However, because the clustering constraint reaches the imposed prior boundary on $\Omega_\mathrm{m}$, these constraints should be interpreted with caution.

The amplitude of the primordial power spectrum, $\ln(10^{10}A_\mathrm{s})$, sets the overall normalisation of the matter power spectrum, directly controlling the abundance and masses of the structures that generate the high-SNR WL peaks. A larger value of $A_s$ increases the amplitude of density fluctuations, leading to more massive haloes and a higher number density of high-SNR peaks. The constraint on $\ln(10^{10}A_\mathrm{s})$ is driven by the angular clustering and peak height distribution, as a higher $\ln (10^{10}A_\mathrm{s})$ produces more WL peaks at fixed $\kappa$ and a more strongly clustered large-scale structure, leading to a stronger two-point signal and more high-SNR peaks.

\subsubsection{Dark energy equation of state}\label{sec:w0wa}

In our forecast, the CPL equation-of-state parameters of DE, $w_0$ and $w_a$, affect the inference through their effects on the expansion history, which enter the lensing kernel, and through their impact on the WL convergence through changes in geometric and growth rates. The DE sets the late-time expansion history, where a less negative $w_0$ corresponds to stronger dark energy dominance, which both suppresses the growth of structure and alters the angular diameter distances to the lens and source planes. Because the simulations do not model dark energy perturbations, the effect of evolving DE parameters on the growth rate is purely through the expansion rate. Together, these effects modify the WL convergence value of a WL peak generated by a halo of fixed mass. In particular, the amplitude of the integrated convergence (Eq.~\ref{eqn:kappa_int_cont}) along the line of sight scales with the product of the matter density contrast, the comoving distances, and the kernel $W(\chi_i)$, all of which depend on $w_0$ and $w_a$. 

As, among the three WL peak statistics, the redshift distribution is the only statistic that directly probes a signal over cosmic time, it is unsurprising that it dominates the constraining power on $w_0$ and $w_a$ in our forecast. The forecast yields $w_0 = -0.98 \pm 0.06$ and $w_a = -0.15 \pm 0.24$. This represents a 6 per cent relative precision on $w_0$ while $w_a$ remains poorly constrained with $\sigma_{w_a} = 0.24$. The parameter $w_a$ controls the time-evolution of the equation of state and is generally less well constrained than $w_0$. This is because the constraining signal for $w_a$ requires resolving the differential evolution of the dark energy contribution across redshift, which remains more uncertain in our forecast.

There is a strong degeneracy direction across all $w_0-w_a$ constraints, as the two parameters affect the forecast in a very similar way. This is also reflected by the figures in Appendix~\ref{app:param_sweep} showing the impact of varying the two parameters individually on the WL peak statistics.

\subsubsection{Hubble parameter}\label{sec:hubble}
The Hubble parameter, $h$, enters the WL peak statistics most directly through the physical matter density, which scales as $\Omega_\mathrm{m} h^2$, and sets the overall amplitude of the WL convergence (Eq.~\ref{eq:kappa_from_sim}). As a consequence, $h$ and $\Omega_\mathrm{m}$ show a negative correlation that is visible in the 2D contour for the height distribution and angular clustering in Fig~\ref{fig:corner_1ss_peaks}, as a larger value of $h$ can compensate for a lower value of $\Omega_\mathrm{m}$ to generate a peak of the same WL convergence value. This anti-correlation is primarily present in the posteriors corresponding to the height distribution and angular clustering, which therefore struggle to constrain $h$ independently of $\Omega_\mathrm{m}$. The redshift distribution is able to break this degeneracy as it provides a tight estimate on $\Omega_\mathrm{m}$, such that the combination is also able to probe $h$, possibly through the impact of $\Omega_\mathrm{m}$ on the geometry that is being probed by the redshift distribution. The qualitative impact of $h$ on the angular clustering and height distribution is very similar to that of $\Omega_\mathrm{m}$, as shown in Fig.~\ref{fig:sweep_h}, albeit with smaller absolute changes. Regarding the redshift distribution, $h$ primarily impacts its amplitude, whereas $\Omega_\mathrm{m}$ also affects the shape, illustrating how including the redshift distribution helps break the $\Omega_\mathrm{m}-h$ degeneracy.

\subsubsection{Baryon density}

In the hypercube, $\Omega_\mathrm{m}$ is varied individually and $\Omega_\mathrm{b}$ is varied through changes in the ratio $\Omega_\mathrm{b}/\Omega_\mathrm{m}$. $\Omega_\mathrm{b}$ at fixed $\Omega_\mathrm{b}/\Omega_\mathrm{m}$ then impacts the forecast through its impact on the expansion history. At fixed $\Omega_\mathrm{m}$, the expansion history is not affected, and changes in $\Omega_\mathrm{b}/\Omega_\mathrm{m}$, instead, likely affect the inference through changes in the primordial power spectrum. Individually, none of the three statistics provides a tight constraint on $\Omega_\mathrm{b}$. However, when combining all three statistics, the constraint on $\Omega_\mathrm{b}$ improves drastically. The physical origin of this improvement is not straightforward to identify, but it reflects the breaking of a degeneracy in the full 10D parameter space that no single statistic can resolve on its own. We note that $\Omega_\mathrm{b}$ is, in any case, already well constrained independently through Big Bang nucleosynthesis \citep[e.g.][]{Fields2020}, and a strong external prior on this parameter could help break the residual degeneracies even further, potentially tightening the constraints on the remaining cosmological parameters in the process.

\subsubsection{Neutrino mass, scalar spectral index, and beyond $w_0w_a$CDM parameters}\label{sec:weakly_constrained}

Several parameters in the hypercube are only weakly constrained by our fiducial forecast, and the marginalised posteriors recover much of the prior range. The sum of neutrino masses $\sum m_\nu c^2$ is not well constrained in our combined forecast compared to competitive results from measurements of, for example, the CMB combined with DESI BAO, which report constraints on the neutrino mass with a marginalised error of 0.078~eV for a $w_0w_a$CDM$+\Sigma m_\nu$ inference \citep[][]{Elbers2025DESI}. The constraining power, however, is still competitive with CMB-independent constraints from DESI. Using the same simulations as us, \citet{Wang2026} showed that persistence strips, a formalism that segments Betti curves by topological persistence, can reach a neutrino mass constraint with an uncertainty down to 0.05~eV.

The scalar spectral index $n_\mathrm{s}$ and its running $\alpha_\mathrm{s}$ affect the shape of the primordial power spectrum and, hence, the matter power spectrum on small scales, which ultimately influence the mass function of haloes and, in turn, the high-SNR peaks. However, given that the range over which these parameters are varied in the hypercube is not very large relative to existing CMB constraints, and the constraining power of the WL peak statistics is limited, we fail to constrain these parameters within our fiducial forecast.

Similarly, the DDM decay rate $\Gamma_\mathrm{dcdm}$ is only weakly constrained, and the marginalised posterior recovers the half-Gaussian prior we impose. This is somewhat surprising, as in \citet{Broxterman2025_WLpeaks} we found that the amplitude of the FLAMINGO variations with DDM rates of 0.12 and 0.24~$H_0/h$ reduced the amplitude of the redshift distribution by factors of $\approx~2$ and $\approx~4$, respectively. Although the latter is outside the range of the hypercube used in this work, naively, given the change in the number of peaks above the SNR = 3 threshold in the 0.12~$H_0/h$ run, we would still expect an additional constraint beyond the prior in this work. The two analyses differ in their peak-height cut and noise properties, although the exact reason we do not recover stronger constraints on $\Gamma_{\mathrm{dcdm}}$ is unclear, it might be the fact that we already impose a strong prior on the value of $\Gamma_{\mathrm{dcdm}}$ and the small variations within this prior do not impact the analysis much. The prior on $\Gamma_{\mathrm{dcdm}}$ is more than 10 times stronger than the weakest DCDM scenario considered in the FLAMINGO suite. For $\Gamma_{\mathrm{dcdm}}$~=~1~km/s/Mpc, which is the width of the half-Gaussian prior we apply, the effect on the linear power spectrum is less than 1 per cent and the effect on the expansion rate is less than 0.1 per cent. An alternative explanation is that our emulators emulate $\Omega_\mathrm{m}$, i.e. the real present-day value of $\Omega_\mathrm{b}+\Omega_\mathrm{cdm}+\Omega_\mathrm{\nu}$ and not $\Omega_\mathrm{m,noDDM}$, which is the parameter that is varied in the hypercube. Some of the sensitivity to $\Gamma_{\mathrm{dcdm}}$ may therefore be absorbed by $\Omega_\mathrm{m}$.

\subsection{Complementarity of WL peak statistics}\label{sec:complementarity}

One of the key results of this work is the complementarity of the three different WL peak statistics. Fig.~\ref{fig:1d_1ss_2stat}, and in more detail, the 2D corner plot in Appendix~\ref{app:corner_peaks}, demonstrates that the combination of the WL peak height distribution, redshift distribution, and angular clustering significantly outperforms each individual statistic for the majority of parameters. Particularly the constraints on $\Omega_\mathrm{b}$, $h$, $w_a$, $\sum m_\nu c^2$, and $\ln(10^{10}A_\mathrm{s})$ improve significantly compared to any individual statistic. 

The height distribution is directly sensitive to the number density of massive haloes as a function of their mass. Individually, the height distribution does not impose strong constraints on any cosmological parameters. Since the height distribution measures the WL convergence values by integrating over the entire source redshift distribution, it averages over the complete growth history, which limits its ability to constrain parameters that affect the growth history in different ways but have similar effects on the amplitude of the WL signal.

Adding the redshift distribution allows us to break some of these degeneracies by probing the evolution of structure formation. Individually, the redshift distribution is only able to constrain well $\Omega_{\mathrm{m}}- w_0$. However, the corner plot shows that combining the statistics helps break degeneracies. Individually, none of the statistics can probe the neutrino mass, but together they can place a clear upper bound, albeit still not competitive with other cosmological probes. Angular clustering provides a qualitatively different source of information by probing the spatial distribution of peaks on the sky. Consequently, it provides clear constraints on $\ln(10^{10}A_\mathrm{s})$.

\citet{Davies2022} found that the clustering of WL peaks is a factor $\approx3$ more sensitive to $h$ and $w_0$ than the peak height distribution is. For $w_0$, we see a qualitatively similar improvement between the red and blue 1D posteriors. However, the redshift distribution outcompetes the height distribution and angular clustering for these parameters. For $h$, we do not see the same improvement. If anything, the height distribution seems to provide slightly better results. The comparison is complicated by the fact that, in our 10D inference, we cannot constrain these parameters within the hypercube bounds, so the prior bounds are therefore important. Also, \citet{Davies2022} combine the clustering of peaks selected at different heights, which they show significantly improves the constraints compared to clustering of peaks at a single cut-off value.

\subsection{Comparison to shear two-point correlation functions}

\begin{figure*}
  \centering
  \includegraphics[width=\textwidth]{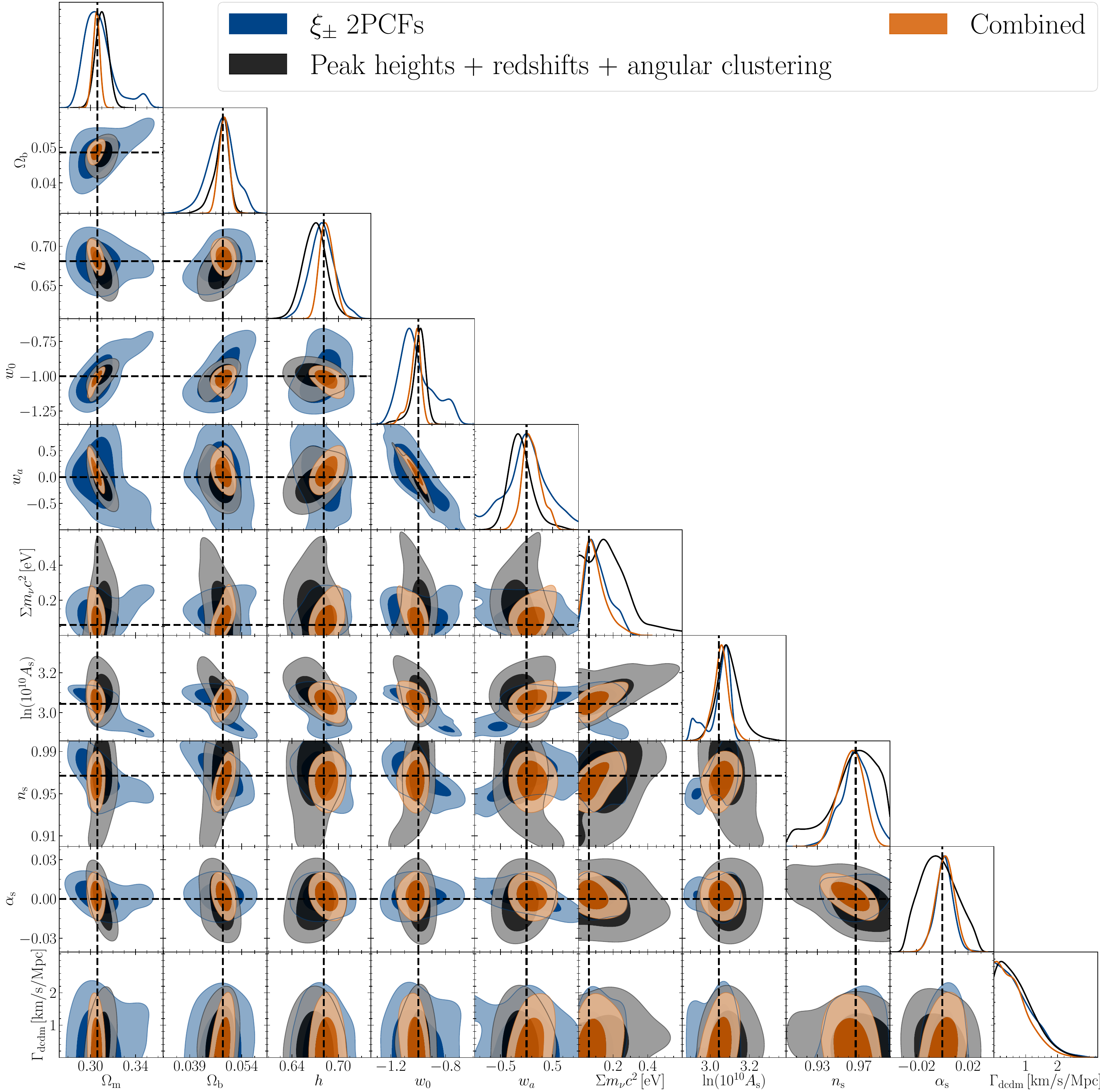}
  \caption{Corner plot of the cosmological parameter forecast using the WL two-point correlation functions (blue) or angular clustering, redshifts and heights of WL peaks (black) of SNR > 3 peaks or a combined inference using all 4 statistics (orange) at 1 arcmin smoothing. The contours indicate the 1- and 2$\sigma$ credible intervals. The combination of WL peak statistics outperforms the $\xi_{+/-}$ 2PCFs for $\Omega_\mathrm{m}, \Omega_\mathrm{b}, w_0$, and $w_a$, while the constraints on $h$, $\ln(10^{10}A_\mathrm{s})$ and $\Gamma_\mathrm{dcdm}$ are similar and the remaining parameters, $\Sigma m_\nu c^2, n_\mathrm{s}$, and $\alpha_\mathrm{s}$ are probed best by the $\xi_{+/-}$ 2PCFs.}
  \label{fig:corner_with_xipmm}
\end{figure*}

As our analysis is idealised, since we do not include marginalisation over nuisance parameters, it is difficult to quantitatively compare our constraints to those of other works, let alone to the precision of current inferences or requirements for \textit{Euclid}. Therefore, to more straightforwardly assess the benefits and downsides of our WL peak forecast, we compare our results to those of a similarly idealised forecast using WL shear two-point correlation functions (2PCFs), $\xi_{\pm}$, which are the standard statistic used by current cosmic shear surveys. The 2PCFs are computed from the same simulations and WL convergence maps and use the same smoothing, priors, and covariance matrix realisations as the fiducial setup. Fig.~\ref{fig:corner_with_xipmm} shows the marginalised 2D posterior distributions for all 10 cosmological parameters for the $\xi_{\pm}$ 2PCFs (blue), the combined WL peak statistics (black), and the combination of all four statistics (orange), i.e. the shear~2pt functions plus the three peak statistics. In general, the combination of WL peak statistics outperforms the $\xi_{\pm}$ 2PCFs. In reality, the cosmic shear signal is measured across different tomographic bins, meaning our forecast on the constraining power of the 2PCFs is a conservative estimate. Additionally, the cosmic shear measurements are usually combined with galaxy clustering and galaxy-galaxy lensing in a 3\texttimes2pt analysis, yielding better constraints than the cosmic shear two-point signal alone. A full comparison to the 3\texttimes2pt, or even 6\texttimes2pt, forecast, however, is beyond the scope of this work.  

For the parameters $\Omega_\mathrm{m}$, $\Omega_\mathrm{b}$, $w_0$, and $w_a$, the combination of WL peak statistics provides a tighter constraint than the $\xi_{+/-}$, while the constraints on $h$, $\ln(10^{10}A_\mathrm{s})$, and $\Gamma_\mathrm{dcdm}$ are of a similar level. Only for the parameters $\Sigma m_\nu c^2$, $n_\mathrm{s}$, and $\alpha_\mathrm{s}$, which were poorly constrained in the WL peak forecast, do the $\xi_{+/-}$ outperform the WL peaks. The constraints on these parameters, however, remain worse than those obtained from other cosmological observables, such as the CMB \citep[see e.g.][]{Planck2020}. Whereas the high-valued WL peaks are sensitive to the most massive haloes and structures, the 2PCFs are instead sensitive to the shape of the matter power spectrum across a wide range of scales, which is why the $\xi_{+/-}$ constrains these parameters better \citep[see, e.g.][]{Hamana2004,Heitmann2009,Yang2025,Broxterman2025_WLpeaks}. 

The joint forecast tightens the contours even further, particularly for $\Omega_\mathrm{m}$, where some parameter degeneracy seems to be broken as the WL peaks and $\xi_{+ /-}$ show different degeneracies in, for example, the $\Omega_\mathrm{m}-h$ and  $\Omega_\mathrm{m}-\ln(10^{10}A_\mathrm{s})$ subpanels.  

The results are qualitatively consistent with \citet{Davies2022}, who also find that WL peaks can probe $S_8$ and $w_0$ more precisely than the shear-2PCFs. However, they find a similar level of constraining power on $\Omega_\mathrm{m}$ and that the shear-2PCFs are able to probe $h$ better than the peak statistics. This could be explained by the fact that they do not include the redshift distribution of WL peaks, which drives our constraints on $\Omega_\mathrm{m}$. As expected, they generally find that combining the statistics leads to even tighter constraints. 

Overall, these results suggest that WL peaks and 2PCFs are complementary probes, as the peaks carry non-Gaussian information that helps break degeneracies in the 2PCF posterior, while the 2PCFs still carry constraints on parameters to which the peak statistics are less sensitive. Their combination, therefore, yields a more powerful cosmological inference than either statistic alone, motivating a joint analysis for future \textit{Euclid}-like surveys, which can be further improved by including additional non-Gaussian statistics \citep[see e.g.][]{Ajani_2023_HOWLS}.

\subsection{Impact of the smoothing scale}\label{sec:smoothing_results}
\begin{figure}
    \centering

    \begin{subfigure}[b]{0.32\textwidth}
        \centering
        \includegraphics[width=\textwidth]{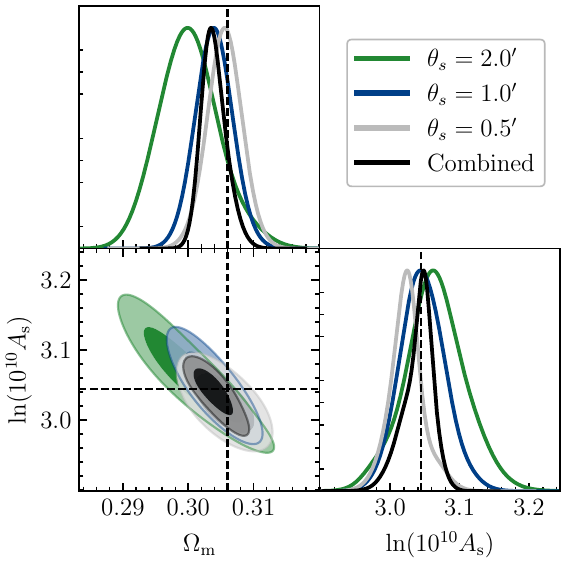}
        \caption{Redshift distribution}
        \label{fig:sub1}
    \end{subfigure}
    \hfill

    \begin{subfigure}[b]{0.32\textwidth}
        \centering
        \includegraphics[width=\textwidth]{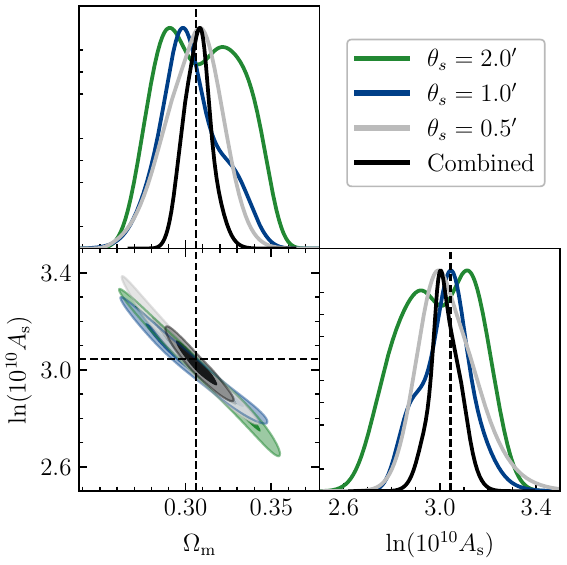}
        \caption{Height distribution}
        \label{fig:sub2}
    \end{subfigure}
    \hfill

    \begin{subfigure}[b]{0.32\textwidth}
        \centering
        \includegraphics[width=\textwidth]{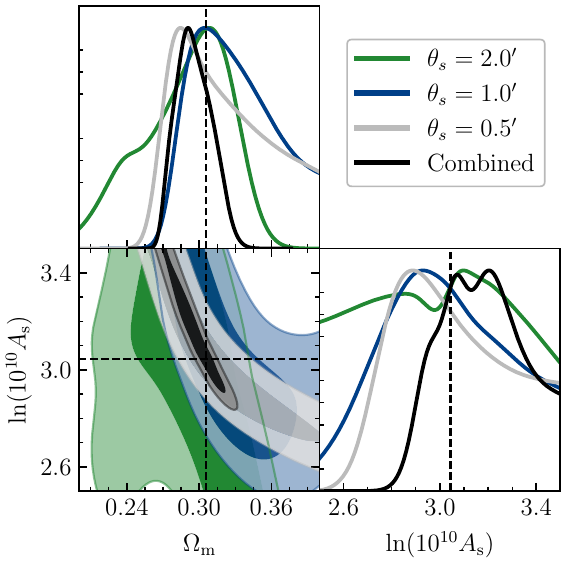}
        \caption{Angular clustering}
        \label{fig:sub3}
    \end{subfigure}

    \caption{Posteriors for the 2 dimensional forecast for $\Omega_\mathrm{m} - \ln (10^{10}A_\mathrm{s})$ using (a) the redshift distribution (top), (b) the height distribution (middle) or (c) the angular clustering (bottom) for 0.5\arcmin (grey), 1.0\arcmin (blue), 2.0\arcmin (green), or the combination of all three (black) smoothing scales. The redshift distribution shows the clearest improvement when combining multiple smoothing scales. The height distribution shows a weaker improvement, whereas the angular clustering shows the characteristic WL banana shape.}
    \label{fig:Om_as_smooths}
\end{figure}

\begin{figure}
    \centering

    \begin{subfigure}[b]{0.32\textwidth}
        \centering
        \includegraphics[width=\textwidth]{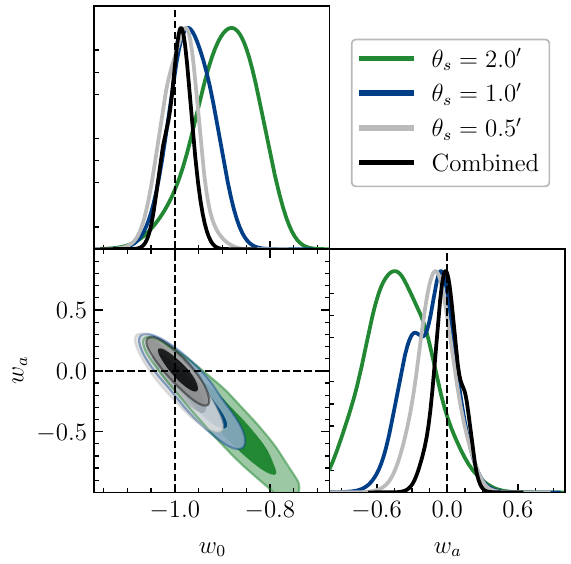}
        \caption{Redshift distribution}
        \label{fig:sub1}
    \end{subfigure}
    \hfill

    \begin{subfigure}[b]{0.32\textwidth}
        \centering
        \includegraphics[width=\textwidth]{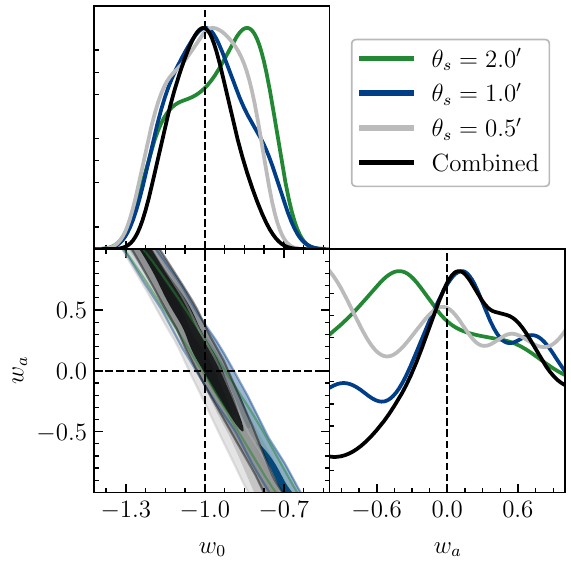}
        \caption{Height distribution}
        \label{fig:sub2}
    \end{subfigure}
    \hfill

    \begin{subfigure}[b]{0.32\textwidth}
        \centering
        \includegraphics[width=\textwidth]{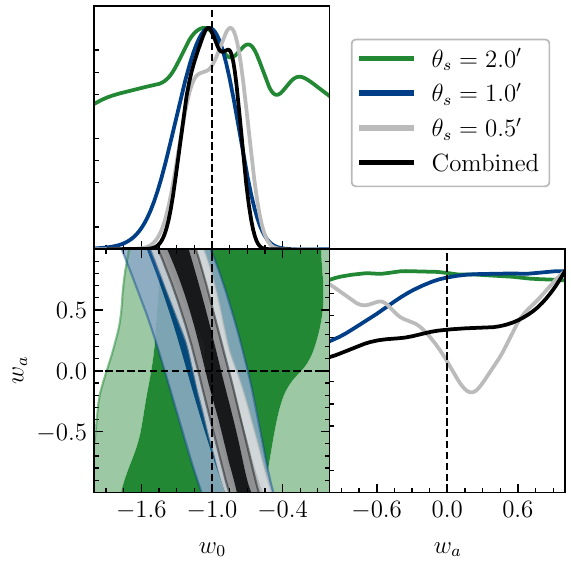}
        \caption{Angular clustering}
        \label{fig:sub3}
    \end{subfigure}

    \caption{Same as Fig.~\ref{fig:Om_as_smooths} but for $w_0-w_a$ posteriors. For the redshift distribution, a smaller smoothing scale yields tighter constraints, while combining smoothing scales yields significantly tighter constraints. The height distribution shows a weaker improvement, and even in the combined case the angular clustering remains insensitive to the time evolution of DE.}
    \label{fig:w0wa_smooths}
\end{figure}

\begin{table*}
\caption{Figure of merit (FoM) and figure of bias (FoB) for the $\Omega_\mathrm{m} - \ln(10^{10}A_\mathrm{s})$ and $w_0 - w_a$ parameter planes when all other cosmological parameters are fixed while varying the smoothing scale for a forecast using the redshift distribution, the height distribution, or angular clustering, at smoothing scales of $\theta_\mathrm{s} = 0.5$\arcmin, $1.0$\arcmin, or $2.0$\arcmin, and for the combination of all three scales. The FoM and FoB corresponding to contours that hit the prior bounds, as shown in Figs.~\ref{fig:Om_as_smooths}~\&~\ref{fig:w0wa_smooths}, should be treated carefully, as the prior truncation may bias the inferred covariance used to compute the FoM and FoB. The redshift distribution provides the tightest constraint on both parameter combinations. The FoM of all statistics and parameter combinations typically increases by a factor $\approx2$ when combining the different scales.}\label{tab:fomsandfobs}
\begin{tabular}{lrlrl}
\hline
& \multicolumn{2}{c}{$\Omega_\mathrm{m}-\ln(10^{10}A_\mathrm{s})$}
& \multicolumn{2}{c}{$w_0-w_a$} \\
\cline{2-3}\cline{4-5}
Statistic and smoothing scale & Figure of merit  & Figure of bias & Figure of merit & Figure of bias \\ \hline
Redshift distribution at $\theta_\mathrm{s} =$ 0.5\arcmin &         18780.5      &  0.87   &  347.2 &   0.90  \\
Redshift distribution at $\theta_\mathrm{s} =$ 1\arcmin &        17850.2      &  0.75  &  243.8 &  0.73   \\
Redshift distribution at $\theta_\mathrm{s} =$ 2\arcmin &      10744.9       &  1.31  &  151.9 &  2.35  \\
Redshift distribution at $\theta_\mathrm{s} =$ 0.5\arcmin + $\theta_\mathrm{s} =$ 1\arcmin + $\theta_\mathrm{s} =$ 2\arcmin &           38363.5   &  1.14   &  649.0 &   0.48   \\
\hline
Height distribution at $\theta_\mathrm{s} =$ 0.5\arcmin &      2868.8       &  0.26   & 59.6   & 1.24   \\
Height distribution at $\theta_\mathrm{s} =$ 1\arcmin &         2842.6      &   0.28  & 57.5 &   0.21  \\
Height distribution at $\theta_\mathrm{s} =$ 2\arcmin &        2359.0       &  0.07  & 39.8  &  1.05  \\
Height distribution at $\theta_\mathrm{s} =$ 0.5\arcmin + $\theta_\mathrm{s} =$ 1\arcmin + $\theta_\mathrm{s} =$ 2\arcmin &          6623.2    &  0.75   &  80.9  &  0.28  \\
\hline
Angular clustering at $\theta_\mathrm{s} =$ 0.5\arcmin &     251.2          &  0.50   &  16.4  &  0.14    \\
Angular clustering at $\theta_\mathrm{s} =$ 1\arcmin &       153.5        &   0.43  &  10.2  &   0.31  \\
Angular clustering at $\theta_\mathrm{s} =$ 2\arcmin &       105.2        &   0.26 &  3.2 &   0.41   \\
Angular clustering at $\theta_\mathrm{s} =$ 0.5\arcmin + $\theta_\mathrm{s} =$ 1\arcmin + $\theta_\mathrm{s} =$ 2\arcmin &           1344.7   &   0.95  &  25.1  &  0.11    \\
\hline
\end{tabular}
\end{table*}

In this section, we explore the impact of the smoothing scale on the cosmological constraints. We chose our fiducial smoothing scale based on the findings of \citet{Liu2015}, who found it to be the most constraining for their inference of the WL peak-height distribution. However, they already showed that combining scales may improve the results, leaving a more dedicated analysis to future work.  

Physically, a different smoothing scale corresponds to a different physical size at each lens redshift. In order to measure an object with the highest significance, the physical aperture corresponding to the smoothing scale should be similar to the object's size. The smoothing will thus affect which WL peaks are detected and at what significance level. By combining scales that are better suited for extracting smaller high-redshift or larger lower-redshift objects, more information could possibly be probed. Additionally, for the smallest scales, extra information may be obtained from different radial profile parts of haloes, as a single low-redshift object may generate multiple peaks.

We fix eight cosmological parameters to their true values and infer only the remaining two. Consequently, the absolute constraining power may differ from the previous figures. We focus on the $\Omega_\mathrm{m} - \ln(10^{10}A_\mathrm{s})$ and $w_0 - w_a$ planes because these are the parameter combinations that upcoming WL inferences will likely be most interested in. At the same time, because we did not re-optimise our emulators for different smoothing scales or for this 2D parameter forecast, we found the results to be less stable to small variations in the emulation set-up. We therefore decided to add the emulator uncertainty as an additional noise component to the covariance matrix $\mathbf{C}$  (Eq.~\ref{eqn:covmat}). Following the same leave-one-out approach described in Appendix~\ref{app:emulator}, we estimate an emulator error covariance matrix, $\Sigma_\mathrm{emu}$, from the residuals between the emulator predictions and the true measurements of the withheld hypercube nodes. The total covariance used in this section is then computed following the combined likelihood function for a multivariate $t$ and Gaussian distribution from \cite{Heavens2026}. The additional term increases the uncertainty, preventing the 2D forecast from becoming artificially overconfident due to emulator imperfections.

From top to bottom, Figs.~\ref{fig:Om_as_smooths} and \ref{fig:w0wa_smooths} show the marginalised 2D posterior distributions for the redshift distribution, height distribution, and angular clustering, respectively, for smoothing scales of $\theta_s = 0.5$ (grey), $1.0$ (blue), and $2.0$ (green) arcmin, as well as the forecast using the three scales combined (black). The results are further quantified in Table~\ref{tab:fomsandfobs}. There, for each statistic at the different smoothing scales, as indicated by the first column, the FoM in the $\Omega_\mathrm{m}-\ln(10^{10}A_\mathrm{s})$ and $w_0-w_a$ planes are listed in the second and fourth columns, respectively. We only interpret the relative changes in the FoMs due to our idealised forecasts. Similarly, the corresponding FoB values are listed in the third and fifth columns. 

The FoB at the 68th-percentile confidence level for this 2D inference is 1.52. Forecasts with a bias below this threshold are therefore less biased than the commonly used $1\sigma$ value to quantify the bias for a single parameter. The FoB values are below the 1.52 threshold in almost every case. Only for the $w_0-w_a$ forecast using the redshift distribution at $\theta_\mathrm{s}=2.0$\arcmin the FoB is 2.35. 

The three statistics show similar improvements when combining the three smoothing scales. For the redshift distribution, combining all three scales improves the FoM by a factor of 2.15 and 2.66 for $\Omega_\mathrm{m}-\ln(10^{10}A_\mathrm{s})$ and $w_0-w_a$, respectively, compared to only using our fiducial 1 arcmin scale. The absolute improvement is the largest of the three statistics. The height distribution shows a relatively similar improvement for $\Omega_\mathrm{m}-\ln(10^{10}A_\mathrm{s})$ of 2.33, while for $w_0-w_a$ the combined FoM improves by a factor of 1.41 relative to the fiducial scale. The relative gain is similar to that of the redshift distribution, even though the absolute constraining power of the height distribution on $w_0-w_a$ remains weaker than that of the redshift distribution at every individual scale. The angular clustering shows the largest relative improvement of all three statistics, with the combined FoM exceeding the fiducial-scale value by a factor of 8.76 for $\Omega_\mathrm{m}-\ln(10^{10}A_\mathrm{s})$ and 2.46 for $w_0-w_a$. As mentioned before, \citet{Davies2022} found that combining peaks selected at multiple SNR thresholds substantially improved angular clustering constraints. Instead, we compared peaks selected from the same SNR threshold but from different smoothing scales, which qualitatively probe the same information. We also find that the results improve significantly when combining multiple, differently selected angular clustering distributions. However, in terms of absolute constraining power, the angular clustering remains the poorest-performing statistic in both parameter planes.

Comparing the different statistics at the fiducial 1 arcmin smoothing scale, the redshift distribution dominates the constraining power in both parameter planes, exceeding the height distribution by a factor of 6 and the angular clustering by a factor of 116 in FoM for $\Omega_\mathrm{m}-\ln(10^{10}A_\mathrm{s})$, and by factors of 4 and 24, respectively, for $w_0-w_a$.

Regarding the sensitivity to the smoothing scale itself, each of the three statistics shows a monotonic trend in which the smallest scale yields the tightest constraints in both parameter planes. Given the number of covariance matrix realisations, it is not possible to include an extra scale in the forecast as the number of data vector bins for the redshift distribution would exceed the number of covariance matrix realisations, even though this might further improve the constraints. However, forecasts of each of the three statistics in both parameter planes, with 5 arcmin smoothing, show that constraints weaken further (not shown), consistent with the trend that smaller smoothing scales yield better constraints.

Compared to the findings of \citet{Liu2015}, who identified $\approx$1 arcmin as close to the optimal smoothing scale for a CFHTLenS-depth inference based solely on the peak number density, our results suggest that, for a \textit{Euclid} inference using SNR$>$3 peaks, a slightly smaller scale is preferred, likely due to (a combination of) the better angular resolution, lower noise levels and higher source number density. The smallest smoothing scale we consider corresponds to only 1.16 pixels for the $N_\mathrm{side} = 8192$ maps that are used. As \textsc{Healpix}' functionality, which we need to smooth and rotate our mass maps, is limited by 32-bit indexing, we did not explore smaller scales, even though they may provide even stronger constraints.

The statistics exhibit qualitatively different directions of degeneracy. The redshift and height distributions both show negatively correlated degeneracies between $\Omega_\mathrm{m}$ and $\ln(10^{10}A_\mathrm{s})$ at every smoothing scale, whereas for the angular clustering forecast, the $\theta_\mathrm{s} = 0.5$ and 1~arcmin $\Omega_\mathrm{m}-\ln(10^{10}A_\mathrm{s})$ posteriors show a banana-shaped degeneracy that is not apparent in the fiducial 10D forecast, which is similar to the banana-shaped posteriors typically found in WL two-point cosmic shear inferences in the $\sigma_8-\Omega_\mathrm{m}$ plane. In our inference, the curvature is a consequence of the non-monotonic $\Omega_\mathrm{m}$ dependence of the clustering signal, as shown before in Fig.~\ref{fig:sweep_omegam}. The clustering amplitude increases with $\Omega_\mathrm{m}$ only up to $\Omega_\mathrm{m} \approx 0.3$, beyond which the impact decreases and a reversed trend is even observed at large scales. At the same time, within our hypercube range, the dependence of clustering amplitude on $\ln(10^{10}A_\mathrm{s})$ is always monotonic, with larger values of $\ln(10^{10}A_\mathrm{s})$ resulting in a stronger clustering signal (not shown). Despite being noisier, the 2\arcmin scale shows a different degeneracy than the other two smoothing scales. Consequently, when combining the scales, the slightly different degeneracies cause a factor $\approx 9$ increase in the FoM compared to the fiducial smoothing scale. It is unclear whether this represents a different physical behaviour or is instead a result of some part of the pipeline. Similar, somewhat confusing behaviour can be seen in the $w_0-w_a$ posteriors for the angular clustering forecast at 2\arcmin, which appear to reverse the degeneracy direction, albeit remaining practically unconstrained within the prior ranges.

More generally, our exploration of sensitivity to the smoothing scale is illustrative rather than exhaustive. Although we typically find that the smaller smoothing scale yields the tightest constraint, we have considered only three discrete scales and only in two restricted parameter planes, and we have ignored baryonic physics, which potentially impacts the smallest scales the strongest. A more complete analysis that combines multiple smoothing scales and multiple statistics across the full 10D forecast is left for future work. Similarly, we did not explore the possibility of mixing different scales for the different statistics.

\section{Conclusions}\label{sec:conclusions}
We have investigated the potential of using WL peak properties to constrain cosmology in a \textit{Euclid}-like setup that indirectly aims to exploit the cosmology dependence of the mass, angular clustering, and redshift distribution of massive structures. We have done so by forward-modelling the final expected signal in a new DMO hypercube, a suite of simulations spanning a ten-dimensional cosmological parameter space varying the matter density, evolving dark energy, neutrino mass, and beyond $w_0w_a$CDM parameters that control the running of the scalar spectral index and the decay rate of cold dark matter. We have run an additional suite of 80 $\Lambda$CDM DMO simulations within the same simulation framework to estimate the covariance matrix at the fixed D3A cosmology. 

Within these simulations, we generate WL convergence maps from which we extract the local maxima, i.e. WL peaks. We study the cosmological constraining potential of the different WL peak properties, namely (the combination of) their angular clustering, redshift distribution, and height distributions. We find that different WL peak statistics probe different regions of the cosmological space. For our fiducial set-up that uses SNR > 3 peaks measured from a map smoothed with 1~arcmin, the WL peak redshift distribution provides the tightest constraints on the matter density, $\Omega_\mathrm{m}$, and on the dark energy equation-of-state parameters, $w_0$ and $w_a$. The angular clustering and peak-height distribution jointly constrain best the amplitude of the primordial power spectrum, $\ln(10^{10}A_\mathrm{s})$. The combination of different statistics can also probe the Hubble parameter, $h$, and the baryon matter density, $\Omega_\mathrm{b}$, through the breaking of parameter degeneracies. None of the considered WL peak statistics, even when combined, provide strong constraints on the dark matter decay rate, the neutrino mass, or the amplitude and running of the scalar spectral index.

We interpret the different cosmological constraining power by considering three pathways through which cosmological forecasts are affected by changes in cosmology, namely through direct impact on the matter density, changes in geometry, or changes in structure formation. Although it is difficult to isolate the effect of a single channel, we find that all channels likely contribute to the constraints, given that the different statistics exhibit distinct degeneracies and sensitivities to cosmological parameters.

We additionally compare the constraining power to that of the commonly used two-point cosmic-shear correlation functions. We find that for most parameters, the combination of WL peak statistics provides equal or better constraining power than the $\xi_{+/-}$. Only for the neutrino mass and the (running of the) scalar spectral index do we find that the shear-2PCFs outperform the WL peaks, which do not constrain these parameters well. When combining both, the constraints improve further.

We additionally explore the impact of the smoothing scale to explore the optimal scale and complementarity of different scales in the 2D $\Omega_\mathrm{m}-\ln(10^{10}A_\mathrm{s})$ and $w_0-w_a$ planes by comparing cosmological forecasts for peaks selected from maps smoothed with 0.5, 1, and 2~arcmin. For both parameter planes, we find the redshift distribution to provide the tightest constraints, with its FoM being a factor $\approx10-100$ larger than the corresponding FoMs for the angular clustering or height distribution. When combining different smoothing scales, which select partially independent sets of objects with different masses and redshifts, we find the FoMs to typically improve by a factor $\approx2$.

The cosmological forecast setup of this paper is idealised, as it ignores the impact of baryonic physics, intrinsic alignment, source clustering, reduced shear, and nuisance parameters that are expected to significantly impact the results. Although the impact of some of these effects and systematics is expected to be minor or qualitatively different for the different WL peak statistics we considered, future analyses will need to better account for these effects to obtain precise and accurate cosmological constraints from upcoming data from \textit{Euclid}, \textit{Roman}, and \textit{Rubin}.

In addition, we restricted our analysis to WL peaks and did not consider other higher-order WL statistics in detail, although including them can further improve the constraints \citep[see e.g.][for the inclusion of WL minima]{Marques2024}.

\section*{Acknowledgements}
This work was supported by the Science and Technology Facilities Council (grant number ST/Y002733/1). This project has received funding from the European Research Council (ERC) under the European Union's Horizon 2020 research and innovation programme (grant agreement No 769130). This work used the DiRAC@Durham facility managed by the Institute for Computational Cosmology on behalf of the STFC DiRAC HPC Facility (\url{www.dirac.ac.uk}). The equipment was funded by BEIS capital funding via STFC capital grants ST/K00042X/1, ST/P002293/1, ST/R002371/1 and ST/S002502/1, Durham University, and STFC operations grant ST/R000832/1. DiRAC is part of the National e-Infrastructure. We have used \textsc{astropy} \citep[][]{Astropy2022}, \textsc{healpy} \citep[][]{Zonca2019}, \textsc{matplotlib} \citep[][]{Hunter2007}, \textsc{numpy} \citep[][]{Harris2020}, and \textsc{swiftsimio} \citep[][]{Borrow2020} in our analysis.

\section*{Data Availability}
The data supporting the figures in this article are available upon reasonable request to the corresponding author. The hypercube will be presented in McCarthy~et~al.~(in preparation) and will eventually be made public.



\bibliographystyle{mnras}
\bibliography{example} 

@preamble{ "\providecommand{\noopsort}[1]{}" }

@ARTICLE{Broxteman2024_WLpeaks,
       author = {{Broxterman}, Jeger C. and {Schaller}, Matthieu and {Schaye}, Joop and {Hoekstra}, Henk and {Kuijken}, Konrad and {Helly}, John C. and {Kugel}, Roi and {Braspenning}, Joey and {Elbers}, Willem and {Frenk}, Carlos S. and {Kwan}, Juliana and {McCarthy}, Ian G. and {Salcido}, Jaime and {van Daalen}, Marcel P. and {Vandenbroucke}, Bert},
        title = "{The FLAMINGO project: baryonic impact on weak gravitational lensing convergence peak counts}",
      journal = {\mnras},
         year = 2024,
        month = apr,
       volume = {529},
       number = {3},
        pages = {2309-2326},
          doi = {10.1093/mnras/stae698},
archivePrefix = {arXiv},
       eprint = {2312.08450},
 primaryClass = {astro-ph.CO},
       adsurl = {https://ui.adsabs.harvard.edu/abs/2024MNRAS.529.2309B}
}

@ARTICLE{Canas2025,
       author = {{Euclid Collaboration: Ca{\~n}as-Herrera}, G. and {Goh}, L.~W.~K. and {Blot}, L. and others},
       title = "{Euclid preparation. XCVI. Cosmology Likelihood for Observables in Euclid (CLOE). 3. Inference and Forecasts}",
      journal = {A\&A, in press, \url{https://doi.org/10.1051/0004-6361/202556861}},
         year = 2025,
        month = oct,
          eid = {arXiv:2510.09153},
        pages = {arXiv:2510.09153},
          doi = {10.48550/arXiv.2510.09153},
archivePrefix = {arXiv},
       eprint = {2510.09153},
 primaryClass = {astro-ph.CO},
       adsurl = {https://ui.adsabs.harvard.edu/abs/2025arXiv251009153E}
}

@ARTICLE{Lesgourgues2011,
       author = {{Lesgourgues}, Julien},
        title = "{The Cosmic Linear Anisotropy Solving System (CLASS) I: Overview}",
      journal = {arXiv e-prints},
         year = 2011,
        month = apr,
          eid = {arXiv:1104.2932},
        pages = {arXiv:1104.2932},
          doi = {10.48550/arXiv.1104.2932},
archivePrefix = {arXiv},
       eprint = {1104.2932},
 primaryClass = {astro-ph.IM},
       adsurl = {https://ui.adsabs.harvard.edu/abs/2011arXiv1104.2932L}
}

@ARTICLE{Zhang2022,
       author = {{Zhang}, Tianyu and {Liu}, Xiangkun and {Wei}, Chengliang and {Li}, Guoliang and {Luo}, Yu and {Kang}, Xi and {Fan}, Zuhui},
        title = "{Effects of Galaxy Intrinsic Alignment on Weak Lensing Peak Statistics}",
      journal = {\apj},
         year = 2022,
        month = dec,
       volume = {940},
       number = {2},
          eid = {96},
        pages = {96},
          doi = {10.3847/1538-4357/ac9a4c},
archivePrefix = {arXiv},
       eprint = {2210.06761},
 primaryClass = {astro-ph.CO},
       adsurl = {https://ui.adsabs.harvard.edu/abs/2022ApJ...940...96Z}
}

@ARTICLE{Marques2024,
       author = {{Marques}, Gabriela A. and {Liu}, Jia and {Shirasaki}, Masato and {Thiele}, Leander and {Grand{\'o}n}, Daniela and {Huffenberger}, Kevin M. and {Cheng}, Sihao and {Harnois-D{\'e}raps}, Joachim and {Osato}, Ken and {Coulton}, William R.},
        title = "{Cosmology from weak lensing peaks and minima with Subaru Hyper Suprime-Cam Survey first-year data}",
      journal = {\mnras},
         year = 2024,
        month = mar,
       volume = {528},
       number = {3},
        pages = {4513-4527},
          doi = {10.1093/mnras/stae098},
archivePrefix = {arXiv},
       eprint = {2308.10866},
 primaryClass = {astro-ph.CO},
       adsurl = {https://ui.adsabs.harvard.edu/abs/2024MNRAS.528.4513M}
}

@ARTICLE{Simone2026,
       author = {{Vinciguerra}, Simone and {Martinet}, Nicolas and {Gatti}, Marco},
        title = "{Comparing explicit likelihood and likelihood-free simulation-based inference for weak lensing cosmic shear}",
      journal = {arXiv e-prints},
         year = 2026,
        month = jul,
          eid = {arXiv:2607.17942},
        pages = {arXiv:2607.17942},
archivePrefix = {arXiv},
       eprint = {2607.17942},
 primaryClass = {astro-ph.CO},
       adsurl = {https://ui.adsabs.harvard.edu/abs/2026arXiv260717942V}
}

@ARTICLE{Reischke2026,
       author = {{Reischke}, Robert and {St{\"o}lzner}, Benjamin and {Joachimi}, Benjamin and {Wright}, Angus H. and {Asgari}, Marika and {Bilicki}, Maciej and {Elisa Chisari}, Nora and {Dvornik}, Andrej and {Georgiou}, Christos and {Giblin}, Benjamin and {Harnois-D{\'e}raps}, Joachim and {Heymans}, Catherine and {Hildebrandt}, Hendrik and {Hoekstra}, Henk and {Joudaki}, Shahab and {Kuijken}, Konrad and {Li}, Shun-Sheng and {Linke}, Laila and {Loureiro}, Arthur and {Mahony}, Constance and {Moscardini}, Lauro and {Napolitano}, Nicola R. and {Porth}, Lucas and {Radovich}, Mario and {Tr{\"o}ster}, Tilman and {von Wietersheim-Kramsta}, Maximilian and {Yan}, Ziang and {Yoon}, Mijin and {Zhang}, Yun-Hao},
        title = "{KiDS-Legacy: Constraining dark energy, neutrino mass, and curvature}",
      journal = {\aap},
         year = 2026,
        month = may,
       volume = {709},
          eid = {A82},
        pages = {A82},
          doi = {10.1051/0004-6361/202558581},
archivePrefix = {arXiv},
       eprint = {2512.11041},
 primaryClass = {astro-ph.CO},
       adsurl = {https://ui.adsabs.harvard.edu/abs/2026A&A...709A..82R}
}

@ARTICLE{Broxterman2025_WLpeaks,
       author = {{Broxterman}, Jeger C. and {Schaller}, Matthieu and {Hoekstra}, Henk and {Schaye}, Joop and {McGibbon}, Robert J. and {Forouhar Moreno}, Victor J. and {Kugel}, Roi and {Elbers}, Willem},
        title = "{The FLAMINGO project: cosmology with the redshift dependence of weak gravitational lensing peaks}",
      journal = {\mnras},
         year = 2025,
        month = apr,
       volume = {538},
       number = {2},
        pages = {755-774},
          doi = {10.1093/mnras/staf357},
archivePrefix = {arXiv},
       eprint = {2412.02736},
 primaryClass = {astro-ph.CO},
       adsurl = {https://ui.adsabs.harvard.edu/abs/2025MNRAS.538..755B}
}

@ARTICLE{Schaye2023_flamingo,
       author = {{Schaye}, Joop and {Kugel}, Roi and {Schaller}, Matthieu and {Helly}, John C. and {Braspenning}, Joey and {Elbers}, Willem and {McCarthy}, Ian G. and {van Daalen}, Marcel P. and {Vandenbroucke}, Bert and {Frenk}, Carlos S. and {Kwan}, Juliana and {Salcido}, Jaime and {Bah{\'e}}, Yannick M. and {Borrow}, Josh and {Chaikin}, Evgenii and {Hahn}, Oliver and {Hu{\v{s}}ko}, Filip and {Jenkins}, Adrian and {Lacey}, Cedric G. and {Nobels}, Folkert S.~J.},
        title = "{The FLAMINGO project: cosmological hydrodynamical simulations for large-scale structure and galaxy cluster surveys}",
      journal = {\mnras},
         year = 2023,
        month = dec,
       volume = {526},
       number = {4},
        pages = {4978-5020},
          doi = {10.1093/mnras/stad2419},
archivePrefix = {arXiv},
       eprint = {2306.04024},
 primaryClass = {astro-ph.CO},
       adsurl = {https://ui.adsabs.harvard.edu/abs/2023MNRAS.526.4978S}
}

@ARTICLE{Kugel2023_flamingo,
       author = {{Kugel}, Roi and {Schaye}, Joop and {Schaller}, Matthieu and {Helly}, John C. and {Braspenning}, Joey and {Elbers}, Willem and {Frenk}, Carlos S. and {McCarthy}, Ian G. and {Kwan}, Juliana and {Salcido}, Jaime and {van Daalen}, Marcel P. and {Vandenbroucke}, Bert and {Bah{\'e}}, Yannick M. and {Borrow}, Josh and {Chaikin}, Evgenii and {Hu{\v{s}}ko}, Filip and {Jenkins}, Adrian and {Lacey}, Cedric G. and {Nobels}, Folkert S.~J. and {Vernon}, Ian},
        title = "{FLAMINGO: calibrating large cosmological hydrodynamical simulations with machine learning}",
      journal = {\mnras},
         year = 2023,
        month = dec,
       volume = {526},
       number = {4},
        pages = {6103-6127},
          doi = {10.1093/mnras/stad2540},
archivePrefix = {arXiv},
       eprint = {2306.05492},
 primaryClass = {astro-ph.CO},
       adsurl = {https://ui.adsabs.harvard.edu/abs/2023MNRAS.526.6103K}
}

@ARTICLE{Manodeep2020_corrfunc,
    author = {{Sinha}, Manodeep and {Garrison}, Lehman H.},
    title = "{CORRFUNC - a suite of blazing fast correlation functions on
    the CPU}",
    journal = {\mnras},
    year = "2020",
    month = "Jan",
    volume = {491},
    number = {2},
    pages = {3022-3041},
    doi = {10.1093/mnras/stz3157},
    adsurl =
    {https://ui.adsabs.harvard.edu/abs/2020MNRAS.491.3022S}
}

@ARTICLE{LandySzalay1993,
       author = {{Landy}, Stephen D. and {Szalay}, Alexander S.},
        title = "{Bias and Variance of Angular Correlation Functions}",
      journal = {\apj},
         year = 1993,
        month = jul,
       volume = {412},
        pages = {64},
          doi = {10.1086/172900},
       adsurl = {https://ui.adsabs.harvard.edu/abs/1993ApJ...412...64L}
}

@ARTICLE{Mellier2025_EuclidSkyOverview,
author = {{Euclid Collaboration: Mellier}, Y. and {Abdurro'uf} and {Acevedo~Barroso}, J.A. and others},
	title = {Euclid - I. Overview of the Euclid mission},
	DOI= "10.1051/0004-6361/202450810",
	url= "https://doi.org/10.1051/0004-6361/202450810",
	journal = {A\&A},
	year = 2025,
	volume = 697,
	pages = "A1",
}

@ARTICLE{Ajani_2023_HOWLS,
       author = {{Euclid Collaboration: Ajani}, V. and {Baldi}, M. and {Barthelemy}, A. and others},
        title = "{\Euclid preparation. XXVIII. Forecasts for ten different higher-order weak lensing statistics}",
      journal = {\aap},
         year = 2023,
        month = jul,
       volume = {675},
          eid = {A120},
        pages = {A120},
          doi = {10.1051/0004-6361/202346017},
archivePrefix = {arXiv},
       eprint = {2301.12890},
 primaryClass = {astro-ph.CO},
       adsurl = {https://ui.adsabs.harvard.edu/abs/2023A&A...675A.120E}
}

@ARTICLE{Fields2020,
       author = {{Fields}, Brian D. and {Olive}, Keith A. and {Yeh}, Tsung-Han and {Young}, Charles},
        title = "{Big-Bang Nucleosynthesis after Planck}",
      journal = {\jcap},
         year = 2020,
        month = mar,
       volume = {2020},
       number = {3},
          eid = {010},
        pages = {010},
          doi = {10.1088/1475-7516/2020/03/010},
archivePrefix = {arXiv},
       eprint = {1912.01132},
 primaryClass = {astro-ph.CO},
       adsurl = {https://ui.adsabs.harvard.edu/abs/2020JCAP...03..010F}
}

@ARTICLE{Schneider2002,
       author = {{Schneider}, P. and {van Waerbeke}, L. and {Mellier}, Y.},
        title = "{B-modes in cosmic shear from source redshift clustering}",
      journal = {\aap},
         year = 2002,
        month = jul,
       volume = {389},
        pages = {729-741},
          doi = {10.1051/0004-6361:20020626},
archivePrefix = {arXiv},
       eprint = {astro-ph/0112441},
 primaryClass = {astro-ph},
       adsurl = {https://ui.adsabs.harvard.edu/abs/2002A&A...389..729S}
}

@ARTICLE{Elbers2025DESI,
       author = {{Elbers}, W. and {Aviles}, A. and {Noriega}, H.~E. and {Chebat}, D. and {Menegas}, A. and {Frenk}, C.~S. and {Garcia-Quintero}, C. and {Gonzalez}, D. and {Ishak}, M. and {Lahav}, O. and {Naidoo}, K. and {Niz}, G. and {Y{\`e}che}, C. and {Abdul-Karim}, M. and {Ahlen}, S. and {Alves}, O. and {Andrade}, U. and {Armengaud}, E. and {Behera}, J. and {BenZvi}, S. and {Bianchi}, D. and {Brieden}, S. and {Brodzeller}, A. and {Brooks}, D. and {Burtin}, E. and {Calderon}, R. and {Canning}, R. and {Carnero Rosell}, A. and {Casas}, L. and {Castander}, F.~J. and {Charles}, M. and {Chaussidon}, E. and {Chaves-Montero}, J. and {Claybaugh}, T. and {Cole}, S. and {Cooper}, A.~P. and {Cuceu}, A. and {Dawson}, K.~S. and {de la Macorra}, A. and {de Mattia}, A. and {Deiosso}, N. and {Dey}, A. and {Dey}, B. and {Ding}, Z. and {Doel}, P. and {Eisenstein}, D.~J. and {Ferraro}, S. and {Font-Ribera}, A. and {Forero-Romero}, J.~E. and {Garrison}, L.~H. and {Gazta{\~n}aga}, E. and {Gil-Mar{\'\i}n}, H. and {Gontcho}, S. Gontcho A. and {Gonzalez-Morales}, A.~X. and {Gutierrez}, G. and {He}, S. and {Herbold}, M. and {Herrera-Alcantar}, H.~K. and {Howlett}, C. and {Huterer}, D. and {Juneau}, S. and {Kehoe}, R. and {Kirkby}, D. and {Kisner}, T. and {Kremin}, A. and {Lamman}, C. and {Landriau}, M. and {Le Guillou}, L. and {Leauthaud}, A. and {Levi}, M.~E. and {Li}, Q. and {Lodha}, K. and {Magneville}, C. and {Manera}, M. and {Martini}, P. and {Matthewson}, W.~L. and {Meisner}, A. and {Mena-Fern{\'a}ndez}, J. and {Miquel}, R. and {Moustakas}, J. and {Nadathur}, S. and {Newman}, J.~A. and {Paillas}, E. and {Palanque-Delabrouille}, N. and {Percival}, W.~J. and {Pieri}, M.~M. and {Poppett}, C. and {Prada}, F. and {P{\'e}rez-R{\`a}fols}, I. and {Rabinowitz}, D. and {Ram{\'\i}rez-P{\'e}rez}, C. and {Rashkovetskyi}, M. and {Ravoux}, C. and {Rivera-Morales}, H. and {Rohlf}, J. and {Ross}, A.~J. and {Rossi}, G. and {Ruhlmann-Kleider}, V. and {Samushia}, L. and {Sanchez}, E. and {Schlegel}, D. and {Schubnell}, M. and {Seo}, H. and {Sinigaglia}, F. and {Sprayberry}, D. and {Tan}, T. and {Tarl{\'e}}, G. and {Taylor}, P. and {Turner}, W. and {Vargas-Maga{\~n}a}, M. and {Verde}, L. and {Walther}, M. and {Weaver}, B.~A. and {Whitford}, A. and {Wolfson}, M. and {Zarrouk}, P. and {Zhao}, C. and {Zhou}, R. and {Zou}, H. and {DESI Collaboration}},
        title = "{Constraints on neutrino physics from DESI DR2 BAO and DR1 full shape}",
      journal = {\prd},
         year = 2025,
        month = oct,
       volume = {112},
       number = {8},
          eid = {083513},
        pages = {083513},
          doi = {10.1103/w9pk-xsk7},
archivePrefix = {arXiv},
       eprint = {2503.14744},
 primaryClass = {astro-ph.CO},
       adsurl = {https://ui.adsabs.harvard.edu/abs/2025PhRvD.112h3513E}
}

@ARTICLE{Vedder2026,
       author = {{Vedder}, Casper and {Bakx}, Thomas and {Chisari}, Nora Elisa and {Hoekstra}, Henk and {Schaller}, Matthieu},
        title = "{Three-point intrinsic alignments of galaxies and haloes in the FLAMINGO simulations}",
      journal = {\mnras},
         year = 2026,
        month = aug,
       volume = {550},
       number = {2},
          eid = {stag1008},
        pages = {stag1008},
          doi = {10.1093/mnras/stag1008},
archivePrefix = {arXiv},
       eprint = {2601.17914},
 primaryClass = {astro-ph.CO},
       adsurl = {https://ui.adsabs.harvard.edu/abs/2026MNRAS.550g1008V}
}

@ARTICLE{Lewis2025_getdist,
       author = {{Lewis}, Antony},
        title = "{GetDist: a Python package for analysing Monte Carlo samples}",
      journal = {\jcap},
         year = 2025,
        month = aug,
       volume = {2025},
       number = {8},
          eid = {025},
        pages = {025},
          doi = {10.1088/1475-7516/2025/08/025},
archivePrefix = {arXiv},
       eprint = {1910.13970},
 primaryClass = {astro-ph.IM},
       adsurl = {https://ui.adsabs.harvard.edu/abs/2025JCAP...08..025L}
}

@ARTICLE{Schaller2024_swift,
       author = {{Schaller}, Matthieu and {Borrow}, Josh and {Draper}, Peter W. and {Ivkovic}, Mladen and {McAlpine}, Stuart and {Vandenbroucke}, Bert and {Bah{\'e}}, Yannick and {Chaikin}, Evgenii and {Chalk}, Aidan B.~G. and {Chan}, Tsang Keung and {Correa}, Camila and {van Daalen}, Marcel and {Elbers}, Willem and {Gonnet}, Pedro and {Hausammann}, Lo{\"\i}c and {Helly}, John and {Hu{\v{s}}ko}, Filip and {Kegerreis}, Jacob A. and {Nobels}, Folkert S.~J. and {Ploeckinger}, Sylvia and {Revaz}, Yves and {Roper}, William J. and {Ruiz-Bonilla}, Sergio and {Sandnes}, Thomas D. and {Uyttenhove}, Yolan and {Willis}, James S. and {Xiang}, Zhen},
        title = "{SWIFT: A modern highly-parallel gravity and smoothed particle hydrodynamics solver for astrophysical and cosmological applications}",
      journal = {\mnras},
         year = 2024,
        month = may,
       volume = {530},
       number = {2},
        pages = {2378-2419},
          doi = {10.1093/mnras/stae922},
archivePrefix = {arXiv},
       eprint = {2305.13380},
 primaryClass = {astro-ph.IM},
       adsurl = {https://ui.adsabs.harvard.edu/abs/2024MNRAS.530.2378S}
}

@software{Hahn2020_monofonic,
       author = {{Hahn}, Oliver and {Michaux}, Micha{\"e}l and {Rampf}, Cornelius and {Uhlemann}, Cora and {Angulo}, Raul E.},
        title = "{MUSIC2-monofonIC: 3LPT initial condition generator}",
 howpublished = {Astrophysics Source Code Library, record ascl:2008.024},
         year = 2020,
        month = aug,
          eid = {ascl:2008.024},
archivePrefix = {ascl},
       eprint = {2008.024},
       adsurl = {https://ui.adsabs.harvard.edu/abs/2020ascl.soft08024H}
}

@ARTICLE{Liu2015,
       author = {{Liu}, Jia and {Petri}, Andrea and {Haiman}, Zolt{\'a}n and {Hui}, Lam and {Kratochvil}, Jan M. and {May}, Morgan},
        title = "{Cosmology constraints from the weak lensing peak counts and the power spectrum in CFHTLenS data}",
      journal = {\prd},
         year = 2015,
        month = mar,
       volume = {91},
       number = {6},
          eid = {063507},
        pages = {063507},
          doi = {10.1103/PhysRevD.91.063507},
archivePrefix = {arXiv},
       eprint = {1412.0757},
 primaryClass = {astro-ph.CO},
       adsurl = {https://ui.adsabs.harvard.edu/abs/2015PhRvD..91f3507L}
}

@ARTICLE{Wang2026,
       author = {{Wang}, Jiaqi and {Elbers}, Willem and {Frenk}, Carlos S. and {Cole}, Shaun and {Yang}, Xiaohu and {McCarthy}, Ian G. and {van de Weygaert}, Rien},
        title = "{Revealing the neutrino mass through persistent homology of the cosmic web}",
      journal = {\mnras},
         year = 2026,
        month = jun,
       volume = {549},
       number = {2},
          eid = {stag937},
        pages = {stag937},
          doi = {10.1093/mnras/stag937},
archivePrefix = {arXiv},
       eprint = {2604.02300},
 primaryClass = {astro-ph.CO},
       adsurl = {https://ui.adsabs.harvard.edu/abs/2026MNRAS.549ag937W}
}

@ARTICLE{Heavens2026,
       author = {{Heavens}, Alan and {Whiteway}, Lorne and {Sellentin}, Elena},
        title = "{On combining estimated and analytic covariance matrices}",
      journal = {The Open Journal of Astrophysics},
         year = 2026,
        month = jul,
       volume = {9},
        pages = {65025},
          doi = {10.33232/001c.165025},
archivePrefix = {arXiv},
       eprint = {2604.19463},
 primaryClass = {astro-ph.CO},
       adsurl = {https://ui.adsabs.harvard.edu/abs/2026OJAp....965025H}
}

@ARTICLE{Michaux2021_monofonic,
       author = {{Michaux}, Micha{\"e}l and {Hahn}, Oliver and {Rampf}, Cornelius and {Angulo}, Raul E.},
        title = "{Accurate initial conditions for cosmological N-body simulations: minimizing truncation and discreteness errors}",
      journal = {\mnras},
         year = 2021,
        month = jan,
       volume = {500},
       number = {1},
        pages = {663-683},
          doi = {10.1093/mnras/staa3149},
archivePrefix = {arXiv},
       eprint = {2008.09588},
 primaryClass = {astro-ph.CO},
       adsurl = {https://ui.adsabs.harvard.edu/abs/2021MNRAS.500..663M}
}

@ARTICLE{Blanchard2020_Euclid,
       author = {{Euclid Collaboration: Blanchard}, A. and {Camera}, S. and {Carbone}, C. and others},
        title = "{Euclid preparation. VII. Forecast validation for Euclid cosmological probes}",
      journal = {\aap},
         year = 2020,
        month = oct,
       volume = {642},
          eid = {A191},
        pages = {A191},
          doi = {10.1051/0004-6361/202038071},
archivePrefix = {arXiv},
       eprint = {1910.09273},
 primaryClass = {astro-ph.CO},
       adsurl = {https://ui.adsabs.harvard.edu/abs/2020A&A...642A.191E}
}

@ARTICLE{Semboloni2013,
       author = {{Semboloni}, Elisabetta and {Hoekstra}, Henk and {Schaye}, Joop},
        title = "{Effect of baryonic feedback on two- and three-point shear statistics: prospects for detection and improved modelling}",
      journal = {\mnras},
         year = 2013,
        month = sep,
       volume = {434},
       number = {1},
        pages = {148-162},
          doi = {10.1093/mnras/stt1013},
archivePrefix = {arXiv},
       eprint = {1210.7303},
 primaryClass = {astro-ph.CO},
       adsurl = {https://ui.adsabs.harvard.edu/abs/2013MNRAS.434..148S}
}

@ARTICLE{Semboloni2011,
       author = {{Semboloni}, Elisabetta and {Hoekstra}, Henk and {Schaye}, Joop and {van Daalen}, Marcel P. and {McCarthy}, Ian G.},
        title = "{Quantifying the effect of baryon physics on weak lensing tomography}",
      journal = {\mnras},
         year = 2011,
        month = nov,
       volume = {417},
       number = {3},
        pages = {2020-2035},
          doi = {10.1111/j.1365-2966.2011.19385.x},
archivePrefix = {arXiv},
       eprint = {1105.1075},
 primaryClass = {astro-ph.CO},
       adsurl = {https://ui.adsabs.harvard.edu/abs/2011MNRAS.417.2020S}
}

@ARTICLE{Bradac2005,
       author = {{Brada{\v{c}}}, M. and {Schneider}, P. and {Lombardi}, M. and {Erben}, T.},
        title = "{Strong and weak lensing united. I. The combined strong and weak lensing cluster mass reconstruction method}",
      journal = {\aap},
         year = 2005,
        month = jul,
       volume = {437},
       number = {1},
        pages = {39-48},
          doi = {10.1051/0004-6361:20042233},
       adsurl = {https://ui.adsabs.harvard.edu/abs/2005A&A...437...39B}
}

@ARTICLE{Gatti2024,
       author = {{Gatti}, M. and {Jeffrey}, N. and {Whiteway}, L. and {Ajani}, V. and {Kacprzak}, T. and {Z{\"u}rcher}, D. and {Chang}, C. and {Jain}, B. and {Blazek}, J. and {Krause}, E. and {Alarcon}, A. and {Amon}, A. and {Bechtol}, K. and {Becker}, M. and {Bernstein}, G. and {Campos}, A. and {Chen}, R. and {Choi}, A. and {Davis}, C. and {Derose}, J. and {Diehl}, H.~T. and {Dodelson}, S. and {Doux}, C. and {Eckert}, K. and {Elvin-Poole}, J. and {Everett}, S. and {Ferte}, A. and {Gruen}, D. and {Gruendl}, R. and {Harrison}, I. and {Hartley}, W.~G. and {Herner}, K. and {Huff}, E.~M. and {Jarvis}, M. and {Kuropatkin}, N. and {Leget}, P.~F. and {MacCrann}, N. and {McCullough}, J. and {Myles}, J. and {Navarro-Alsina}, A. and {Pandey}, S. and {Prat}, J. and {Raveri}, M. and {Rollins}, R.~P. and {Roodman}, A. and {Sanchez}, C. and {Secco}, L.~F. and {Sevilla-Noarbe}, I. and {Sheldon}, E. and {Shin}, T. and {Troxel}, M. and {Tutusaus}, I. and {Varga}, T.~N. and {Yanny}, B. and {Yin}, B. and {Zhang}, Y. and {Zuntz}, J. and {Allam}, S.~S. and {Alves}, O. and {Aguena}, M. and {Bacon}, D. and {Bertin}, E. and {Brooks}, D. and {Burke}, D.~L. and {Rosell}, A. Carnero and {Carretero}, J. and {Cawthon}, R. and {da Costa}, L.~N. and {Davis}, T.~M. and {De Vicente}, J. and {Desai}, S. and {Doel}, P. and {Garc{\'\i}a-Bellido}, J. and {Giannini}, G. and {Gutierrez}, G. and {Ferrero}, I. and {Frieman}, J. and {Hinton}, S.~R. and {Hollowood}, D.~L. and {Honscheid}, K. and {James}, D.~J. and {Kuehn}, K. and {Lahav}, O. and {Marshall}, J.~L. and {Mena-Fern{\'a}ndez}, J. and {Miquel}, R. and {Ogando}, R.~L.~C. and {Palmese}, A. and {Pereira}, M.~E.~S. and {Malag{\'o}n}, A.~A. Plazas and {Rodriguez-Monroy}, M. and {Samuroff}, S. and {Sanchez}, E. and {Schubnell}, M. and {Smith}, M. and {Sobreira}, F. and {Suchyta}, E. and {Swanson}, M.~E.~C. and {Tarle}, G. and {Weaverdyck}, N. and {Wiseman}, P. and {DES Collaboration}},
        title = "{Detection of the significant impact of source clustering on higher order statistics with DES Year 3 weak gravitational lensing data}",
      journal = {\mnras},
         year = 2024,
        month = jan,
       volume = {527},
       number = {1},
        pages = {L115-L121},
          doi = {10.1093/mnrasl/slad143},
archivePrefix = {arXiv},
       eprint = {2307.13860},
 primaryClass = {astro-ph.CO},
       adsurl = {https://ui.adsabs.harvard.edu/abs/2024MNRAS.527L.115G}
}

@ARTICLE{Lee2026,
       author = {{Lee}, Max E. and {Haiman}, Zolt{\'a}n and {Pandey}, Shivam and {Genel}, Shy},
        title = "{The Effect of Intrinsic Alignments on Weak-lensing Statistics in Hydrodynamical Simulations}",
      journal = {\apj},
         year = 2026,
        month = jan,
       volume = {996},
       number = {1},
          eid = {36},
        pages = {36},
          doi = {10.3847/1538-4357/ae1ca7},
archivePrefix = {arXiv},
       eprint = {2504.12460},
 primaryClass = {astro-ph.CO},
       adsurl = {https://ui.adsabs.harvard.edu/abs/2026ApJ...996...36L}
}

@ARTICLE{Lu2021,
       author = {{Lu}, Tianhuan and {Haiman}, Zolt{\'a}n},
        title = "{The impact of baryons on cosmological inference from weak lensing statistics}",
      journal = {\mnras},
         year = 2021,
        month = sep,
       volume = {506},
       number = {3},
        pages = {3406-3417},
          doi = {10.1093/mnras/stab1978},
archivePrefix = {arXiv},
       eprint = {2104.04165},
 primaryClass = {astro-ph.CO},
       adsurl = {https://ui.adsabs.harvard.edu/abs/2021MNRAS.506.3406L}
}

@ARTICLE{Chevallier2001,
       author = {{Chevallier}, Michel and {Polarski}, David},
        title = "{Accelerating Universes with Scaling Dark Matter}",
      journal = {International Journal of Modern Physics D},
         year = 2001,
        month = jan,
       volume = {10},
       number = {2},
        pages = {213-223},
          doi = {10.1142/S0218271801000822},
archivePrefix = {arXiv},
       eprint = {gr-qc/0009008},
 primaryClass = {gr-qc},
       adsurl = {https://ui.adsabs.harvard.edu/abs/2001IJMPD..10..213C}
}

@ARTICLE{Linder2003,
       author = {{Linder}, Eric V.},
        title = "{Exploring the Expansion History of the Universe}",
      journal = {\prl},
         year = 2003,
        month = mar,
       volume = {90},
       number = {9},
          eid = {091301},
        pages = {091301},
          doi = {10.1103/PhysRevLett.90.091301},
archivePrefix = {arXiv},
       eprint = {astro-ph/0208512},
 primaryClass = {astro-ph},
       adsurl = {https://ui.adsabs.harvard.edu/abs/2003PhRvL..90i1301L}
}

@ARTICLE{Elbers2025,
       author = {{Elbers}, Willem and {Frenk}, Carlos S. and {Jenkins}, Adrian and {Li}, Baojiu and {Helly}, John C. and {Kugel}, Roi and {Schaller}, Matthieu and {Schaye}, Joop and {Braspenning}, Joey and {Kwan}, Juliana and {McCarthy}, Ian G. and {Salcido}, Jaime and {van Daalen}, Marcel P. and {Vandenbroucke}, Bert and {Pascoli}, Silvia},
        title = "{The FLAMINGO project: the coupling between baryonic feedback and cosmology in light of the S$_{8}$ tension}",
      journal = {\mnras},
         year = 2025,
        month = feb,
       volume = {537},
       number = {2},
        pages = {2160-2178},
          doi = {10.1093/mnras/staf093},
archivePrefix = {arXiv},
       eprint = {2403.12967},
 primaryClass = {astro-ph.CO},
       adsurl = {https://ui.adsabs.harvard.edu/abs/2025MNRAS.537.2160E}
}

@ARTICLE{Gorski2005,
       author = {{G{\'o}rski}, K.~M. and {Hivon}, E. and {Banday}, A.~J. and {Wandelt}, B.~D. and {Hansen}, F.~K. and {Reinecke}, M. and {Bartelmann}, M.},
        title = "{HEALPix: A Framework for High-Resolution Discretization and Fast Analysis of Data Distributed on the Sphere}",
      journal = {\apj},
         year = 2005,
        month = apr,
       volume = {622},
       number = {2},
        pages = {759-771},
          doi = {10.1086/427976},
archivePrefix = {arXiv},
       eprint = {astro-ph/0409513},
 primaryClass = {astro-ph},
       adsurl = {https://ui.adsabs.harvard.edu/abs/2005ApJ...622..759G}
}

@ARTICLE{Hunter2007,
       author = {{Hunter}, John D.},
        title = "{Matplotlib: A 2D Graphics Environment}",
      journal = {Computing in Science and Engineering},
         year = 2007,
        month = jan,
       volume = {9},
       number = {3},
        pages = {90-95},
          doi = {10.1109/MCSE.2007.55},
       adsurl = {https://ui.adsabs.harvard.edu/abs/2007CSE.....9...90H}
}

@ARTICLE{Pantos2026,
       author = {{Pantos}, Ioannis and {Perivolaropoulos}, Leandros},
        title = "{Status of the S$_{8}$ tension: A 2026 review of probe discrepancies}",
      journal = {Physics of the Dark Universe},
         year = 2026,
        month = jun,
       volume = {52},
          eid = {102286},
        pages = {102286},
          doi = {10.1016/j.dark.2026.102286},
archivePrefix = {arXiv},
       eprint = {2602.12238},
 primaryClass = {astro-ph.CO},
       adsurl = {https://ui.adsabs.harvard.edu/abs/2026PDU....5202286P}
}

@ARTICLE{Yang2025,
       author = {{Yang}, Yanhui and {Bird}, Simeon and {Ho}, Ming-Feng},
        title = "{Ten-parameter simulation suite for cosmological emulation beyond {\ensuremath{\Lambda}}CDM}",
      journal = {\prd},
         year = 2025,
        month = apr,
       volume = {111},
       number = {8},
          eid = {083529},
        pages = {083529},
          doi = {10.1103/PhysRevD.111.083529},
archivePrefix = {arXiv},
       eprint = {2501.06296},
 primaryClass = {astro-ph.CO},
       adsurl = {https://ui.adsabs.harvard.edu/abs/2025PhRvD.111h3529Y}
}

@ARTICLE{Heitmann2009,
       author = {{Heitmann}, Katrin and {Higdon}, David and {White}, Martin and {Habib}, Salman and {Williams}, Brian J. and {Lawrence}, Earl and {Wagner}, Christian},
        title = "{The Coyote Universe. II. Cosmological Models and Precision Emulation of the Nonlinear Matter Power Spectrum}",
      journal = {\apj},
         year = 2009,
        month = nov,
       volume = {705},
       number = {1},
        pages = {156-174},
          doi = {10.1088/0004-637X/705/1/156},
archivePrefix = {arXiv},
       eprint = {0902.0429},
 primaryClass = {astro-ph.CO},
       adsurl = {https://ui.adsabs.harvard.edu/abs/2009ApJ...705..156H}
}

@ARTICLE{Cheng2020,
       author = {{Cheng}, Sihao and {Ting}, Yuan-Sen and {M{\'e}nard}, Brice and {Bruna}, Joan},
        title = "{A new approach to observational cosmology using the scattering transform}",
      journal = {\mnras},
         year = 2020,
        month = dec,
       volume = {499},
       number = {4},
        pages = {5902-5914},
          doi = {10.1093/mnras/staa3165},
archivePrefix = {arXiv},
       eprint = {2006.08561},
 primaryClass = {astro-ph.CO},
       adsurl = {https://ui.adsabs.harvard.edu/abs/2020MNRAS.499.5902C}
}

@ARTICLE{DES2026_wowa,
       author = {{DES Collaboration} and {Abbott}, T.~M.~C. and {Adamow}, M. and {Aguena}, M. and {Alarcon}, A. and {Allam}, S. and {Alves}, O. and {Amon}, A. and {Anbajagane}, D. and {Andrade-Oliveira}, F. and {Armstrong}, P. and {Avila}, S. and {Beas-Gonzalez}, J. and {Bechtol}, K. and {Becker}, M.~R. and {Bernstein}, G.~M. and {Bertin}, E. and {Blazek}, J. and {Bocquet}, S. and {Brooks}, D. and {Brout}, D. and {Burke}, D.~L. and {Camacho}, H. and {Camacho-Ciurana}, G. and {Camilleri}, R. and {Campailla}, G. and {Campos}, A. and {Carnero Rosell}, A. and {Carr}, A. and {Carretero}, J. and {Castander}, F.~J. and {Cawthon}, R. and {Chan}, K.~C. and {Chang}, C. and {Chen}, R. and {Coloma-Nadal}, J.~M. and {Conselice}, C. and {Costanzi}, M. and {Crocce}, M. and {d'Assignies}, W. and {da Costa}, L.~N. and {da Silva Pereira}, M.~E. and {Davis}, T.~M. and {De Vicente}, J. and {DePoy}, D.~L. and {DeRose}, J. and {Desai}, S. and {Diehl}, H.~T. and {Dodelson}, S. and {Doel}, P. and {Doux}, C. and {Drlica-Wagner}, A. and {Eifler}, T.~F. and {Elvin-Poole}, J. and {Everett}, S. and {Evrard}, A.~E. and {Ferrero}, I. and {Fert{\'e}}, A. and {Flaugher}, B. and {Fosalba}, P. and {de Souza}, D. Francis and {Frieman}, J. and {Galbany}, L. and {Garc{\'\i}a-Bellido}, J. and {Gatti}, M. and {Giannini}, G. and {Giles}, P. and {Glazebrook}, K. and {Gruen}, D. and {Gruendl}, R.~A. and {Gutierrez}, G. and {Harrison}, I. and {Hartley}, W.~G. and {Herner}, K. and {Hinton}, S.~R. and {Hollowood}, D.~L. and {Honscheid}, K. and {Huff}, E.~M. and {Huterer}, D. and {Jain}, B. and {James}, D.~J. and {Jarvis}, M. and {Jeffrey}, N. and {Jeltema}, T. and {Kent}, S. and {Kessler}, R. and {Kovacs}, A. and {Koyama}, K. and {Krause}, E. and {Kron}, R. and {Kuehn}, K. and {Lahav}, O. and {Lee}, J. and {Lee}, S. and {Legnani}, E. and {Li}, T.~S. and {Liddle}, A.~R. and {Lidman}, C. and {Lin}, H. and {Lin}, M. and {MacCrann}, N. and {Marshall}, J.~L. and {Mau}, S. and {McMahon}, R.~G. and {Mena-Fern{\'a}ndez}, J. and {Menanteau}, F. and {Miquel}, R. and {Mohr}, J.~J. and {Muir}, J. and {Myles}, J. and {M{\"o}ller}, A. and {Nichol}, R.~C. and {Ogando}, R.~L.~C. and {Percival}, W.~J. and {Petravick}, D. and {Pieres}, A. and {Plazas Malag{\'o}n}, A.~A. and {Popovic}, B. and {Porredon}, A. and {Prat}, J. and {Qu}, H. and {Raveri}, M. and {Rebou{\c{c}}as}, J. and {Riquelme}, W. and {Rodriguez-Monroy}, M. and {Rogozenski}, P. and {Romer}, A.~K. and {Roodman}, A. and {Rosenfeld}, R. and {Ross}, A.~J. and {Rykoff}, E.~S. and {Sako}, M. and {Samuroff}, S. and {S{\'a}nchez}, C. and {Sanchez}, E. and {Sanchez Cid}, D. and {Schutt}, T. and {Scolnic}, D. and {Sevilla-Noarbe}, I. and {Shah}, N. and {Shah}, P. and {Sheldon}, E. and {Smith}, M. and {Soares-Santos}, M. and {Suchyta}, E. and {Sullivan}, M. and {Swanson}, M.~E.~C. and {S{\'a}nchez}, B.~O. and {Tabbutt}, M. and {Tarle}, G. and {Taylor}, G. and {Thomas}, D. and {To}, C. and {Toribio San Cipriano}, L. and {Toy}, M. and {Troxel}, M.~A. and {Tucker}, D.~L. and {Vikram}, V. and {Vincenzi}, M. and {Weaverdyck}, N. and {Weller}, J. and {Whyley}, A. and {Wilkinson}, R.~D. and {Wiseman}, P. and {Yamamoto}, M. and {Yanny}, B. and {Yin}, B. and {Zhang}, Y. and {Zuntz}, J.},
        title = "{Constraints on Dynamical Dark Energy from Multiple Probes in the Full Dark Energy Survey}",
      journal = {arXiv e-prints},
         year = 2026,
        month = may,
          eid = {arXiv:2605.27221},
        pages = {arXiv:2605.27221},
          doi = {10.48550/arXiv.2605.27221},
archivePrefix = {arXiv},
       eprint = {2605.27221},
 primaryClass = {astro-ph.CO},
       adsurl = {https://ui.adsabs.harvard.edu/abs/2026arXiv260527221D}
}

@ARTICLE{Zonca2019,
       author = {{Zonca}, Andrea and {Singer}, Leo and {Lenz}, Daniel and {Reinecke}, Martin and {Rosset}, Cyrille and {Hivon}, Eric and {Gorski}, Krzysztof},
        title = "{healpy: equal area pixelization and spherical harmonics transforms for data on the sphere in Python}",
      journal = {The Journal of Open Source Software},
         year = 2019,
        month = mar,
       volume = {4},
       number = {35},
          eid = {1298},
        pages = {1298},
          doi = {10.21105/joss.01298},
       adsurl = {https://ui.adsabs.harvard.edu/abs/2019JOSS....4.1298Z}
}

@ARTICLE{Harris2020,
       author = {{Harris}, Charles R. and {Millman}, K. Jarrod and {van der Walt}, St{\'e}fan J. and {Gommers}, Ralf and {Virtanen}, Pauli and {Cournapeau}, David and {Wieser}, Eric and {Taylor}, Julian and {Berg}, Sebastian and {Smith}, Nathaniel J. and {Kern}, Robert and {Picus}, Matti and {Hoyer}, Stephan and {van Kerkwijk}, Marten H. and {Brett}, Matthew and {Haldane}, Allan and {del R{\'\i}o}, Jaime Fern{\'a}ndez and {Wiebe}, Mark and {Peterson}, Pearu and {G{\'e}rard-Marchant}, Pierre and {Sheppard}, Kevin and {Reddy}, Tyler and {Weckesser}, Warren and {Abbasi}, Hameer and {Gohlke}, Christoph and {Oliphant}, Travis E.},
        title = "{Array programming with NumPy}",
      journal = {\nat},
         year = 2020,
        month = sep,
       volume = {585},
       number = {7825},
        pages = {357-362},
          doi = {10.1038/s41586-020-2649-2},
archivePrefix = {arXiv},
       eprint = {2006.10256},
 primaryClass = {cs.MS},
       adsurl = {https://ui.adsabs.harvard.edu/abs/2020Natur.585..357H}
}

@ARTICLE{Borrow2020,
       author = {{Borrow}, Josh and {Borrisov}, Alexei},
        title = "{swiftsimio: A Python library for reading SWIFT data}",
      journal = {The Journal of Open Source Software},
         year = 2020,
        month = aug,
       volume = {5},
       number = {52},
          eid = {2430},
        pages = {2430},
          doi = {10.21105/joss.02430},
       adsurl = {https://ui.adsabs.harvard.edu/abs/2020JOSS....5.2430B}
}

@ARTICLE{Astropy2022,
       author = {{Astropy Collaboration} and {Price-Whelan}, Adrian M. and {Lim}, Pey Lian and {Earl}, Nicholas and {Starkman}, Nathaniel and {Bradley}, Larry and {Shupe}, David L. and {Patil}, Aarya A. and {Corrales}, Lia and {Brasseur}, C.~E. and {N{\"o}the}, Maximilian and {Donath}, Axel and {Tollerud}, Erik and {Morris}, Brett M. and {Ginsburg}, Adam and {Vaher}, Eero and {Weaver}, Benjamin A. and {Tocknell}, James and {Jamieson}, William and {van Kerkwijk}, Marten H. and {Robitaille}, Thomas P. and {Merry}, Bruce and {Bachetti}, Matteo and {G{\"u}nther}, H. Moritz and {Aldcroft}, Thomas L. and {Alvarado-Montes}, Jaime A. and {Archibald}, Anne M. and {B{\'o}di}, Attila and {Bapat}, Shreyas and {Barentsen}, Geert and {Baz{\'a}n}, Juanjo and {Biswas}, Manish and {Boquien}, M{\'e}d{\'e}ric and {Burke}, D.~J. and {Cara}, Daria and {Cara}, Mihai and {Conroy}, Kyle E. and {Conseil}, Simon and {Craig}, Matthew W. and {Cross}, Robert M. and {Cruz}, Kelle L. and {D'Eugenio}, Francesco and {Dencheva}, Nadia and {Devillepoix}, Hadrien A.~R. and {Dietrich}, J{\"o}rg P. and {Eigenbrot}, Arthur Davis and {Erben}, Thomas and {Ferreira}, Leonardo and {Foreman-Mackey}, Daniel and {Fox}, Ryan and {Freij}, Nabil and {Garg}, Suyog and {Geda}, Robel and {Glattly}, Lauren and {Gondhalekar}, Yash and {Gordon}, Karl D. and {Grant}, David and {Greenfield}, Perry and {Groener}, Austen M. and {Guest}, Steve and {Gurovich}, Sebastian and {Handberg}, Rasmus and {Hart}, Akeem and {Hatfield-Dodds}, Zac and {Homeier}, Derek and {Hosseinzadeh}, Griffin and {Jenness}, Tim and {Jones}, Craig K. and {Joseph}, Prajwel and {Kalmbach}, J. Bryce and {Karamehmetoglu}, Emir and {Ka{\l}uszy{\'n}ski}, Miko{\l}aj and {Kelley}, Michael S.~P. and {Kern}, Nicholas and {Kerzendorf}, Wolfgang E. and {Koch}, Eric W. and {Kulumani}, Shankar and {Lee}, Antony and {Ly}, Chun and {Ma}, Zhiyuan and {MacBride}, Conor and {Maljaars}, Jakob M. and {Muna}, Demitri and {Murphy}, N.~A. and {Norman}, Henrik and {O'Steen}, Richard and {Oman}, Kyle A. and {Pacifici}, Camilla and {Pascual}, Sergio and {Pascual-Granado}, J. and {Patil}, Rohit R. and {Perren}, Gabriel I. and {Pickering}, Timothy E. and {Rastogi}, Tanuj and {Roulston}, Benjamin R. and {Ryan}, Daniel F. and {Rykoff}, Eli S. and {Sabater}, Jose and {Sakurikar}, Parikshit and {Salgado}, Jes{\'u}s and {Sanghi}, Aniket and {Saunders}, Nicholas and {Savchenko}, Volodymyr and {Schwardt}, Ludwig and {Seifert-Eckert}, Michael and {Shih}, Albert Y. and {Jain}, Anany Shrey and {Shukla}, Gyanendra and {Sick}, Jonathan and {Simpson}, Chris and {Singanamalla}, Sudheesh and {Singer}, Leo P. and {Singhal}, Jaladh and {Sinha}, Manodeep and {Sip{\H{o}}cz}, Brigitta M. and {Spitler}, Lee R. and {Stansby}, David and {Streicher}, Ole and {{\v{S}}umak}, Jani and {Swinbank}, John D. and {Taranu}, Dan S. and {Tewary}, Nikita and {Tremblay}, Grant R. and {de Val-Borro}, Miguel and {Van Kooten}, Samuel J. and {Vasovi{\'c}}, Zlatan and {Verma}, Shresth and {de Miranda Cardoso}, Jos{\'e} Vin{\'\i}cius and {Williams}, Peter K.~G. and {Wilson}, Tom J. and {Winkel}, Benjamin and {Wood-Vasey}, W.~M. and {Xue}, Rui and {Yoachim}, Peter and {Zhang}, Chen and {Zonca}, Andrea and {Astropy Project Contributors}},
        title = "{The Astropy Project: Sustaining and Growing a Community-oriented Open-source Project and the Latest Major Release (v5.0) of the Core Package}",
      journal = {\apj},
         year = 2022,
        month = aug,
       volume = {935},
       number = {2},
          eid = {167},
        pages = {167},
          doi = {10.3847/1538-4357/ac7c74},
archivePrefix = {arXiv},
       eprint = {2206.14220},
 primaryClass = {astro-ph.IM},
       adsurl = {https://ui.adsabs.harvard.edu/abs/2022ApJ...935..167A}
}

@ARTICLE{Foreman2012,
       author = {{Foreman-Mackey}, Daniel and {Hogg}, David W. and {Lang}, Dustin and {Goodman}, Jonathan},
        title = "{emcee: The MCMC Hammer}",
      journal = {\pasp},
         year = 2013,
        month = mar,
       volume = {125},
       number = {925},
        pages = {306},
          doi = {10.1086/670067},
archivePrefix = {arXiv},
       eprint = {1202.3665},
 primaryClass = {astro-ph.IM},
       adsurl = {https://ui.adsabs.harvard.edu/abs/2013PASP..125..306F}
}

@ARTICLE{Grandon2024,
       author = {{Grand{\'o}n}, Daniela and {Marques}, Gabriela A. and {Thiele}, Leander and {Cheng}, Sihao and {Shirasaki}, Masato and {Liu}, Jia},
        title = "{Impact of baryonic feedback on HSC-Y1 weak lensing non-Gaussian statistics}",
      journal = {\prd},
         year = 2024,
        month = nov,
       volume = {110},
       number = {10},
          eid = {103539},
        pages = {103539},
          doi = {10.1103/PhysRevD.110.103539},
archivePrefix = {arXiv},
       eprint = {2403.03807},
 primaryClass = {astro-ph.CO},
       adsurl = {https://ui.adsabs.harvard.edu/abs/2024PhRvD.110j3539G}
}

@ARTICLE{Weiss2019,
       author = {{Weiss}, Andreas J. and {Schneider}, Aurel and {Sgier}, Raphael and {Kacprzak}, Tomasz and {Amara}, Adam and {Refregier}, Alexandre},
        title = "{Effects of baryons on weak lensing peak statistics}",
      journal = {\jcap},
         year = 2019,
        month = oct,
       volume = {2019},
       number = {10},
          eid = {011},
        pages = {011},
          doi = {10.1088/1475-7516/2019/10/011},
archivePrefix = {arXiv},
       eprint = {1905.11636},
 primaryClass = {astro-ph.CO},
       adsurl = {https://ui.adsabs.harvard.edu/abs/2019JCAP...10..011W}
}

@ARTICLE{deJong2013,
       author = {{\noopsort{De Jong}}{de Jong}, Jelte T.~A. and {Verdoes Kleijn}, Gijs A. and {Kuijken}, Konrad H. and {Valentijn}, Edwin A.},
        title = "{The Kilo-Degree Survey}",
      journal = {Experimental Astronomy},
         year = 2013,
        month = jan,
       volume = {35},
       number = {1-2},
        pages = {25-44},
          doi = {10.1007/s10686-012-9306-1},
archivePrefix = {arXiv},
       eprint = {1206.1254},
 primaryClass = {astro-ph.CO},
       adsurl = {https://ui.adsabs.harvard.edu/abs/2013ExA....35...25D}
}

@INPROCEEDINGS{HSC2012,
       author = {{Miyazaki}, Satoshi and {Komiyama}, Yutaka and {Nakaya}, Hidehiko and {Kamata}, Yukiko and {Doi}, Yoshi and {Hamana}, Takashi and {Karoji}, Hiroshi and {Furusawa}, Hisanori and {Kawanomoto}, Satoshi and {Morokuma}, Tomoki and {Ishizuka}, Yuki and {Nariai}, Kyoji and {Tanaka}, Yoko and {Uraguchi}, Fumihiro and {Utsumi}, Yousuke and {Obuchi}, Yoshiyuki and {Okura}, Yuki and {Oguri}, Masamune and {Takata}, Tadafumi and {Tomono}, Daigo and {Kurakami}, Tomio and {Namikawa}, Kazuhito and {Usuda}, Tomonori and {Yamanoi}, Hitomi and {Terai}, Tsuyoshi and {Uekiyo}, Hatsue and {Yamada}, Yoshihiko and {Koike}, Michitaro and {Aihara}, Hiro and {Fujimori}, Yuki and {Mineo}, Sogo and {Miyatake}, Hironao and {Yasuda}, Naoki and {Nishizawa}, Jun and {Saito}, Tomoki and {Tanaka}, Manobu and {Uchida}, Tomohisa and {Katayama}, Nobu and {Wang}, Shiang-Yu and {Chen}, Hsin-Yo and {Lupton}, Robert and {Loomis}, Craig and {Bickerton}, Steve and {Price}, Paul and {Gunn}, Jim and {Suzuki}, Hisanori and {Miyazaki}, Yasuhito and {Muramatsu}, Masaharu and {Yamamoto}, Koei and {Endo}, Makoto and {Ezaki}, Yutaka and {Itoh}, Noboru and {Miwa}, Yoshinori and {Yokota}, Hideo and {Matsuda}, Toru and {Ebinuma}, Ryuichi and {Takeshi}, Kunio},
        title = "{Hyper Suprime-Cam}",
    booktitle = {Ground-based and Airborne Instrumentation for Astronomy IV},
         year = 2012,
       editor = {{McLean}, Ian S. and {Ramsay}, Suzanne K. and {Takami}, Hideki},
       series = {Society of Photo-Optical Instrumentation Engineers (SPIE) Conference Series},
       volume = {8446},
        month = sep,
          eid = {84460Z},
        pages = {84460Z},
          doi = {10.1117/12.926844},
       adsurl = {https://ui.adsabs.harvard.edu/abs/2012SPIE.8446E..0ZM}
}

@ARTICLE{DES2005,
       author = {{The Dark Energy Survey Collaboration}},
        title = "{The Dark Energy Survey}",
      journal = {arXiv e-prints},
         year = 2005,
        month = oct,
          eid = {astro-ph/0510346},
        pages = {astro-ph/0510346},
          doi = {10.48550/arXiv.astro-ph/0510346},
archivePrefix = {arXiv},
       eprint = {astro-ph/0510346},
 primaryClass = {astro-ph},
       adsurl = {https://ui.adsabs.harvard.edu/abs/2005astro.ph.10346T}
}

@ARTICLE{Zurhcer2021,
       author = {{Z{\"u}rcher}, Dominik and {Fluri}, Janis and {Sgier}, Raphael and {Kacprzak}, Tomasz and {Refregier}, Alexandre},
        title = "{Cosmological forecast for non-Gaussian statistics in large-scale weak lensing surveys}",
      journal = {\jcap},
         year = 2021,
        month = jan,
       volume = {2021},
       number = {1},
          eid = {028},
        pages = {028},
          doi = {10.1088/1475-7516/2021/01/028},
archivePrefix = {arXiv},
       eprint = {2006.12506},
 primaryClass = {astro-ph.CO},
       adsurl = {https://ui.adsabs.harvard.edu/abs/2021JCAP...01..028Z}
}

@ARTICLE{Hildebrandt2020,
       author = {{Hildebrandt}, H. and {K{\"o}hlinger}, F. and {van den Busch}, J.~L. and {Joachimi}, B. and {Heymans}, C. and {Kannawadi}, A. and {Wright}, A.~H. and {Asgari}, M. and {Blake}, C. and {Hoekstra}, H. and {Joudaki}, S. and {Kuijken}, K. and {Miller}, L. and {Morrison}, C.~B. and {Tr{\"o}ster}, T. and {Amon}, A. and {Archidiacono}, M. and {Brieden}, S. and {Choi}, A. and {de Jong}, J.~T.~A. and {Erben}, T. and {Giblin}, B. and {Mead}, A. and {Peacock}, J.~A. and {Radovich}, M. and {Schneider}, P. and {Sif{\'o}n}, C. and {Tewes}, M.},
        title = "{KiDS+VIKING-450: Cosmic shear tomography with optical and infrared data}",
      journal = {\aap},
         year = 2020,
        month = jan,
       volume = {633},
          eid = {A69},
        pages = {A69},
          doi = {10.1051/0004-6361/201834878},
archivePrefix = {arXiv},
       eprint = {1812.06076},
 primaryClass = {astro-ph.CO},
       adsurl = {https://ui.adsabs.harvard.edu/abs/2020A&A...633A..69H}
}

@ARTICLE{Liu2014,
       author = {{Liu}, Jia and {Haiman}, Zolt{\'a}n and {Hui}, Lam and {Kratochvil}, Jan M. and {May}, Morgan},
        title = "{Impact of magnification and size bias on the weak lensing power spectrum and peak statistics}",
      journal = {\prd},
         year = 2014,
        month = jan,
       volume = {89},
       number = {2},
          eid = {023515},
        pages = {023515},
          doi = {10.1103/PhysRevD.89.023515},
archivePrefix = {arXiv},
       eprint = {1310.7517},
 primaryClass = {astro-ph.CO},
       adsurl = {https://ui.adsabs.harvard.edu/abs/2014PhRvD..89b3515L}
}

@ARTICLE{Sellentin2016,
       author = {{Sellentin}, Elena and {Heavens}, Alan F.},
        title = "{Parameter inference with estimated covariance matrices}",
      journal = {\mnras},
         year = 2016,
        month = feb,
       volume = {456},
       number = {1},
        pages = {L132-L136},
          doi = {10.1093/mnrasl/slv190},
archivePrefix = {arXiv},
       eprint = {1511.05969},
 primaryClass = {astro-ph.CO},
       adsurl = {https://ui.adsabs.harvard.edu/abs/2016MNRAS.456L.132S}
}

@ARTICLE{DESyr62026,
       author = {{DES Collaboration} and {Abbott}, T.~M.~C. and {Adamow}, M. and {Aguena}, M. and {Alarcon}, A. and {Allam}, S.~S. and {Alves}, O. and {Amon}, A. and {Anbajagane}, D. and {Andrade-Oliveira}, F. and {Avila}, S. and {Bacon}, D. and {Baxter}, E.~J. and {Beas-Gonzalez}, J. and {Bechtol}, K. and {Becker}, M.~R. and {Bernstein}, G.~M. and {Bertin}, E. and {Blazek}, J. and {Bocquet}, S. and {Brooks}, D. and {Brout}, D. and {Camacho}, H. and {Camacho-Ciurana}, G. and {Camilleri}, R. and {Campailla}, G. and {Campos}, A. and {Carnero Rosell}, A. and {Carrasco Kind}, M. and {Carretero}, J. and {Carrilho}, P. and {Castander}, F.~J. and {Cawthon}, R. and {Chang}, C. and {Choi}, A. and {Coloma-Nadal}, J.~M. and {Costanzi}, M. and {Crocce}, M. and {d'Assignies}, W. and {da Costa}, L.~N. and {da Silva Pereira}, M.~E. and {Davis}, T.~M. and {De Vicente}, J. and {DeRose}, J. and {Diehl}, H.~T. and {Dodelson}, S. and {Doel}, P. and {Doux}, C. and {Drlica-Wagner}, A. and {Eifler}, T.~F. and {Elvin-Poole}, J. and {Estrada}, J. and {Everett}, S. and {Evrard}, A.~E. and {Fang}, J. and {Farahi}, A. and {Fert{\'e}}, A. and {Flaugher}, B. and {Fosalba}, P. and {Frieman}, J. and {Garc{\'\i}a-Bellido}, J. and {Gatti}, M. and {Gaztanaga}, E. and {Giannini}, G. and {Giles}, P. and {Glazebrook}, K. and {Gorsuch}, M. and {Gruen}, D. and {Gruendl}, R.~A. and {Gschwend}, J. and {Gutierrez}, G. and {Harrison}, I. and {Hartley}, W.~G. and {Henning}, E. and {Herner}, K. and {Hinton}, S.~R. and {Hollowood}, D.~L. and {Honscheid}, K. and {Huff}, E.~M. and {Huterer}, D. and {Jain}, B. and {James}, D.~J. and {Jarvis}, M. and {Jeffrey}, N. and {Jeltema}, T. and {Kacprzak}, T. and {Kent}, S. and {Kovacs}, A. and {Krause}, E. and {Kron}, R. and {Kuehn}, K. and {Lahav}, O. and {Lee}, S. and {Legnani}, E. and {Lidman}, C. and {Lin}, H. and {MacCrann}, N. and {Manera}, M. and {Manning}, T. and {Marshall}, J.~L. and {Mau}, S. and {McCullough}, J. and {Mena-Fern{\'a}ndez}, J. and {Menanteau}, F. and {Miquel}, R. and {Mohr}, J.~J. and {Muir}, J. and {Myles}, J. and {Nichol}, R.~C. and {Nord}, B. and {O'Donnell}, J.~H. and {Ogando}, R.~L.~C. and {Palmese}, A. and {Paterno}, M. and {Peoples}, J. and {Percival}, W.~J. and {Petravick}, D. and {Pieres}, A. and {Plazas Malag{\'o}n}, A.~A. and {Porredon}, A. and {Pourtsidou}, A. and {Prat}, J. and {Preston}, C. and {Raveri}, M. and {Riquelme}, W. and {Rodriguez-Monroy}, M. and {Rogozenski}, P. and {Romer}, A.~K. and {Roodman}, A. and {Rosenfeld}, R. and {Ross}, A.~J. and {Rozo}, E. and {Rykoff}, E.~S. and {Samuroff}, S. and {S{\'a}nchez}, C. and {Sanchez}, E. and {Sanchez Cid}, D. and {Schutt}, T. and {Sevilla-Noarbe}, I. and {Sheldon}, E. and {Sherman}, N. and {Shin}, T. and {Smith}, M. and {Soares-Santos}, M. and {Suchyta}, E. and {Swanson}, M.~E.~C. and {Tabbutt}, M. and {Tarle}, G. and {Thomas}, D. and {To}, C. and {Tong}, A. and {Toribio San Cipriano}, L. and {Troxel}, M.~A. and {Tsedrik}, M. and {Tucker}, D.~L. and {Vikram}, V. and {Walker}, A.~R. and {Weaverdyck}, N. and {Wechsler}, R.~H. and {Weinberg}, D.~H. and {Weller}, J. and {Wetzell}, V. and {Whyley}, A. and {Wilkinson}, R.~D. and {Wiseman}, P. and {Wu}, H.-Y. and {Yamamoto}, M. and {Yanny}, B. and {Yin}, B. and {Zacharegkas}, G. and {Zhang}, Y. and {Zuntz}, J.},
        title = "{Dark Energy Survey Year 6 Results: Cosmological Constraints from Galaxy Clustering and Weak Lensing}",
      journal = {arXiv e-prints},
         year = 2026,
        month = jan,
          eid = {arXiv:2601.14559},
        pages = {arXiv:2601.14559},
          doi = {10.48550/arXiv.2601.14559},
archivePrefix = {arXiv},
       eprint = {2601.14559},
 primaryClass = {astro-ph.CO},
       adsurl = {https://ui.adsabs.harvard.edu/abs/2026arXiv260114559D}
}

@ARTICLE{Bartelmann2001,
       author = {{Bartelmann}, M. and {Schneider}, P.},
        title = "{Weak gravitational lensing}",
      journal = {\physrep},
         year = 2001,
        month = jan,
       volume = {340},
       number = {4-5},
        pages = {291-472},
          doi = {10.1016/S0370-1573(00)00082-X},
archivePrefix = {arXiv},
       eprint = {astro-ph/9912508},
 primaryClass = {astro-ph},
       adsurl = {https://ui.adsabs.harvard.edu/abs/2001PhR...340..291B}
}

@ARTICLE{Kilbinger2015,
       author = {{Kilbinger}, Martin},
        title = "{Cosmology with cosmic shear observations: a review}",
      journal = {Reports on Progress in Physics},
         year = 2015,
        month = jul,
       volume = {78},
       number = {8},
          eid = {086901},
        pages = {086901},
          doi = {10.1088/0034-4885/78/8/086901},
archivePrefix = {arXiv},
       eprint = {1411.0115},
 primaryClass = {astro-ph.CO},
       adsurl = {https://ui.adsabs.harvard.edu/abs/2015RPPh...78h6901K}
}

@ARTICLE{DESI2025_DR2cosmo,
       author = {{Abdul Karim}, M. and {Aguilar}, J. and {Ahlen}, S. and {Alam}, S. and {Allen}, L. and {Prieto}, C. Allende and {Alves}, O. and {Anand}, A. and {Andrade}, U. and {Armengaud}, E. and {Aviles}, A. and {Bailey}, S. and {Baltay}, C. and {Bansal}, P. and {Bault}, A. and {Behera}, J. and {BenZvi}, S. and {Bianchi}, D. and {Blake}, C. and {Brieden}, S. and {Brodzeller}, A. and {Brooks}, D. and {Buckley-Geer}, E. and {Burtin}, E. and {Calderon}, R. and {Canning}, R. and {Rosell}, A. Carnero and {Carrilho}, P. and {Casas}, L. and {Castander}, F.~J. and {Charles}, M. and {Chaussidon}, E. and {Chaves-Montero}, J. and {Chebat}, D. and {Chen}, X. and {Claybaugh}, T. and {Cole}, S. and {Cooper}, A.~P. and {Cuceu}, A. and {Dawson}, K.~S. and {de la Macorra}, A. and {de Mattia}, A. and {Deiosso}, N. and {Della Costa}, J. and {Demina}, R. and {Dey}, A. and {Dey}, B. and {Ding}, Z. and {Doel}, P. and {Edelstein}, J. and {Eisenstein}, D.~J. and {Elbers}, W. and {Fagrelius}, P. and {Fanning}, K. and {Fern{\'a}ndez-Garc{\'\i}a}, E. and {Ferraro}, S. and {Font-Ribera}, A. and {Forero-Romero}, J.~E. and {Frenk}, C.~S. and {Garcia-Quintero}, C. and {Garrison}, L.~H. and {Gazta{\~n}aga}, E. and {Gil-Mar{\'\i}n}, H. and {Gontcho A Gontcho}, S. and {Gonzalez}, D. and {Gonzalez-Morales}, A.~X. and {Gordon}, C. and {Green}, D. and {Gutierrez}, G. and {Guy}, J. and {Hadzhiyska}, B. and {Hahn}, C. and {He}, S. and {Herbold}, M. and {Herrera-Alcantar}, H.~K. and {Ho}, M.-F. and {Honscheid}, K. and {Howlett}, C. and {Huterer}, D. and {Ishak}, M. and {Juneau}, S. and {Kamble}, N.~V. and {Kara{\c{c}}ayl{\i}}, N.~G. and {Kehoe}, R. and {Kent}, S. and {Kim}, A.~G. and {Kirkby}, D. and {Kisner}, T. and {Koposov}, S.~E. and {Kremin}, A. and {Krolewski}, A. and {Lahav}, O. and {Lamman}, C. and {Landriau}, M. and {Lang}, D. and {Lasker}, J. and {Le Goff}, J.~M. and {Le Guillou}, L. and {Leauthaud}, A. and {Levi}, M.~E. and {Li}, Q. and {Li}, T.~S. and {Lodha}, K. and {Lokken}, M. and {Lozano-Rodr{\'\i}guez}, F. and {Magneville}, C. and {Manera}, M. and {Martini}, P. and {Matthewson}, W.~L. and {Meisner}, A. and {Mena-Fern{\'a}ndez}, J. and {Menegas}, A. and {Mergulh{\~a}o}, T. and {Miquel}, R. and {Moustakas}, J. and {Mu{\~n}oz-Guti{\'e}rrez}, A. and {Mu{\~n}oz-Santos}, D. and {Myers}, A.~D. and {Nadathur}, S. and {Naidoo}, K. and {Napolitano}, L. and {Newman}, J.~A. and {Niz}, G. and {Noriega}, H.~E. and {Paillas}, E. and {Palanque-Delabrouille}, N. and {Pan}, J. and {Peacock}, J.~A. and {Pellejero Ibanez}, M. and {Percival}, W.~J. and {P{\'e}rez-Fern{\'a}ndez}, A. and {P{\'e}rez-R{\`a}fols}, I. and {Pieri}, M.~M. and {Poppett}, C. and {Prada}, F. and {Rabinowitz}, D. and {Raichoor}, A. and {Ram{\'\i}rez-P{\'e}rez}, C. and {Rashkovetskyi}, M. and {Ravoux}, C. and {Rich}, J. and {Rocher}, A. and {Rockosi}, C. and {Rohlf}, J. and {Rom{\'a}n-Herrera}, J.~O. and {Ross}, A.~J. and {Rossi}, G. and {Ruggeri}, R. and {Ruhlmann-Kleider}, V. and {Samushia}, L. and {Sanchez}, E. and {Sanders}, N. and {Schlegel}, D. and {Schubnell}, M. and {Seo}, H. and {Shafieloo}, A. and {Sharples}, R. and {Silber}, J. and {Sinigaglia}, F. and {Sprayberry}, D. and {Tan}, T. and {Tarl{\'e}}, G. and {Taylor}, P. and {Turner}, W. and {Ure{\~n}a-L{\'o}pez}, L.~A. and {Vaisakh}, R. and {Valdes}, F. and {Valogiannis}, G. and {Vargas-Maga{\~n}a}, M. and {Verde}, L. and {Walther}, M. and {Weaver}, B.~A. and {Weinberg}, D.~H. and {White}, M. and {Wolfson}, M. and {Y{\`e}che}, C. and {Yu}, J. and {Zaborowski}, E.~A. and {Zarrouk}, P. and {Zhai}, Z. and {Zhang}, H. and {Zhao}, C. and {Zhao}, G.~B. and {Zhou}, R. and {Zou}, H. and {DESI Collaboration}},
        title = "{DESI DR2 results. II. Measurements of baryon acoustic oscillations and cosmological constraints}",
      journal = {\prd},
         year = 2025,
        month = oct,
       volume = {112},
       number = {8},
          eid = {083515},
        pages = {083515},
          doi = {10.1103/tr6y-kpc6},
archivePrefix = {arXiv},
       eprint = {2503.14738},
 primaryClass = {astro-ph.CO},
       adsurl = {https://ui.adsabs.harvard.edu/abs/2025PhRvD.112h3515A}
}

@ARTICLE{DESI2025_DR1_fullshape,
       author = {{Adame}, A.~G. and {Aguilar}, J. and {Ahlen}, S. and {Alam}, S. and {Alexander}, D.~M. and {Allende Prieto}, C. and {Alvarez}, M. and {Alves}, O. and {Anand}, A. and {Andrade}, U. and {Armengaud}, E. and {Avila}, S. and {Aviles}, A. and {Awan}, H. and {Bahr-Kalus}, B. and {Bailey}, S. and {Baltay}, C. and {Bault}, A. and {Behera}, J. and {BenZvi}, S. and {Beutler}, F. and {Bianchi}, D. and {Blake}, C. and {Blum}, R. and {Bonici}, M. and {Brieden}, S. and {Brodzeller}, A. and {Brooks}, D. and {Buckley-Geer}, E. and {Burtin}, E. and {Calderon}, R. and {Canning}, R. and {Carnero Rosell}, A. and {Cereskaite}, R. and {Cervantes-Cota}, J.~L. and {Chabanier}, S. and {Chaussidon}, E. and {Chaves-Montero}, J. and {Chebat}, D. and {Chen}, S. and {Chen}, X. and {Claybaugh}, T. and {Cole}, S. and {Cuceu}, A. and {Davis}, T.~M. and {Dawson}, K. and {de la Macorra}, A. and {de Mattia}, A. and {Deiosso}, N. and {Dey}, A. and {Dey}, B. and {Ding}, Z. and {Doel}, P. and {Edelstein}, J. and {Eftekharzadeh}, S. and {Eisenstein}, D.~J. and {Elbers}, W. and {Elliott}, A. and {Fagrelius}, P. and {Fanning}, K. and {Ferraro}, S. and {Ereza}, J. and {Findlay}, N. and {Flaugher}, B. and {Font-Ribera}, A. and {Forero-S{\'a}nchez}, D. and {Forero-Romero}, J.~E. and {Frenk}, C.~S. and {Garcia-Quintero}, C. and {Garrison}, L.~H. and {Gazta{\~n}aga}, E. and {Gil-Mar{\'\i}n}, H. and {Gontcho}, S. Gontcho A. and {Gonzalez-Morales}, A.~X. and {Gonzalez-Perez}, V. and {Gordon}, C. and {Green}, D. and {Gruen}, D. and {Gsponer}, R. and {Gutierrez}, G. and {Guy}, J. and {Hadzhiyska}, B. and {Hahn}, C. and {Hanif}, M.~M.~S. and {Herrera-Alcantar}, H.~K. and {Honscheid}, K. and {Howlett}, C. and {Huterer}, D. and {Ir{\v{s}}i{\v{c}}}, V. and {Ishak}, M. and {Joyce}, R. and {Juneau}, S. and {Kara{\c{c}}ayl{\i}}, N.~G. and {Kehoe}, R. and {Kent}, S. and {Kirkby}, D. and {Kong}, H. and {Koposov}, S.~E. and {Kremin}, A. and {Krolewski}, A. and {Lahav}, O. and {Lai}, Y. and {Lan}, T.-W. and {Landriau}, M. and {Lang}, D. and {Lasker}, J. and {Le Goff}, J.~M. and {Le Guillou}, L. and {Leauthaud}, A. and {Levi}, M.~E. and {Li}, T.~S. and {Lodha}, K. and {Magneville}, C. and {Manera}, M. and {Margala}, D. and {Martini}, P. and {Matthewson}, W. and {Maus}, M. and {McDonald}, P. and {Medina-Varela}, L. and {Meisner}, A. and {Mena-Fern{\'a}ndez}, J. and {Miquel}, R. and {Moon}, J. and {Moore}, S. and {Moustakas}, J. and {Mudur}, N. and {Mueller}, E. and {Mu{\~n}oz-Guti{\'e}rrez}, A. and {Myers}, A.~D. and {Nadathur}, S. and {Napolitano}, L. and {Neveux}, R. and {Newman}, J.~A. and {Nguyen}, N.~M. and {Nie}, J. and {Niz}, G. and {Noriega}, H.~E. and {Padmanabhan}, N. and {Paillas}, E. and {Palanque-Delabrouille}, N. and {Pan}, J. and {Penmetsa}, S. and {Percival}, W.~J. and {Pieri}, M.~M. and {Pinon}, M. and {Poppett}, C. and {Porredon}, A. and {Prada}, F. and {P{\'e}rez-Fern{\'a}ndez}, A. and {P{\'e}rez-R{\`a}fols}, I. and {Rabinowitz}, D. and {Raichoor}, A. and {Ram{\'\i}rez-P{\'e}rez}, C. and {Ramirez-Solano}, S. and {Rashkovetskyi}, M. and {Ravoux}, C. and {Rezaie}, M. and {Rich}, J. and {Rocher}, A. and {Rockosi}, C. and {Roe}, N.~A. and {Rosado-Marin}, A. and {Ross}, A.~J. and {Rossi}, G. and {Ruggeri}, R. and {Ruhlmann-Kleider}, V. and {Samushia}, L. and {Sanchez}, E. and {Saulder}, C. and {Schlafly}, E.~F. and {Schlegel}, D. and {Schubnell}, M. and {Seo}, H. and {Shafieloo}, A. and {Sharples}, R. and {Silber}, J. and {Slosar}, A. and {Smith}, A. and {Sprayberry}, D. and {Tan}, T. and {Tarl{\'e}}, G. and {Taylor}, P. and {Trusov}, S. and {Vaisakh}, R. and {Valcin}, D. and {Valdes}, F. and {Valogiannis}, G. and {Vargas-Maga{\~n}a}, M. and {Verde}, L. and {Walther}, M. and {Wang}, B. and {Wang}, M.~S. and {Weaver}, B.~A. and {Weaverdyck}, N. and {Wechsler}, R.~H. and {Weinberg}, D.~H. and {White}, M. and {Wilson}, M.~J. and {Yi}, L.},
        title = "{DESI 2024 VII: cosmological constraints from the full-shape modeling of clustering measurements}",
      journal = {\jcap},
         year = 2025,
        month = jul,
       volume = {2025},
       number = {7},
          eid = {028},
        pages = {028},
          doi = {10.1088/1475-7516/2025/07/028},
archivePrefix = {arXiv},
       eprint = {2411.12022},
 primaryClass = {astro-ph.CO},
       adsurl = {https://ui.adsabs.harvard.edu/abs/2025JCAP...07..028A}
}

@ARTICLE{Wright2025,
       author = {{Wright}, Angus H. and {St{\"o}lzner}, Benjamin and {Asgari}, Marika and {Bilicki}, Maciej and {Giblin}, Benjamin and {Heymans}, Catherine and {Hildebrandt}, Hendrik and {Hoekstra}, Henk and {Joachimi}, Benjamin and {Kuijken}, Konrad and {Li}, Shun-Sheng and {Reischke}, Robert and {von Wietersheim-Kramsta}, Maximilian and {Yoon}, Mijin and {Burger}, Pierre and {Chisari}, Nora Elisa and {de Jong}, Jelte and {Dvornik}, Andrej and {Georgiou}, Christos and {Harnois-D{\'e}raps}, Joachim and {Jalan}, Priyanka and {William}, Anjitha John and {Joudaki}, Shahab and {Lesci}, Giorgio Francesco and {Linke}, Laila and {Loureiro}, Arthur and {Mahony}, Constance and {Maturi}, Matteo and {Miller}, Lance and {Moscardini}, Lauro and {Napolitano}, Nicola R. and {Porth}, Lucas and {Radovich}, Mario and {Schneider}, Peter and {Tr{\"o}ster}, Tilman and {Valentijn}, Edwin and {Wittje}, Anna and {Yan}, Ziang and {Zhang}, Yun-Hao},
        title = "{KiDS-Legacy: Cosmological constraints from cosmic shear with the complete Kilo-Degree Survey}",
      journal = {\aap},
         year = 2025,
        month = nov,
       volume = {703},
          eid = {A158},
        pages = {A158},
          doi = {10.1051/0004-6361/202554908},
archivePrefix = {arXiv},
       eprint = {2503.19441},
 primaryClass = {astro-ph.CO},
       adsurl = {https://ui.adsabs.harvard.edu/abs/2025A&A...703A.158W}
}

@ARTICLE{Planck2020,
       author = {{Planck Collaboration} and {Aghanim}, N. and {Akrami}, Y. and {Ashdown}, M. and {Aumont}, J. and {Baccigalupi}, C. and {Ballardini}, M. and {Banday}, A.~J. and {Barreiro}, R.~B. and {Bartolo}, N. and {Basak}, S. and {Battye}, R. and {Benabed}, K. and {Bernard}, J.-P. and {Bersanelli}, M. and {Bielewicz}, P. and {Bock}, J.~J. and {Bond}, J.~R. and {Borrill}, J. and {Bouchet}, F.~R. and {Boulanger}, F. and {Bucher}, M. and {Burigana}, C. and {Butler}, R.~C. and {Calabrese}, E. and {Cardoso}, J.-F. and {Carron}, J. and {Challinor}, A. and {Chiang}, H.~C. and {Chluba}, J. and {Colombo}, L.~P.~L. and {Combet}, C. and {Contreras}, D. and {Crill}, B.~P. and {Cuttaia}, F. and {de Bernardis}, P. and {de Zotti}, G. and {Delabrouille}, J. and {Delouis}, J.-M. and {Di Valentino}, E. and {Diego}, J.~M. and {Dor{\'e}}, O. and {Douspis}, M. and {Ducout}, A. and {Dupac}, X. and {Dusini}, S. and {Efstathiou}, G. and {Elsner}, F. and {En{\ss}lin}, T.~A. and {Eriksen}, H.~K. and {Fantaye}, Y. and {Farhang}, M. and {Fergusson}, J. and {Fernandez-Cobos}, R. and {Finelli}, F. and {Forastieri}, F. and {Frailis}, M. and {Fraisse}, A.~A. and {Franceschi}, E. and {Frolov}, A. and {Galeotta}, S. and {Galli}, S. and {Ganga}, K. and {G{\'e}nova-Santos}, R.~T. and {Gerbino}, M. and {Ghosh}, T. and {Gonz{\'a}lez-Nuevo}, J. and {G{\'o}rski}, K.~M. and {Gratton}, S. and {Gruppuso}, A. and {Gudmundsson}, J.~E. and {Hamann}, J. and {Handley}, W. and {Hansen}, F.~K. and {Herranz}, D. and {Hildebrandt}, S.~R. and {Hivon}, E. and {Huang}, Z. and {Jaffe}, A.~H. and {Jones}, W.~C. and {Karakci}, A. and {Keih{\"a}nen}, E. and {Keskitalo}, R. and {Kiiveri}, K. and {Kim}, J. and {Kisner}, T.~S. and {Knox}, L. and {Krachmalnicoff}, N. and {Kunz}, M. and {Kurki-Suonio}, H. and {Lagache}, G. and {Lamarre}, J.-M. and {Lasenby}, A. and {Lattanzi}, M. and {Lawrence}, C.~R. and {Le Jeune}, M. and {Lemos}, P. and {Lesgourgues}, J. and {Levrier}, F. and {Lewis}, A. and {Liguori}, M. and {Lilje}, P.~B. and {Lilley}, M. and {Lindholm}, V. and {L{\'o}pez-Caniego}, M. and {Lubin}, P.~M. and {Ma}, Y.-Z. and {Mac{\'\i}as-P{\'e}rez}, J.~F. and {Maggio}, G. and {Maino}, D. and {Mandolesi}, N. and {Mangilli}, A. and {Marcos-Caballero}, A. and {Maris}, M. and {Martin}, P.~G. and {Martinelli}, M. and {Mart{\'\i}nez-Gonz{\'a}lez}, E. and {Matarrese}, S. and {Mauri}, N. and {McEwen}, J.~D. and {Meinhold}, P.~R. and {Melchiorri}, A. and {Mennella}, A. and {Migliaccio}, M. and {Millea}, M. and {Mitra}, S. and {Miville-Desch{\^e}nes}, M.-A. and {Molinari}, D. and {Montier}, L. and {Morgante}, G. and {Moss}, A. and {Natoli}, P. and {N{\o}rgaard-Nielsen}, H.~U. and {Pagano}, L. and {Paoletti}, D. and {Partridge}, B. and {Patanchon}, G. and {Peiris}, H.~V. and {Perrotta}, F. and {Pettorino}, V. and {Piacentini}, F. and {Polastri}, L. and {Polenta}, G. and {Puget}, J.-L. and {Rachen}, J.~P. and {Reinecke}, M. and {Remazeilles}, M. and {Renzi}, A. and {Rocha}, G. and {Rosset}, C. and {Roudier}, G. and {Rubi{\~n}o-Mart{\'\i}n}, J.~A. and {Ruiz-Granados}, B. and {Salvati}, L. and {Sandri}, M. and {Savelainen}, M. and {Scott}, D. and {Shellard}, E.~P.~S. and {Sirignano}, C. and {Sirri}, G. and {Spencer}, L.~D. and {Sunyaev}, R. and {Suur-Uski}, A.-S. and {Tauber}, J.~A. and {Tavagnacco}, D. and {Tenti}, M. and {Toffolatti}, L. and {Tomasi}, M. and {Trombetti}, T. and {Valenziano}, L. and {Valiviita}, J. and {Van Tent}, B. and {Vibert}, L. and {Vielva}, P. and {Villa}, F. and {Vittorio}, N. and {Wandelt}, B.~D. and {Wehus}, I.~K. and {White}, M. and {White}, S.~D.~M. and {Zacchei}, A. and {Zonca}, A.},
        title = "{Planck 2018 results. VI. Cosmological parameters}",
      journal = {\aap},
         year = 2020,
        month = sep,
       volume = {641},
          eid = {A6},
        pages = {A6},
          doi = {10.1051/0004-6361/201833910},
archivePrefix = {arXiv},
       eprint = {1807.06209},
 primaryClass = {astro-ph.CO},
       adsurl = {https://ui.adsabs.harvard.edu/abs/2020A&A...641A...6P}
}

@ARTICLE{Heydenreich2021,
       author = {{Heydenreich}, Sven and {Br{\"u}ck}, Benjamin and {Harnois-D{\'e}raps}, Joachim},
        title = "{Persistent homology in cosmic shear: Constraining parameters with topological data analysis}",
      journal = {\aap},
         year = 2021,
        month = apr,
       volume = {648},
          eid = {A74},
        pages = {A74},
          doi = {10.1051/0004-6361/202039048},
archivePrefix = {arXiv},
       eprint = {2007.13724},
 primaryClass = {astro-ph.CO},
       adsurl = {https://ui.adsabs.harvard.edu/abs/2021A&A...648A..74H}
}

@ARTICLE{Hoekstra2008,
       author = {{Hoekstra}, Henk and {Jain}, Bhuvnesh},
        title = "{Weak Gravitational Lensing and Its Cosmological Applications}",
      journal = {Annual Review of Nuclear and Particle Science},
         year = 2008,
        month = nov,
       volume = {58},
       number = {1},
        pages = {99-123},
          doi = {10.1146/annurev.nucl.58.110707.171151},
archivePrefix = {arXiv},
       eprint = {0805.0139},
 primaryClass = {astro-ph},
       adsurl = {https://ui.adsabs.harvard.edu/abs/2008ARNPS..58...99H}
}

@ARTICLE{Allys2020,
       author = {{Allys}, E. and {Marchand}, T. and {Cardoso}, J.-F. and {Villaescusa-Navarro}, F. and {Ho}, S. and {Mallat}, S.},
        title = "{New interpretable statistics for large-scale structure analysis and generation}",
      journal = {\prd},
         year = 2020,
        month = nov,
       volume = {102},
       number = {10},
          eid = {103506},
        pages = {103506},
          doi = {10.1103/PhysRevD.102.103506},
archivePrefix = {arXiv},
       eprint = {2006.06298},
 primaryClass = {astro-ph.CO},
       adsurl = {https://ui.adsabs.harvard.edu/abs/2020PhRvD.102j3506A}
}

@ARTICLE{Vinciguerra2025,
       author = {{Euclid Collaboration: Vinciguerra}, S. and {Bouch{\`e}}, F. and {Martinet}, N. and others},
       title = "{Euclid preparation. LXXXV. Towards a DR1 application of higher-order weak lensing statistics}",
      journal = {\aap},
         year = 2026,
        month = mar,
       volume = {707},
          eid = {A235},
        pages = {A235},
          doi = {10.1051/0004-6361/202557573},
archivePrefix = {arXiv},
       eprint = {2510.04953},
 primaryClass = {astro-ph.CO},
       adsurl = {https://ui.adsabs.harvard.edu/abs/2026A&A...707A.235E}
}

@ARTICLE{Marques2019,
       author = {{Marques}, Gabriela A. and {Liu}, Jia and {Zorrilla Matilla}, Jos{\'e} Manuel and {Haiman}, Zolt{\'a}n and {Bernui}, Armando and {Novaes}, Camila P.},
        title = "{Constraining neutrino mass with weak lensing Minkowski Functionals}",
      journal = {\jcap},
         year = 2019,
        month = jun,
       volume = {2019},
       number = {6},
          eid = {019},
        pages = {019},
          doi = {10.1088/1475-7516/2019/06/019},
archivePrefix = {arXiv},
       eprint = {1812.08206},
 primaryClass = {astro-ph.CO},
       adsurl = {https://ui.adsabs.harvard.edu/abs/2019JCAP...06..019M}
}

@ARTICLE{Feldbrugge2019,
       author = {{Feldbrugge}, Job and {van Engelen}, Matti and {van de Weygaert}, Rien and {Pranav}, Pratyush and {Vegter}, Gert},
        title = "{Stochastic homology of Gaussian vs. non-Gaussian random fields: graphs towards Betti numbers and persistence diagrams}",
      journal = {\jcap},
         year = 2019,
        month = sep,
       volume = {2019},
       number = {9},
          eid = {052},
        pages = {052},
          doi = {10.1088/1475-7516/2019/09/052},
archivePrefix = {arXiv},
       eprint = {1908.01619},
 primaryClass = {astro-ph.CO},
       adsurl = {https://ui.adsabs.harvard.edu/abs/2019JCAP...09..052F}
}

@ARTICLE{Jain2000,
       author = {{Jain}, Bhuvnesh and {Van Waerbeke}, Ludovic},
        title = "{Statistics of Dark Matter Halos from Gravitational Lensing}",
      journal = {\apjl},
         year = 2000,
        month = feb,
       volume = {530},
       number = {1},
        pages = {L1-L4},
          doi = {10.1086/312480},
archivePrefix = {arXiv},
       eprint = {astro-ph/9910459},
 primaryClass = {astro-ph},
       adsurl = {https://ui.adsabs.harvard.edu/abs/2000ApJ...530L...1J}
}

@BOOK{Rasmussen2006,
       author = {{Rasmussen}, Carl Edward and {Williams}, Christopher K.~I.},
        title = "{Gaussian Processes for Machine Learning}",
         year = 2006,
       adsurl = {https://ui.adsabs.harvard.edu/abs/2006gpml.book.....R}
}

@ARTICLE{Wang2009,
       author = {{Wang}, Sheng and {Haiman}, Zolt{\'a}n and {May}, Morgan},
        title = "{Constraining Cosmology with High-Convergence Regions in Weak Lensing Surveys}",
      journal = {\apj},
         year = 2009,
        month = jan,
       volume = {691},
       number = {1},
        pages = {547-559},
          doi = {10.1088/0004-637X/691/1/547},
archivePrefix = {arXiv},
       eprint = {0809.4052},
 primaryClass = {astro-ph},
       adsurl = {https://ui.adsabs.harvard.edu/abs/2009ApJ...691..547W}
}

@ARTICLE{Hamana2004,
       author = {{Hamana}, Takashi and {Takada}, Masahiro and {Yoshida}, Naoki},
        title = "{Searching for massive clusters in weak lensing surveys}",
      journal = {\mnras},
         year = 2004,
        month = may,
       volume = {350},
       number = {3},
        pages = {893-913},
          doi = {10.1111/j.1365-2966.2004.07691.x},
archivePrefix = {arXiv},
       eprint = {astro-ph/0310607},
 primaryClass = {astro-ph},
       adsurl = {https://ui.adsabs.harvard.edu/abs/2004MNRAS.350..893H}
}

@ARTICLE{Kratochvil2010,
       author = {{Kratochvil}, Jan M. and {Haiman}, Zolt{\'a}n and {May}, Morgan},
        title = "{Probing cosmology with weak lensing peak counts}",
      journal = {\prd},
         year = 2010,
        month = feb,
       volume = {81},
       number = {4},
          eid = {043519},
        pages = {043519},
          doi = {10.1103/PhysRevD.81.043519},
archivePrefix = {arXiv},
       eprint = {0907.0486},
 primaryClass = {astro-ph.CO},
       adsurl = {https://ui.adsabs.harvard.edu/abs/2010PhRvD..81d3519K}
}

@ARTICLE{Dietrich2010,
       author = {{Dietrich}, J.~P. and {Hartlap}, J.},
        title = "{Cosmology with the shear-peak statistics}",
      journal = {\mnras},
         year = 2010,
        month = feb,
       volume = {402},
       number = {2},
        pages = {1049-1058},
          doi = {10.1111/j.1365-2966.2009.15948.x},
archivePrefix = {arXiv},
       eprint = {0906.3512},
 primaryClass = {astro-ph.CO},
       adsurl = {https://ui.adsabs.harvard.edu/abs/2010MNRAS.402.1049D}
}

@ARTICLE{Li2023,
       author = {{Li}, Ziwei and {Liu}, Xiangkun and {Fan}, Zuhui},
        title = "{Weak-lensing peak statistics - steepness versus height}",
      journal = {\mnras},
         year = 2023,
        month = apr,
       volume = {520},
       number = {4},
        pages = {6382-6400},
          doi = {10.1093/mnras/stad534},
archivePrefix = {arXiv},
       eprint = {2302.08255},
 primaryClass = {astro-ph.CO},
       adsurl = {https://ui.adsabs.harvard.edu/abs/2023MNRAS.520.6382L}
}

@ARTICLE{Marian2013,
       author = {{Marian}, Laura and {Smith}, Robert E. and {Hilbert}, Stefan and {Schneider}, Peter},
        title = "{The cosmological information of shear peaks: beyond the abundance}",
      journal = {\mnras},
         year = 2013,
        month = jun,
       volume = {432},
       number = {2},
        pages = {1338-1350},
          doi = {10.1093/mnras/stt552},
archivePrefix = {arXiv},
       eprint = {1301.5001},
 primaryClass = {astro-ph.CO},
       adsurl = {https://ui.adsabs.harvard.edu/abs/2013MNRAS.432.1338M}
}

@ARTICLE{Davies2022,
       author = {{Davies}, Christopher T. and {Cautun}, Marius and {Giblin}, Benjamin and {Li}, Baojiu and {Harnois-D{\'e}raps}, Joachim and {Cai}, Yan-Chuan},
        title = "{Cosmological forecasts with the clustering of weak lensing peaks}",
      journal = {\mnras},
         year = 2022,
        month = jul,
       volume = {513},
       number = {4},
        pages = {4729-4746},
          doi = {10.1093/mnras/stac1204},
archivePrefix = {arXiv},
       eprint = {2110.10164},
 primaryClass = {astro-ph.CO},
       adsurl = {https://ui.adsabs.harvard.edu/abs/2022MNRAS.513.4729D}
}

@ARTICLE{Yang2011,
       author = {{Yang}, Xiuyuan and {Kratochvil}, Jan M. and {Wang}, Sheng and {Lim}, Eugene A. and {Haiman}, Zolt{\'a}n and {May}, Morgan},
        title = "{Cosmological information in weak lensing peaks}",
      journal = {\prd},
         year = 2011,
        month = aug,
       volume = {84},
       number = {4},
          eid = {043529},
        pages = {043529},
          doi = {10.1103/PhysRevD.84.043529},
archivePrefix = {arXiv},
       eprint = {1109.6333},
 primaryClass = {astro-ph.CO},
       adsurl = {https://ui.adsabs.harvard.edu/abs/2011PhRvD..84d3529Y}
}

@ARTICLE{Davies2019,
       author = {{Davies}, Christopher T. and {Cautun}, Marius and {Li}, Baojiu},
        title = "{The self-similarity of weak lensing peaks}",
      journal = {\mnras},
         year = 2019,
        month = oct,
       volume = {488},
       number = {4},
        pages = {5833-5851},
          doi = {10.1093/mnras/stz2157},
archivePrefix = {arXiv},
       eprint = {1905.01710},
 primaryClass = {astro-ph.CO},
       adsurl = {https://ui.adsabs.harvard.edu/abs/2019MNRAS.488.5833D}
}

@ARTICLE{Lahav2024,
       author = {{Lahav}, Ofer and {Liddle}, Andrew R},
        title = "{The Cosmological Parameters (2023)}",
      journal = {arXiv e-prints},
         year = 2024,
        month = mar,
          eid = {arXiv:2403.15526},
        pages = {arXiv:2403.15526},
          doi = {10.48550/arXiv.2403.15526},
archivePrefix = {arXiv},
       eprint = {2403.15526},
 primaryClass = {astro-ph.CO},
       adsurl = {https://ui.adsabs.harvard.edu/abs/2024arXiv240315526L}
}

@ARTICLE{Kugel2024,
       author = {{Kugel}, Roi and {Schaye}, Joop and {Schaller}, Matthieu and {McCarthy}, Ian G. and {Braspenning}, Joey and {Helly}, John C. and {Forouhar Moreno}, Victor J. and {McGibbon}, Robert J.},
        title = "{The FLAMINGO project: a comparison of galaxy cluster samples selected on mass, X-ray luminosity, Compton-Y parameter, or galaxy richness}",
      journal = {\mnras},
         year = 2024,
        month = nov,
       volume = {534},
       number = {3},
        pages = {2378-2396},
          doi = {10.1093/mnras/stae2218},
archivePrefix = {arXiv},
       eprint = {2406.03180},
 primaryClass = {astro-ph.CO},
       adsurl = {https://ui.adsabs.harvard.edu/abs/2024MNRAS.534.2378K}
}

@ARTICLE{Kugel2025,
       author = {{Kugel}, Roi and {Schaye}, Joop and {Schaller}, Matthieu and {Forouhar Moreno}, Victor J. and {McGibbon}, Robert J.},
        title = "{The FLAMINGO Project: An assessment of the systematic errors in the predictions of models for galaxy cluster counts used to infer cosmological parameters}",
      journal = {\mnras},
         year = 2025,
        month = feb,
       volume = {537},
       number = {2},
        pages = {2179-2197},
          doi = {10.1093/mnras/staf111},
archivePrefix = {arXiv},
       eprint = {2408.17217},
 primaryClass = {astro-ph.CO},
       adsurl = {https://ui.adsabs.harvard.edu/abs/2025MNRAS.537.2179K}
}

@ARTICLE{Spergel2015,
       author = {{Spergel}, D. and {Gehrels}, N. and {Baltay}, C. and {Bennett}, D. and {Breckinridge}, J. and {Donahue}, M. and {Dressler}, A. and {Gaudi}, B.~S. and {Greene}, T. and {Guyon}, O. and {Hirata}, C. and {Kalirai}, J. and {Kasdin}, N.~J. and {Macintosh}, B. and {Moos}, W. and {Perlmutter}, S. and {Postman}, M. and {Rauscher}, B. and {Rhodes}, J. and {Wang}, Y. and {Weinberg}, D. and {Benford}, D. and {Hudson}, M. and {Jeong}, W. -S. and {Mellier}, Y. and {Traub}, W. and {Yamada}, T. and {Capak}, P. and {Colbert}, J. and {Masters}, D. and {Penny}, M. and {Savransky}, D. and {Stern}, D. and {Zimmerman}, N. and {Barry}, R. and {Bartusek}, L. and {Carpenter}, K. and {Cheng}, E. and {Content}, D. and {Dekens}, F. and {Demers}, R. and {Grady}, K. and {Jackson}, C. and {Kuan}, G. and {Kruk}, J. and {Melton}, M. and {Nemati}, B. and {Parvin}, B. and {Poberezhskiy}, I. and {Peddie}, C. and {Ruffa}, J. and {Wallace}, J.~K. and {Whipple}, A. and {Wollack}, E. and {Zhao}, F.},
        title = "{Wide-Field InfrarRed Survey Telescope-Astrophysics Focused Telescope Assets WFIRST-AFTA 2015 Report}",
      journal = {arXiv e-prints},
         year = 2015,
        month = mar,
          eid = {arXiv:1503.03757},
        pages = {arXiv:1503.03757},
          doi = {10.48550/arXiv.1503.03757},
archivePrefix = {arXiv},
       eprint = {1503.03757},
 primaryClass = {astro-ph.IM},
       adsurl = {https://ui.adsabs.harvard.edu/abs/2015arXiv150303757S}
}

@ARTICLE{LSST2009,
       author = {{LSST Science Collaboration} and {Abell}, Paul A. and {Allison}, Julius and {Anderson}, Scott F. and {Andrew}, John R. and {Angel}, J. Roger P. and {Armus}, Lee and {Arnett}, David and {Asztalos}, S.~J. and {Axelrod}, Tim S. and {Bailey}, Stephen and {Ballantyne}, D.~R. and {Bankert}, Justin R. and {Barkhouse}, Wayne A. and {Barr}, Jeffrey D. and {Barrientos}, L. Felipe and {Barth}, Aaron J. and {Bartlett}, James G. and {Becker}, Andrew C. and {Becla}, Jacek and {Beers}, Timothy C. and {Bernstein}, Joseph P. and {Biswas}, Rahul and {Blanton}, Michael R. and {Bloom}, Joshua S. and {Bochanski}, John J. and {Boeshaar}, Pat and {Borne}, Kirk D. and {Bradac}, Marusa and {Brandt}, W.~N. and {Bridge}, Carrie R. and {Brown}, Michael E. and {Brunner}, Robert J. and {Bullock}, James S. and {Burgasser}, Adam J. and {Burge}, James H. and {Burke}, David L. and {Cargile}, Phillip A. and {Chandrasekharan}, Srinivasan and {Chartas}, George and {Chesley}, Steven R. and {Chu}, You-Hua and {Cinabro}, David and {Claire}, Mark W. and {Claver}, Charles F. and {Clowe}, Douglas and {Connolly}, A.~J. and {Cook}, Kem H. and {Cooke}, Jeff and {Cooray}, Asantha and {Covey}, Kevin R. and {Culliton}, Christopher S. and {de Jong}, Roelof and {de Vries}, Willem H. and {Debattista}, Victor P. and {Delgado}, Francisco and {Dell'Antonio}, Ian P. and {Dhital}, Saurav and {Di Stefano}, Rosanne and {Dickinson}, Mark and {Dilday}, Benjamin and {Djorgovski}, S.~G. and {Dobler}, Gregory and {Donalek}, Ciro and {Dubois-Felsmann}, Gregory and {Durech}, Josef and {Eliasdottir}, Ardis and {Eracleous}, Michael and {Eyer}, Laurent and {Falco}, Emilio E. and {Fan}, Xiaohui and {Fassnacht}, Christopher D. and {Ferguson}, Harry C. and {Fernandez}, Yanga R. and {Fields}, Brian D. and {Finkbeiner}, Douglas and {Figueroa}, Eduardo E. and {Fox}, Derek B. and {Francke}, Harold and {Frank}, James S. and {Frieman}, Josh and {Fromenteau}, Sebastien and {Furqan}, Muhammad and {Galaz}, Gaspar and {Gal-Yam}, A. and {Garnavich}, Peter and {Gawiser}, Eric and {Geary}, John and {Gee}, Perry and {Gibson}, Robert R. and {Gilmore}, Kirk and {Grace}, Emily A. and {Green}, Richard F. and {Gressler}, William J. and {Grillmair}, Carl J. and {Habib}, Salman and {Haggerty}, J.~S. and {Hamuy}, Mario and {Harris}, Alan W. and {Hawley}, Suzanne L. and {Heavens}, Alan F. and {Hebb}, Leslie and {Henry}, Todd J. and {Hileman}, Edward and {Hilton}, Eric J. and {Hoadley}, Keri and {Holberg}, J.~B. and {Holman}, Matt J. and {Howell}, Steve B. and {Infante}, Leopoldo and {Ivezic}, Zeljko and {Jacoby}, Suzanne H. and {Jain}, Bhuvnesh and {R} and {Jedicke} and {Jee}, M. James and {Garrett Jernigan}, J. and {Jha}, Saurabh W. and {Johnston}, Kathryn V. and {Jones}, R. Lynne and {Juric}, Mario and {Kaasalainen}, Mikko and {Styliani} and {Kafka} and {Kahn}, Steven M. and {Kaib}, Nathan A. and {Kalirai}, Jason and {Kantor}, Jeff and {Kasliwal}, Mansi M. and {Keeton}, Charles R. and {Kessler}, Richard and {Knezevic}, Zoran and {Kowalski}, Adam and {Krabbendam}, Victor L. and {Krughoff}, K. Simon and {Kulkarni}, Shrinivas and {Kuhlman}, Stephen and {Lacy}, Mark and {Lepine}, Sebastien and {Liang}, Ming and {Lien}, Amy and {Lira}, Paulina and {Long}, Knox S. and {Lorenz}, Suzanne and {Lotz}, Jennifer M. and {Lupton}, R.~H. and {Lutz}, Julie and {Macri}, Lucas M. and {Mahabal}, Ashish A. and {Mandelbaum}, Rachel and {Marshall}, Phil and {May}, Morgan and {McGehee}, Peregrine M. and {Meadows}, Brian T. and {Meert}, Alan and {Milani}, Andrea and {Miller}, Christopher J. and {Miller}, Michelle and {Mills}, David and {Minniti}, Dante and {Monet}, David and {Mukadam}, Anjum S. and {Nakar}, Ehud and {Neill}, Douglas R. and {Newman}, Jeffrey A. and {Nikolaev}, Sergei and {Nordby}, Martin and {O'Connor}, Paul and {Oguri}, Masamune and {Oliver}, John and {Olivier}, Scot S. and {Olsen}, Julia K. and {Olsen}, Knut and {Olszewski}, Edward W. and {Oluseyi}, Hakeem and {Padilla}, Nelson D. and {Parker}, Alex and {Pepper}, Joshua and {Peterson}, John R. and {Petry}, Catherine and {Pinto}, Philip A. and {Pizagno}, James L. and {Popescu}, Bogdan and {Prsa}, Andrej and {Radcka}, Veljko and {Raddick}, M. Jordan and {Rasmussen}, Andrew and {Rau}, Arne and {Rho}, Jeonghee and {Rhoads}, James E. and {Richards}, Gordon T. and {Ridgway}, Stephen T. and {Robertson}, Brant E. and {Roskar}, Rok and {Saha}, Abhijit and {Sarajedini}, Ata and {Scannapieco}, Evan and {Schalk}, Terry and {Schindler}, Rafe and {Schmidt}, Samuel},
        title = "{LSST Science Book, Version 2.0}",
      journal = {arXiv e-prints},
         year = 2009,
        month = dec,
          eid = {arXiv:0912.0201},
        pages = {arXiv:0912.0201},
          doi = {10.48550/arXiv.0912.0201},
archivePrefix = {arXiv},
       eprint = {0912.0201},
 primaryClass = {astro-ph.IM},
       adsurl = {https://ui.adsabs.harvard.edu/abs/2009arXiv0912.0201L}
}

@ARTICLE{Bilicki2021,
       author = {{Bilicki}, M. and {Dvornik}, A. and {Hoekstra}, H. and {Wright}, A.~H. and {Chisari}, N.~E. and {Vakili}, M. and {Asgari}, M. and {Giblin}, B. and {Heymans}, C. and {Hildebrandt}, H. and {Holwerda}, B.~W. and {Hopkins}, A. and {Johnston}, H. and {Kannawadi}, A. and {Kuijken}, K. and {Nakoneczny}, S.~J. and {Shan}, H.~Y. and {Sonnenfeld}, A. and {Valentijn}, E.},
        title = "{Bright galaxy sample in the Kilo-Degree Survey Data Release 4. Selection, photometric redshifts, and physical properties}",
      journal = {\aap},
         year = 2021,
        month = sep,
       volume = {653},
          eid = {A82},
        pages = {A82},
          doi = {10.1051/0004-6361/202140352},
archivePrefix = {arXiv},
       eprint = {2101.06010},
 primaryClass = {astro-ph.GA},
       adsurl = {https://ui.adsabs.harvard.edu/abs/2021A&A...653A..82B}
}

@ARTICLE{Abbott2022,
       author = {{Abbott}, T.~M.~C. and {Aguena}, M. and {Alarcon}, A. and {Allam}, S. and {Alves}, O. and {Amon}, A. and {Andrade-Oliveira}, F. and {Annis}, J. and {Avila}, S. and {Bacon}, D. and {Baxter}, E. and {Bechtol}, K. and {Becker}, M.~R. and {Bernstein}, G.~M. and {Bhargava}, S. and {Birrer}, S. and {Blazek}, J. and {Brandao-Souza}, A. and {Bridle}, S.~L. and {Brooks}, D. and {Buckley-Geer}, E. and {Burke}, D.~L. and {Camacho}, H. and {Campos}, A. and {Carnero Rosell}, A. and {Carrasco Kind}, M. and {Carretero}, J. and {Castander}, F.~J. and {Cawthon}, R. and {Chang}, C. and {Chen}, A. and {Chen}, R. and {Choi}, A. and {Conselice}, C. and {Cordero}, J. and {Costanzi}, M. and {Crocce}, M. and {da Costa}, L.~N. and {da Silva Pereira}, M.~E. and {Davis}, C. and {Davis}, T.~M. and {De Vicente}, J. and {DeRose}, J. and {Desai}, S. and {Di Valentino}, E. and {Diehl}, H.~T. and {Dietrich}, J.~P. and {Dodelson}, S. and {Doel}, P. and {Doux}, C. and {Drlica-Wagner}, A. and {Eckert}, K. and {Eifler}, T.~F. and {Elsner}, F. and {Elvin-Poole}, J. and {Everett}, S. and {Evrard}, A.~E. and {Fang}, X. and {Farahi}, A. and {Fernandez}, E. and {Ferrero}, I. and {Fert{\'e}}, A. and {Fosalba}, P. and {Friedrich}, O. and {Frieman}, J. and {Garc{\'\i}a-Bellido}, J. and {Gatti}, M. and {Gaztanaga}, E. and {Gerdes}, D.~W. and {Giannantonio}, T. and {Giannini}, G. and {Gruen}, D. and {Gruendl}, R.~A. and {Gschwend}, J. and {Gutierrez}, G. and {Harrison}, I. and {Hartley}, W.~G. and {Herner}, K. and {Hinton}, S.~R. and {Hollowood}, D.~L. and {Honscheid}, K. and {Hoyle}, B. and {Huff}, E.~M. and {Huterer}, D. and {Jain}, B. and {James}, D.~J. and {Jarvis}, M. and {Jeffrey}, N. and {Jeltema}, T. and {Kovacs}, A. and {Krause}, E. and {Kron}, R. and {Kuehn}, K. and {Kuropatkin}, N. and {Lahav}, O. and {Leget}, P.-F. and {Lemos}, P. and {Liddle}, A.~R. and {Lidman}, C. and {Lima}, M. and {Lin}, H. and {MacCrann}, N. and {Maia}, M.~A.~G. and {Marshall}, J.~L. and {Martini}, P. and {McCullough}, J. and {Melchior}, P. and {Mena-Fern{\'a}ndez}, J. and {Menanteau}, F. and {Miquel}, R. and {Mohr}, J.~J. and {Morgan}, R. and {Muir}, J. and {Myles}, J. and {Nadathur}, S. and {Navarro-Alsina}, A. and {Nichol}, R.~C. and {Ogando}, R.~L.~C. and {Omori}, Y. and {Palmese}, A. and {Pandey}, S. and {Park}, Y. and {Paz-Chinch{\'o}n}, F. and {Petravick}, D. and {Pieres}, A. and {Plazas Malag{\'o}n}, A.~A. and {Porredon}, A. and {Prat}, J. and {Raveri}, M. and {Rodriguez-Monroy}, M. and {Rollins}, R.~P. and {Romer}, A.~K. and {Roodman}, A. and {Rosenfeld}, R. and {Ross}, A.~J. and {Rykoff}, E.~S. and {Samuroff}, S. and {S{\'a}nchez}, C. and {Sanchez}, E. and {Sanchez}, J. and {Sanchez Cid}, D. and {Scarpine}, V. and {Schubnell}, M. and {Scolnic}, D. and {Secco}, L.~F. and {Serrano}, S. and {Sevilla-Noarbe}, I. and {Sheldon}, E. and {Shin}, T. and {Smith}, M. and {Soares-Santos}, M. and {Suchyta}, E. and {Swanson}, M.~E.~C. and {Tabbutt}, M. and {Tarle}, G. and {Thomas}, D. and {To}, C. and {Troja}, A. and {Troxel}, M.~A. and {Tucker}, D.~L. and {Tutusaus}, I. and {Varga}, T.~N. and {Walker}, A.~R. and {Weaverdyck}, N. and {Wechsler}, R. and {Weller}, J. and {Yanny}, B. and {Yin}, B. and {Zhang}, Y. and {Zuntz}, J. and {DES Collaboration}},
        title = "{Dark Energy Survey Year 3 results: Cosmological constraints from galaxy clustering and weak lensing}",
      journal = {\prd},
         year = 2022,
        month = jan,
       volume = {105},
       number = {2},
          eid = {023520},
        pages = {023520},
          doi = {10.1103/PhysRevD.105.023520},
archivePrefix = {arXiv},
       eprint = {2105.13549},
 primaryClass = {astro-ph.CO},
       adsurl = {https://ui.adsabs.harvard.edu/abs/2022PhRvD.105b3520A}
}

@article{scikit-learn,
  title={Scikit-learn: Machine Learning in {P}ython},
  author={Pedregosa, F. and Varoquaux, G. and Gramfort, A. and Michel, V.
          and Thirion, B. and Grisel, O. and Blondel, M. and Prettenhofer, P.
          and Weiss, R. and Dubourg, V. and Vanderplas, J. and Passos, A. and
          Cournapeau, D. and Brucher, M. and Perrot, M. and Duchesnay, E.},
  journal={Journal of Machine Learning Research},
  volume={12},
  pages={2825--2830},
  year={2011}
}

@ARTICLE{Seitz1995,
       author = {{Seitz}, C. and {Schneider}, P.},
        title = "{Steps towards nonlinear cluster inversion through gravitational distortions II. Generalization of the Kaiser and Squires method.}",
      journal = {\aap},
         year = 1995,
        month = may,
       volume = {297},
        pages = {287},
          doi = {10.48550/arXiv.astro-ph/9408050},
archivePrefix = {arXiv},
       eprint = {astro-ph/9408050},
 primaryClass = {astro-ph},
       adsurl = {https://ui.adsabs.harvard.edu/abs/1995A&A...297..287S}
}

@ARTICLE{Rykoff2014,
       author = {{Rykoff}, E.~S. and {Rozo}, E. and {Busha}, M.~T. and {Cunha}, C.~E. and {Finoguenov}, A. and {Evrard}, A. and {Hao}, J. and {Koester}, B.~P. and {Leauthaud}, A. and {Nord}, B. and {Pierre}, M. and {Reddick}, R. and {Sadibekova}, T. and {Sheldon}, E.~S. and {Wechsler}, R.~H.},
        title = "{redMaPPer. I. Algorithm and SDSS DR8 Catalog}",
      journal = {\apj},
         year = 2014,
        month = apr,
       volume = {785},
       number = {2},
          eid = {104},
        pages = {104},
          doi = {10.1088/0004-637X/785/2/104},
archivePrefix = {arXiv},
       eprint = {1303.3562},
 primaryClass = {astro-ph.CO},
       adsurl = {https://ui.adsabs.harvard.edu/abs/2014ApJ...785..104R}
}

@ARTICLE{Chen2025,
       author = {{Chen}, Kai-Feng and {Chiu}, I.-Non and {Oguri}, Masamune and {Lin}, Yen-Ting and {Miyatake}, Hironao and {Miyazaki}, Satoshi and {More}, Surhud and {Hamana}, Takashi and {Rau}, Markus M. and {Sunayama}, Tomomi and {Sugiyama}, Sunao and {Takada}, Masahiro},
        title = "{Weak-Lensing Shear-Selected Galaxy Clusters from the Hyper Suprime-Cam Subaru Strategic Program: I. Cluster Catalog, Selection Function and Mass─Observable Relation}",
      journal = {The Open Journal of Astrophysics},
         year = 2025,
        month = jan,
       volume = {8},
          eid = {2},
        pages = {2},
          doi = {10.33232/001c.128019},
archivePrefix = {arXiv},
       eprint = {2406.11966},
 primaryClass = {astro-ph.CO},
       adsurl = {https://ui.adsabs.harvard.edu/abs/2025OJAp....8E...2C}
}

@ARTICLE{Chiu2024,
       author = {{Chiu}, I.-Non and {Chen}, Kai-Feng and {Oguri}, Masamune and {Rau}, Markus M. and {Hamana}, Takashi and {Lin}, Yen-Ting and {Miyatake}, Hironao and {Miyazaki}, Satoshi and {More}, Surhud and {Sunayama}, Tomomi and {Sugiyama}, Sunao and {Takada}, Masahiro},
        title = "{Weak-Lensing Shear-Selected Galaxy Clusters from the Hyper Suprime-Cam Subaru Strategic Program: II. Cosmological Constraints from the Cluster Abundance}",
      journal = {The Open Journal of Astrophysics},
         year = 2024,
        month = oct,
       volume = {7},
          eid = {90},
        pages = {90},
          doi = {10.33232/001c.124537},
archivePrefix = {arXiv},
       eprint = {2406.11970},
 primaryClass = {astro-ph.CO},
       adsurl = {https://ui.adsabs.harvard.edu/abs/2024OJAp....7E..90C}
}

@ARTICLE{Williamson2026,
       author = {{Williamson}, J. and {Makinen}, T.~L. and {Porqueres}, N. and {Jeffrey}, N. and {Heavens}, A. and {Gatti}, M. and {Wandelt}, B.~D. and {Whiteway}, L. and {Prat}, J. and {Alarcon}, A. and {Amon}, A. and {Bechtol}, K. and {Becker}, M.~R. and {Bernstein}, G.~M. and {Campos}, A. and {Carnero Rosell}, A. and {Chen}, R. and {Choi}, A. and {DeRose}, J. and {Doux}, C. and {Drlica-Wagner}, A. and {Eckert}, K. and {Everett}, S. and {Fert{\'e}}, A. and {Gong}, Z. and {Gruen}, D. and {Gruendl}, R.~A. and {Herner}, K. and {Jarvis}, M. and {Kacprzak}, T. and {Lahav}, O. and {McCullough}, J. and {Myles}, J. and {Navarro-Alsina}, A. and {Pandey}, S. and {Raveri}, M. and {Rollins}, R.~P. and {Rykoff}, E.~S. and {S{\'a}nchez}, C. and {Secco}, L.~F. and {Sevilla-Noarbe}, I. and {Sheldon}, E. and {Shin}, T. and {Thomsen}, A. and {Troxel}, M.~A. and {Tutusaus}, I. and {Varga}, T.~N. and {Yanny}, B. and {Yin}, B. and {Zuntz}, J. and {Abbott}, T.~M.~C. and {Aguena}, M. and {Andrade-Oliveira}, F. and {Brooks}, D. and {Camilleri}, R. and {Carretero}, J. and {Cawthon}, R. and {Crocce}, M. and {da Costa}, L.~N. and {Davis}, T.~M. and {De Vicente}, J. and {Desai}, S. and {Diehl}, H.~T. and {Flaugher}, B. and {Frieman}, J. and {Garc{\'\i}a-Bellido}, J. and {Gutierrez}, G. and {Hinton}, S.~R. and {Hollowood}, D.~L. and {Kuehn}, K. and {Marshall}, J.~L. and {Mena-Fern{\'a}ndez}, J. and {Miquel}, R. and {Mohr}, J.~J. and {Muir}, J. and {Plazas Malag{\'o}n}, A.~A. and {Porredon}, A. and {Roodman}, A. and {Sanchez}, E. and {Sanchez Cid}, D. and {Suchyta}, E. and {Swanson}, M.~E.~C. and {To}, C. and {Tucker}, D.~L. and {Vikram}, V. and {Walker}, A.~R. and {Weaverdyck}, N. and {Weller}, J.},
        title = "{Dark Energy Survey Year 3 results: optimized $w$CDM simulation-based inference with weak lensing map-level hybrid statistics}",
      journal = {arXiv e-prints},
         year = 2026,
        month = jun,
          eid = {arXiv:2606.11309},
        pages = {arXiv:2606.11309},
          doi = {10.48550/arXiv.2606.11309},
archivePrefix = {arXiv},
       eprint = {2606.11309},
 primaryClass = {astro-ph.CO},
       adsurl = {https://ui.adsabs.harvard.edu/abs/2026arXiv260611309W}
}

@ARTICLE{Linke2026,
       author = {{Linke}, L. and {Porth}, L. and {Burger}, P. and {Harnois-D{\'e}raps}, J. and {Heydenreich}, S. and {Schneider}, P. and {Asgari}, M. and {Bilicki}, M. and {Georgiou}, C. and {Heymans}, C. and {Hildebrandt}, H. and {Hoekstra}, H. and {Jalan}, P. and {Joachimi}, B. and {Joudaki}, S. and {Kuijken}, K. and {Li}, S. and {Moscardini}, L. and {Radovich}, M. and {Reischke}, R. and {St{\"o}lzner}, B. and {Wright}, A.~H. and {Yan}, Z. and {Zhang}, Y.-H.},
        title = "{KiDS-Legacy: Joint analysis of second- and third-order cosmic shear}",
      journal = {arXiv e-prints},
         year = 2026,
        month = jun,
          eid = {arXiv:2606.12389},
        pages = {arXiv:2606.12389},
          doi = {10.48550/arXiv.2606.12389},
archivePrefix = {arXiv},
       eprint = {2606.12389},
 primaryClass = {astro-ph.CO},
       adsurl = {https://ui.adsabs.harvard.edu/abs/2026arXiv260612389L}
}

@ARTICLE{Marini2024,
       author = {{Marini}, I. and {Popesso}, P. and {Lamer}, G. and {Dolag}, K. and {Biffi}, V. and {Vladutescu-Zopp}, S. and {Dev}, A. and {Toptun}, V. and {Bulbul}, E. and {Comparat}, J. and {Malavasi}, N. and {Merloni}, A. and {Mroczkowski}, T. and {Ponti}, G. and {Seppi}, R. and {Shreeram}, S. and {Zhang}, Y.},
        title = "{Detecting galaxy groups populating the local Universe in the eROSITA era}",
      journal = {\aap},
         year = 2024,
        month = sep,
       volume = {689},
          eid = {A7},
        pages = {A7},
          doi = {10.1051/0004-6361/202450442},
archivePrefix = {arXiv},
       eprint = {2404.12719},
 primaryClass = {astro-ph.CO},
       adsurl = {https://ui.adsabs.harvard.edu/abs/2024A&A...689A...7M}
}

@ARTICLE{Limber1953,
       author = {{Limber}, D. Nelson},
        title = "{The Analysis of Counts of the Extragalactic Nebulae in Terms of a Fluctuating Density Field.}",
      journal = {\apj},
         year = 1953,
        month = jan,
       volume = {117},
        pages = {134},
          doi = {10.1086/145672},
       adsurl = {https://ui.adsabs.harvard.edu/abs/1953ApJ...117..134L}
}

@ARTICLE{Kodwani2019,
       author = {{Kodwani}, Darsh and {Alonso}, David and {Ferreira}, Pedro},
        title = "{The effect on cosmological parameter estimation of a parameter dependent covariance matrix}",
      journal = {The Open Journal of Astrophysics},
         year = 2019,
        month = mar,
       volume = {2},
       number = {1},
          eid = {3},
        pages = {3},
          doi = {10.21105/astro.1811.11584},
archivePrefix = {arXiv},
       eprint = {1811.11584},
 primaryClass = {astro-ph.CO},
       adsurl = {https://ui.adsabs.harvard.edu/abs/2019OJAp....2E...3K}
}



\appendix

\section{Emulator construction and validation}\label{app:emulator}

\begin{figure*}
  \centering
  \includegraphics[width=\textwidth]{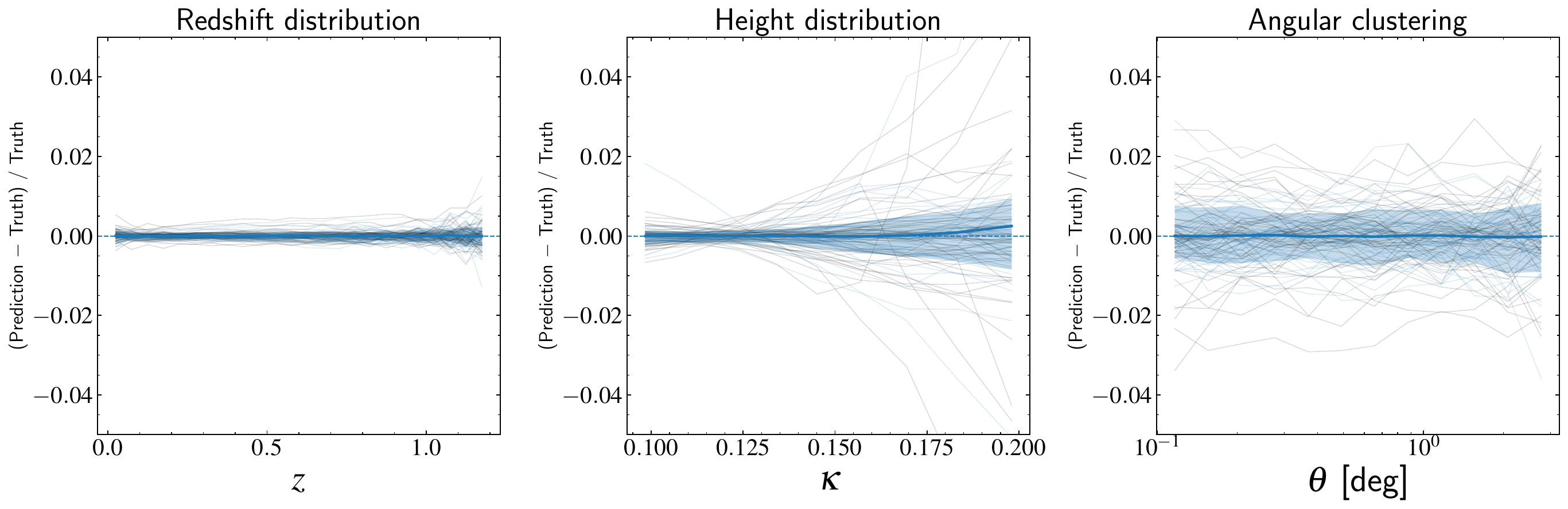}
  \caption{Leave-one-out emulator test showing the fractional accuracy of the predicted statistics by training on all but one hypercube node and predicting the statistics at the cosmology of the final hypercube node for the SNR $>3$ WL peak redshift distribution (left), height distribution (middle) and angular clustering (right), corresponding to the statistics shown in Fig.~\ref{fig:fiducial_statistics}. The emulators show per cent-level accuracy across the complete ranges for which the statistics are used in the forecast.}
  \label{fig:loo_test}
\end{figure*}

In this appendix, we provide additional details on the construction and accuracy of the emulators used for the forecasting. We use Gaussian process (GP) emulators that are built using \textsc{scikit-learn}'s \texttt{GaussianProcessRegressor} class \citep{scikit-learn}. All emulators use a Mat\'{e}rn-$\tfrac{3}{2}$ kernel with a trainable amplitude \citep[see e.g.][]{Rasmussen2006}. We add a white-noise kernel to absorb stochasticity, as can be seen in, for example, the high-$\kappa$ tail of the peak height distributions in Fig.~\ref{fig:fiducial_statistics}.

The kernel hyperparameters are optimised by maximising the GP log-marginal likelihood, with the optimisation repeated 10 times with random restarts to avoid local maxima. The different WL peak summary statistics are emulated separately at each smoothing scale, and the resulting emulators are concatenated before forecasting. The different statistics and their corresponding emulator architectures are described below.

\subsection{Redshift distribution}
SNR > 3 peaks are selected and binned in redshift using 25 linearly spaced bins over $z \in [0,\,1.25]$, with bin widths normalised to yield the density $\mathrm{d}N/\mathrm{d}z$. We chose this SNR cut because in \citet{Broxterman2025_WLpeaks}, we showed WL peaks above the corresponding $\kappa$-cut can be robustly assigned to a single massive object. Also, lower-valued peaks are more sensitive to the noise and instrument systematics, as discussed in Appendix~B of \citet{Broxteman2024_WLpeaks}. In this way, we aim at a potentially robust cosmological inference. Note, however, that we apply stronger noise, so the SNR differs. The binning corresponds to the FLAMINGO light-cone redshift resolution and is similar to the expected redshift uncertainty on the massive objects that generate the high-valued WL peaks. For additional details, see the discussion in Section~3.5 of \citet{Broxterman2025_WLpeaks}. Before emulating the distributions, we apply a logarithmic transform $\ln(\mathrm{d}N/\mathrm{d}z + \epsilon)$ (with $\epsilon = 10^{-12}$) to stabilise the variance at low counts. The data vectors are then emulated using principal component analysis (PCA) separately for the shape of the redshift distribution and the normalisation, i.e., the total number of peaks above the SNR > 3 threshold. We emulate the first 15 PCA components. When predicting the redshift distribution, the shape is reconstructed from an inverse PCA transform and multiplied by the amplitude prediction to yield the full data vector.

\subsection{Height distribution}
Peaks above the SNR threshold are binned into 10 logarithmically spaced SNR bins, with the upper bin edge set at the 99th percentile of the $\kappa$ distribution across all training simulations. We do this because preliminary tests showed it yielded better predictions than including all high-end $\kappa$ tail peaks, likely due to cosmic variance. At the same time, we do not expect these peaks to carry much of the cosmological information, given the relative increase in cosmic variance, as shown in the spread among the covariance matrix realisations in Fig.~\ref{fig:fiducial_statistics}. Nevertheless, this may be improved in future work. We again normalise the counts to a density $n(\nu) = \mathrm{d}N/\mathrm{d}\kappa$, and apply a logarithmic transform $\ln[n(\nu) + \epsilon]$ before emulation. For the height distribution, we use the same 15-component PCA + amplitude GPE as described for the peak redshift distribution.

\subsection{Angular clustering}
The angular clustering of the peaks is measured in 12 logarithmically-spaced bins between $\theta =  0.1-3.0\,\mathrm{deg}$. We chose this range because it corresponds to the angular range we also use for our $\xi_{+/-}$ 2PCFs. To reduce the dynamic range and improve the emulation accuracy, we transform the correlation function distributions as $\tilde{\omega}(\theta) =\ln [ \theta\,\omega(\theta) ]$. For $\tilde{\omega}(\theta)$, we adopt a similar PCA GP emulator.

\subsection{Leave-one-out test}
We assess the emulator accuracy by performing a leave-one-out (LOO) test. Each of the $2\times50$ training simulations is withheld in turn, and an emulator is trained on the remaining 99 nodes. We then evaluate the prediction at the cosmology of the remaining node and compare it to the actual measurements. For the fiducial forecast setup considered in this work, Fig.~\ref{fig:loo_test} shows the emulator's fractional accuracy on the log-transformed quantities. The left panel shows the redshift distribution, the central panel the height distribution, and the right panel the angular clustering. For each statistic, averaged over the entire emulated range, the emulator typically has an accuracy of within 1 per cent. The mean 1$\sigma$ accuracy averaged across all bins is 0.09, 0.40, and 0.58 per cent for the redshift distribution, height distribution, and angular clustering, respectively. Additionally, for the clustering, although the simulations are predicted with per-cent-level accuracy on average, there are individual simulations in which the angular clustering is mispredicted by up to $\approx3$ per cent across all angular scales.

We stress that we designed the emulators to perform well in the fiducial forecast setup, i.e., with 1-arcmin smoothing while inferring all 10 cosmological parameters simultaneously. We did not change the architecture for the 0.5- or 2.0-arcmin smoothing tests. Inspection of the emulator LOO accuracy tests shows that the emulators do not perform much worse at those smoothing scales.

\section{Correlation matrix}\label{app:corr_matrix}
\begin{figure}
  \centering
  \includegraphics[width=\hsize]{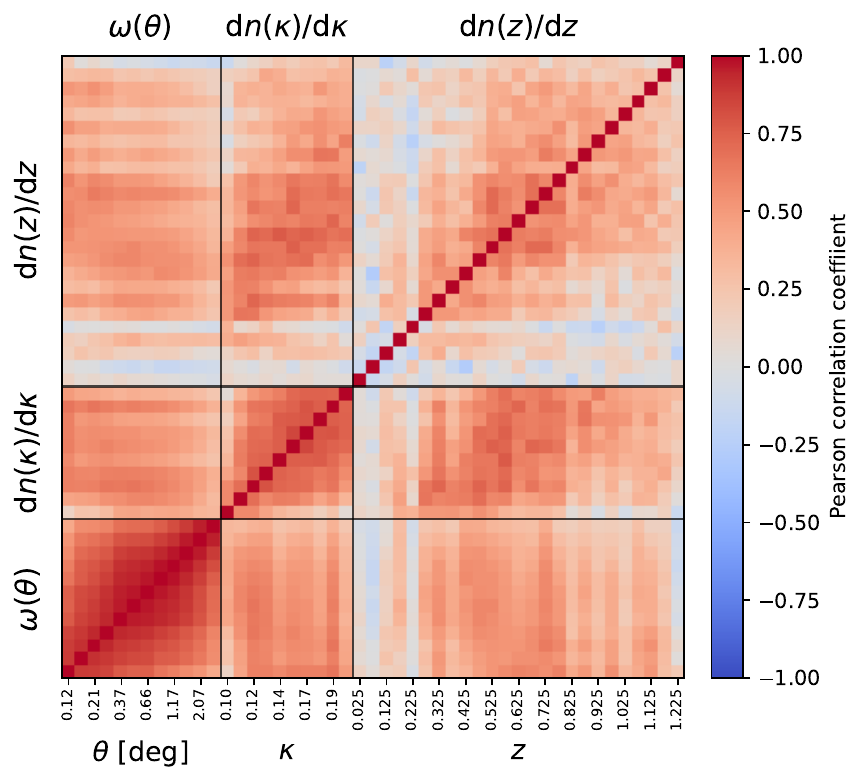}
  \caption{Matrix of Pearson correlation coefficients for the inference for the fiducial inference setup. The solid black curves separate the bins corresponding to the different WL peak statistics. Bottom left: angular clustering of SNR > 3 peaks. Centre middle: height distribution. Top right: Redshift distribution of SNR > 3 WL peaks. The different clustering bins are strongly correlated, whereas the other statistics and cross-correlations show weaker correlations.}
  \label{fig:corr_mat}
\end{figure}

In this appendix, we present the correlation matrix for our fiducial forecast setup. Fig.~\ref{fig:corr_mat} shows the Pearson correlation coefficient values, $\rho_{i,j} = C_{ij} / (\sigma_i \sigma_j)$, where  $C_{ij}$ is the $i,j$-th component of the covariance matrix and $\sigma_i$ and $\sigma_j$ denote the standard deviation of the $i$-th and $j$-th component, respectively, for the forecast corresponding to the black contours in Figs.~\ref{fig:1d_1ss_2stat}~\&~\ref{fig:corner_1ss_peaks}. The auto- and cross-correlation blocks corresponding to separate (combinations of) statistics are separated by solid black lines. From bottom left to top right, the figure shows the clustering, $\omega(\theta)$, height distribution, $\mathrm{d}n(\kappa)/\mathrm{d}\kappa$, and redshift distribution, $\mathrm{d}n(z)/\mathrm{d}z$. The x-axis indicates the angular, WL convergence, or redshift bin values. The clustering shows strong correlations, particularly between individual low-angular-separation bins and between individual high-separation bins, whereas the other statistics and cross-correlations show milder correlations, though there is still typically some moderate correlation.

\section{Cosmological constraints from WL peaks}\label{app:corner_peaks}
\begin{figure*}
  \centering
  \includegraphics[width=\textwidth]{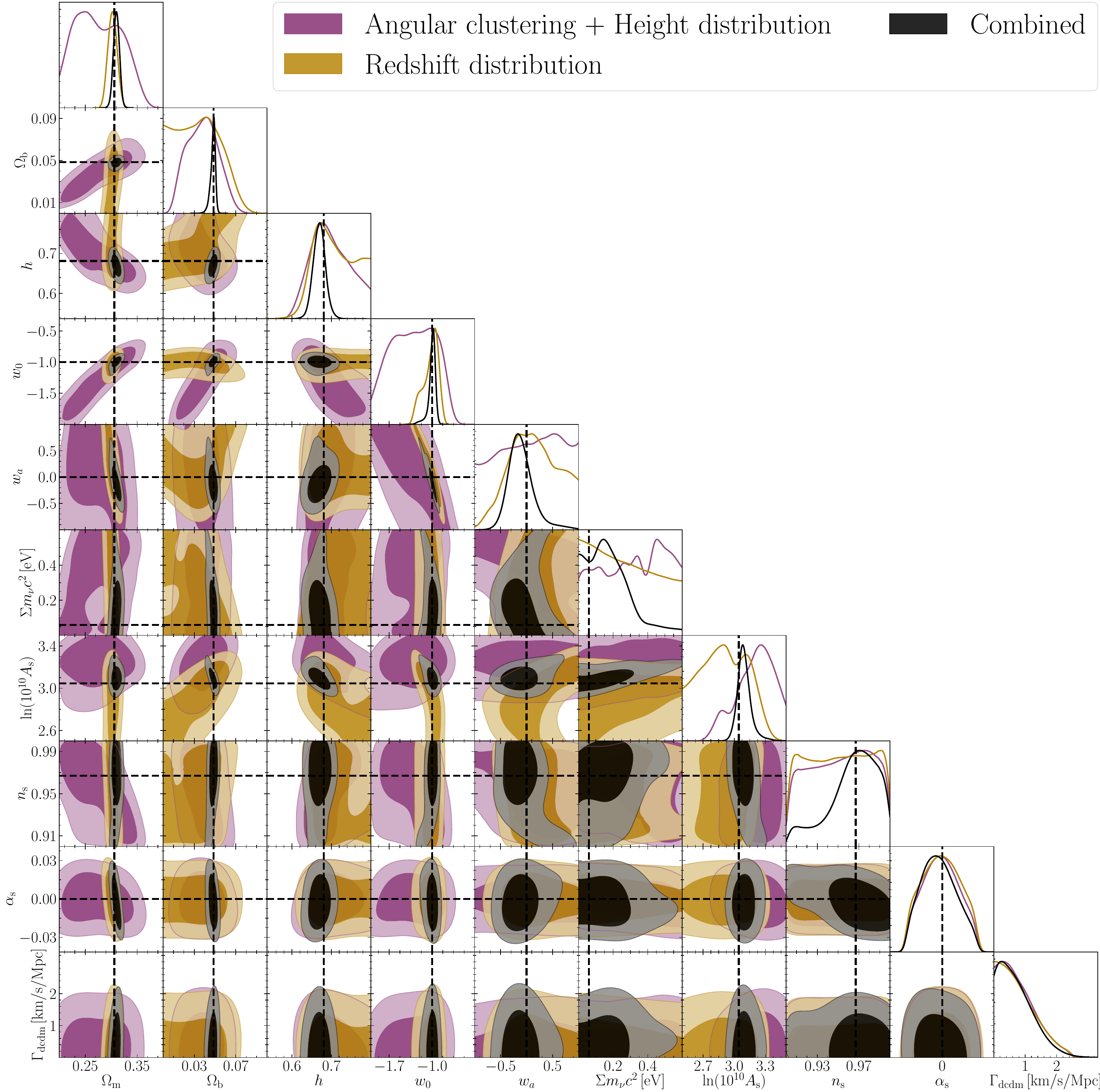}
  \caption{Corner plot of the cosmological parameter forecast using the angular clustering and height distributions (purple), redshift distribution (orange) or the combination of the three statistics (black) of WL peaks. The contours indicate the 1- and 2$\sigma$ credible intervals. Different statistics are sensitive to different parameters, with the redshift distribution being most sensitive to $\Omega_\mathrm{m}$, $w_0$, $w_a$, and the angular clustering and height distribution jointly probing $\ln(10^{10}A_\mathrm{s})$. Additionally, due to the breaking of degeneracies, the combination of statistics is also able to probe $\Omega_\mathrm{b}$ and $h$, whereas $\sum m_\nu c^2$, $n_\mathrm{s}$, $\alpha_\mathrm{s}$, and $\Gamma_\mathrm{dcdm}$ remain unconstrained.}
  \label{fig:corner_1ss_peaks}
\end{figure*}

In this appendix, we present the full corner plot of the 2D posteriors for the analysis of the three different WL peak statistics jointly, the redshift distribution individually and the angular clustering and height distribution combined, with 1 arcmin smoothing, whose individual 1D parameter posteriors were already shown in Fig.~\ref{fig:1d_1ss_2stat}. Fig.~\ref{fig:corner_1ss_peaks} shows the posteriors corresponding to the forecast using the redshift distribution (orange), height distribution and angular clustering (purple) or all three distributions (black).

Individually, the WL peak statistics cannot provide strong constraints over the entire 10D parameter space. $\Omega_\mathrm{m}$, $w_0$, and $w_a$ are most tightly constrained by the redshift distribution. The peak height distribution and angular clustering constrain $\ln(10^{10} A_\mathrm{s})$ best, whereas $\Omega_\mathrm{b}$ and $h$ are primarily constrained due to breaking of degeneracies in the combined forecast, as can be seen in the $\Omega_\mathrm{m}-\Omega_\mathrm{b}$ and $\Omega_\mathrm{m}-h$ panels. $\Sigma m_\nu c^2,  n_\mathrm{s}, \alpha_\mathrm{s},$ and $\Gamma_{\mathrm{dcdm}}$ remain unconstrained or recover the imposed prior even in the combined case.

The corner plot illustrates the complementarity of combining the different statistics to break degeneracies and obtain tighter constraints. In particular, in the $\Omega_\mathrm{m}$--$\Omega_\mathrm{b}$ plane, no individual statistic can simultaneously constrain both parameters, whereas their combination yields a clearly localised posterior, which seems to result from the combination of the angular clustering and the redshift distribution.

Some of the 2D contours, particularly when parameters are not well constrained, show multimodal posteriors, which may indicate an insufficiently regular emulator. This particularly holds for the angular clustering and height distribution combination, as for instance seen in the $\Omega_\mathrm{m}-\Sigma m_\nu c^2$, where Appendix~\ref{app:emulator} already showed that individual nodes can be misestimated on average by several per cent. This behaviour could indicate that improvements in emulation or additional hypercube nodes are necessary to obtain accurate predictions across the entire 10D cosmological parameter space.

\section{Impact of individual cosmological parameters on the WL peak statistics}\label{app:param_sweep}
\begin{figure*}
  \centering
  \includegraphics[width=\textwidth]{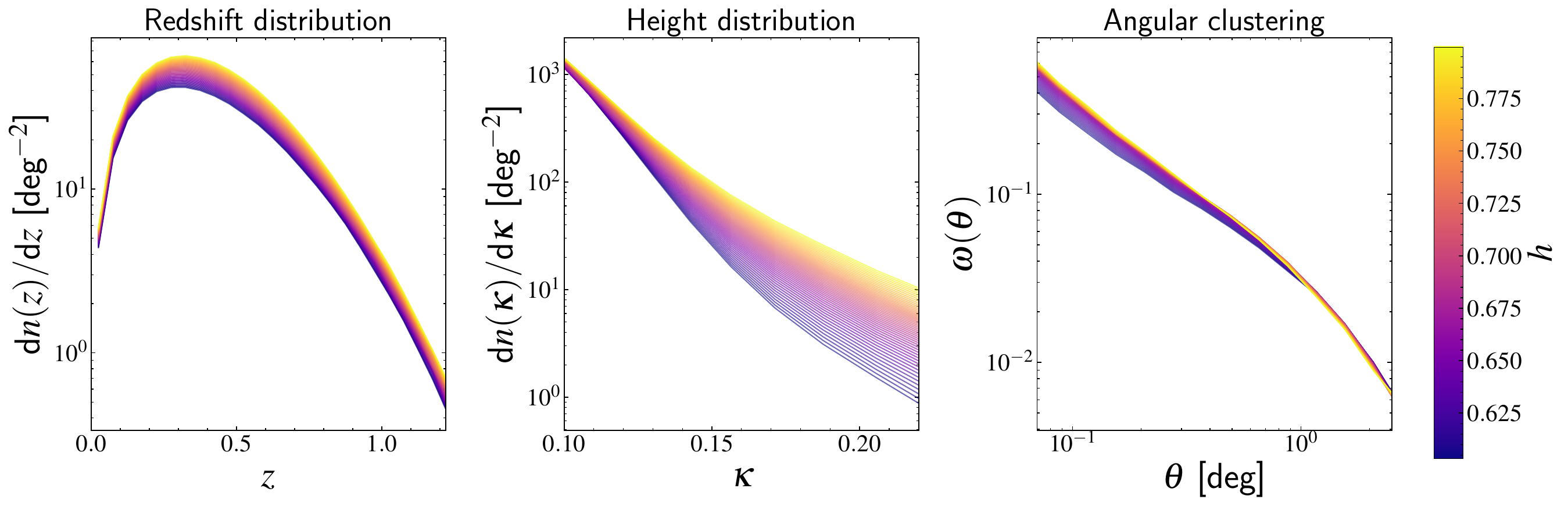}
  \caption{Same as Fig.~\ref{fig:sweep_omegam} but varying $h$. Larger values of $h$ generally enhance the amplitude of each of the three signals. Additionally, the height distribution is enhanced preferentially at large $\kappa$ and the clustering signal at small $\theta$.}
  \label{fig:sweep_h}
\end{figure*}

\begin{figure*}
  \centering
  \includegraphics[width=\textwidth]{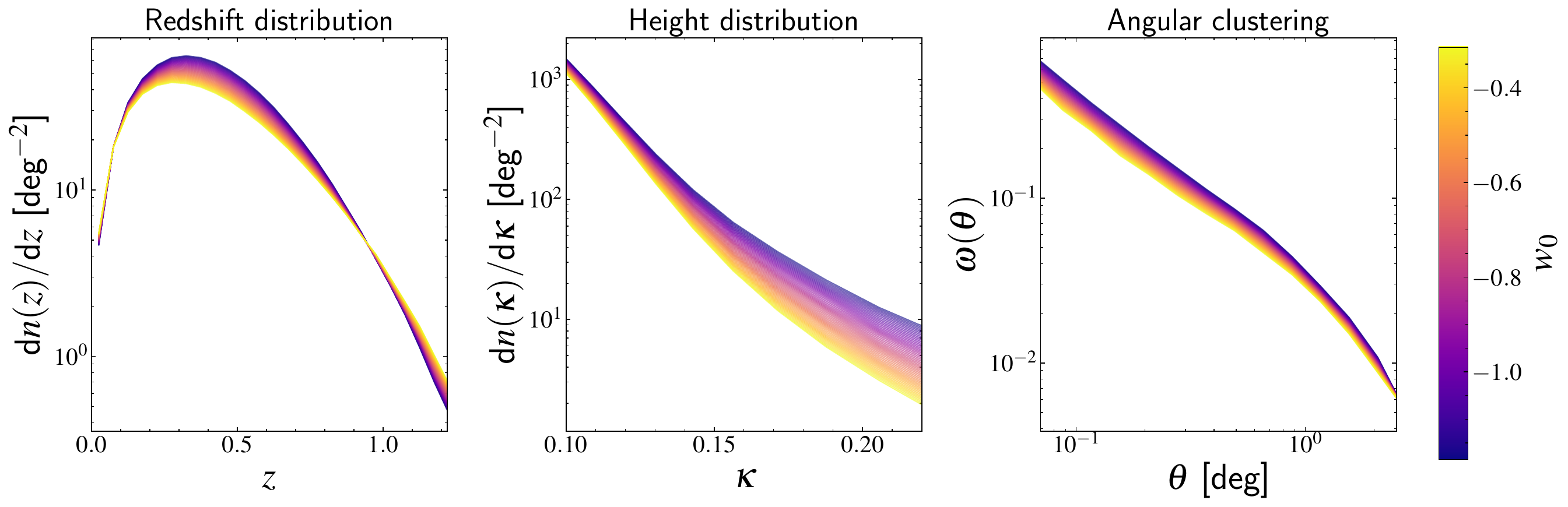}
  \caption{Same as Fig.~\ref{fig:sweep_omegam} but varying $w_0$. Stronger evolving DE, corresponding to less negative values of $w_0$, suppresses each of the considered statistics, with the suppression most pronounced at low redshift, high $\kappa$ and small $\theta$.}
  \label{fig:sweep_w0}
\end{figure*}

\begin{figure*}
  \centering
  \includegraphics[width=\textwidth]{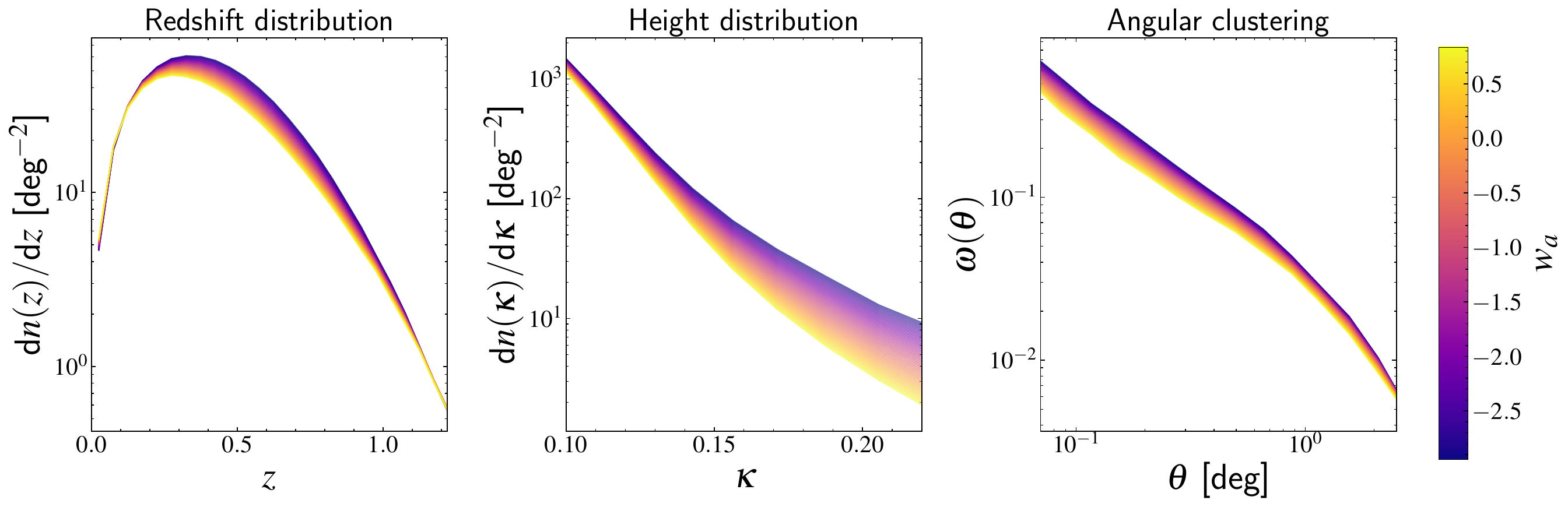}
  \caption{Same as Fig.~\ref{fig:sweep_omegam} but varying $w_a$. The impact of varying $w_a$ is very similar to that of varying $w_0$ as shown in Fig.~\ref{fig:sweep_w0}.}
  \label{fig:sweep_wa}
\end{figure*}

\begin{figure*}
  \centering
  \includegraphics[width=\textwidth]{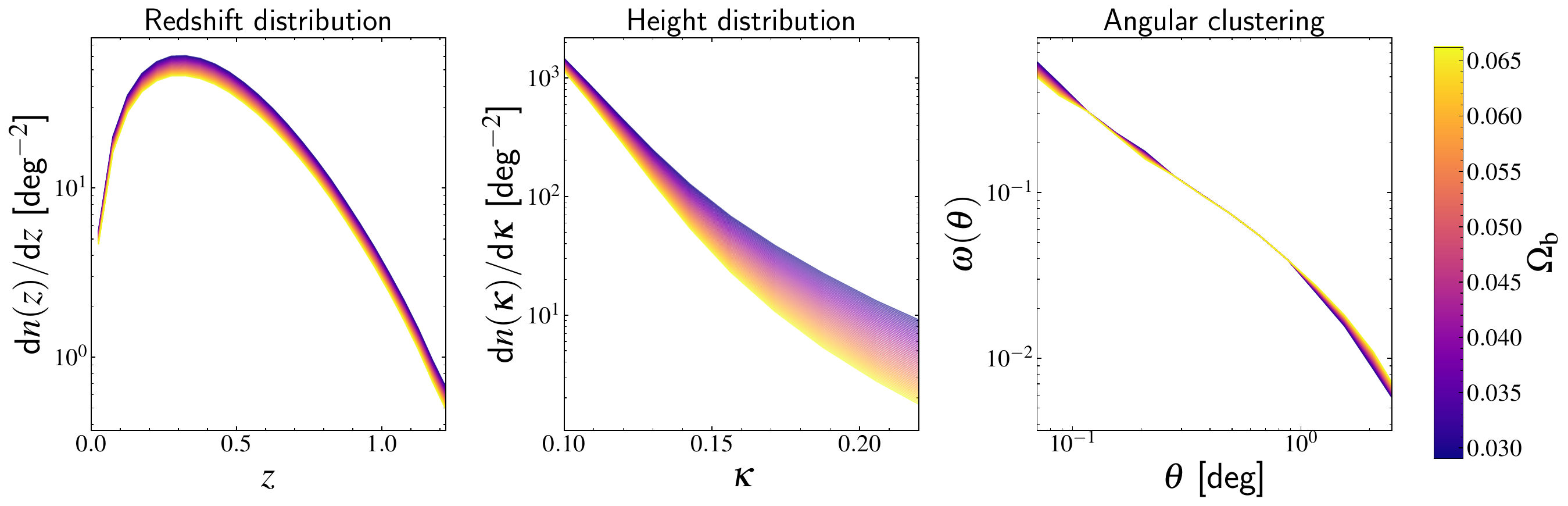}
  \caption{Same as Fig.~\ref{fig:sweep_omegam} but varying $\Omega_\mathrm{b}$. The impact of varying $\Omega_\mathrm{b}$ is most distinct on the height distribution, where lower values of $\Omega_\mathrm{b}$ lead to more higher-valued WL peaks at fixed $\kappa$.}
  \label{fig:sweep_ob}
\end{figure*}

In this appendix, we show the impact of additional individual cosmological parameters on the WL peak statistics using emulator predictions, analogous to the distributions shown for varying $\Omega_\mathrm{m}$ in Fig.~\ref{fig:sweep_omegam}. Nine out of the ten parameters that are input to the emulator are set to their mean value, and the tenth parameter is varied within its range within the hypercube. Figs.~\ref{fig:sweep_h},~\ref{fig:sweep_w0},~\ref{fig:sweep_wa},~and~\ref{fig:sweep_ob}, respectively, show the redshift distribution (left), height distribution (centre) and angular clustering (right) for varying $h$, $w_0$, $w_a$, and $\Omega_\mathrm{b}$ as indicated by the colour bar. Individually varying all remaining hypercube parameters results in smaller variations than seen for these parameters.

Varying $h$ affects each of the WL peak statistics. The amplitude of the redshift distribution has a positive correlation with $h$. The height distribution is more sensitive to $h$, as larger values of $h$ systematically increase the abundance of peaks at fixed $\kappa$, with larger $\kappa$ values being preferred for larger $h$ values. For the angular clustering distributions, larger $h$ increases the signal for $\theta < 1$~deg.

Stronger evolving DE, with larger values of $w_0$ or $w_a$, leads to a suppression of each of the three WL peak statistics. The behaviour of $w_0$ and $w_a$ is very similar across the three statistics. Only above $z\approx1$ do larger values of $w_0$ lead to a small enhancement of the number of peaks, whereas for the $w_a$ values considered, there is only a suppression. 

Varying $\Omega_\mathrm{b}$ only has a minor impact on the redshift distribution and angular clustering, whereas the height distribution shows a stronger dependence, with higher values of $\Omega_\mathrm{b}$ leading to a suppression of the number of high-valued peaks.

\bsp	
\label{lastpage}
\end{document}